\documentclass{aa}  

\usepackage{graphicx}
\usepackage{txfonts}
\usepackage{xcolor}
\usepackage{tabularx}
\usepackage{appendix,natbib}
\usepackage{amsmath}
\usepackage{subfigure}
\usepackage[breaklinks=true]{hyperref} 

\usepackage{pifont}
\def\apjl{ApJL}

\def\aap{A\&A}

\def\apjs{ApJS}

\def \dm{dark matter}

\def \RC{rotation curve}
\def \RCs{rotation curves}
\def \sfgs {star-forming galaxies}

\def\hz{high-$z \ $}

\begin{document} 
	\title{Dark matter fraction in disk-like galaxies over the past 10 Gyr 
	}

	\author{Gauri Sharma, 
		\inst{1,2,3,4,5,6} \fnmsep\thanks{Contact: gsharma@unistra.fr; gsharma@uwc.ac.za;}
		\and  Glenn van de Ven\inst{7}
		\and  Paolo Salucci\inst{2,3,4}
		\and Marco Martorano \inst{8}
	}
	
	\institute{Department of Physics and Astronomy, University of the Western Cape, Cape Town 7535, South Africa
		\and
		SISSA International School for Advanced Studies, Via Bonomea 265, I-34136 Trieste, Italy
		\and
		INFN-Sezione di Trieste, via Valerio 2, I-34127 Trieste, Italy
		\and
		IFPU Institute for Fundamental Physics of the Universe, Via Beirut, 2, 34151 Trieste, Italy
		\and 
		University of Strasbourg, CNRS UMR 7550, Observatoire astronomique de Strasbourg, F-67000 Strasbourg, France      
		\and
		University of Strasbourg Institute for Advanced Study, 5 all\'ee du G\'en\'eral Rouvillois, F-67083 Strasbourg, France
		\and
		Department of Astrophysics, University of Vienna, T\"urkenschanzstra{\ss}e 17, 1180 Vienna, Austria       
		\and 
		Sterrenkundig Observatorium, Universiteit Gent, Krijgslaan 281 S9, 9000 Gent, Belgium
	}
	
	\date{Received 10/10/2023; accepted 10/05/2025}

	\abstract
	{We present an observational study of the dark matter fraction in star-forming disk-like galaxies up to redshift $z \sim 2.5$ selected from publicly available integral field spectroscopic surveys: KMOS$^{\rm 3D}$, KGES, and KROSS. To model the $H\alpha$ kinematics of these galaxies, we employed 3D forward modeling, which incorporates beam-smearing and inclination corrections and yields rotation curves. Subsequently, we corrected these  rotation curves for gas pressure gradients, resulting in circular velocity curves or `intrinsic' rotation curves. Our final sample comprises 263 rotationally supported galaxies with redshifts ranging from $0.6 \leq z < 2.5$, stellar masses within the range $9.0 \leq \log(M_{\rm star}\ [\mathrm{M_\odot}]) < 11.5$, and star formation rates between $0.49 \leq \log\left(\mathrm{SFR}\ [\mathrm{M_{\odot}\ yr^{-1}}]\right) \leq 2.5$. We estimated the dark matter fraction of these galaxies by subtracting the baryonic mass from the total mass, where the total mass is derived from the intrinsic rotation curves. We provide novel observational evidence suggesting that at a fixed redshift, the dark matter fraction gradually increases with radius such that the outskirts of galaxies are dark matter dominated, similarly to local star-forming disk galaxies. This observed dark matter fraction exhibits a decreasing trend with increasing redshift, and on average, the fraction within the effective radius (up to the outskirts) remains above 50\%, similar to the galaxies in the local Universe. We investigated the relationships between dark matter, baryon surface density, and the circular velocity of galaxies. We observed that low stellar mass galaxies, with $\log(M_{\rm star}\ [\mathrm{M_\odot}]) \leq 10.0$, undergo a higher degree of evolution, which may be attributed to the hierarchical merging of galaxies. We discuss several sources of uncertainties and current limitations in the field as well as their impact on the measurements of the dark matter fraction and its trend across galactic scales and cosmic time.
	}
	
	\maketitle
	%

			\section{Introduction}
			\label{sec:Intro}

			Within the current standard model of structure formation, dark matter (DM), an enigmatic form of matter devoid of any electromagnetic interaction, is believed to represent most of the matter in the Universe \citep[e.g.,][]{Peebles1993}. Despite its abundance, DM remains elusive and challenging to detect directly. Its existence is essentially inferred from the gravitational attraction it exerts on visible matter. In particular, flat outer galactic rotation curves have historically served as compelling evidence for the discrepancy between observed dynamics and the amount of baryons, leading to the hypothesis that galaxies are embedded within vast, extended halos of DM that largely surpass their luminous components in mass \citep[e.g., ][and references therein]{rubin1980, bosma1981, PS2019}.  Within the standard cold DM framework, cosmological simulations have corroborated that DM halos provide a gravitational scaffolding that allows galaxies to form and maintain their structures \citep[][and references therein]{Sims_review_2022}, although so-called small-scale challenges remain (e.g., \citealt{Bullock2017}, \citealt{Sales2022}).
			
			The amount of DM within galaxies varies depending on their baryonic mass, size, and environment, underscoring its critical influence on their overall dynamics and evolution. Observations of nearby star-forming galaxies indeed yield a DM fraction between 40\% and 50\% in their inner region within the projected half-light radius ($R_{\rm e}$) and between 70\% and 90\% within $3 R_{\rm e}$, which encompasses most of the galaxy's light \citep{Kassin2006, Martinsson2013, Courteau2015}. Observations of early-type galaxies often reveal lower DM fractions in the inner regions ($\lesssim 20\%$ within $R_{\rm e}$, \citealt{ATLAS3D_2013}), which tend to rise with increasing stellar mass in galaxies with masses $M_{\rm star} \geq 10^{10} \ \rm M_\odot$, but similarly high fractions in the outskirts ($70-90\%$ within $5 R_{\rm e}$, \citealt{William2020}). One reason for this may be that early-type galaxies may have experienced extensive baryonic processes over their lifetimes. These processes can affect both the distribution of baryons and that of DM, as simulations have shown that baryonic processes can expel DM over a time span of several gigayears \citep{Pontzen2012}. In particular, feedback processes such as supernova explosions, stellar winds, and active galactic nuclei stir the interstellar medium of galaxies and launch powerful gas outflows, which induce fluctuations in the gravitational potential that can in turn affect DM and diminish its fraction in the inner region of all type of galaxies \citep[e.g., ][]{Pontzen2014, Dutton2016, El-Zant2016, JF2020a, Dekel2021, Li2023}. These processes can indeed dynamically heat up the DM and lead to the formation of constant DM density cores rather than steep cusps. 
			Therefore, characterizing the DM fraction and its evolution with cosmic time not only enables better understanding of the influence of the DM distribution on galaxy formation and evolution, but it also provides valuable insights into the physical processes that govern galaxies and contributes to testing of the implementation of feedback processes in simulations. 
			
			In the past decade, integral field units (IFUs) in galaxy surveys have opened up new possibilities for studying the spatially resolved kinematics of \hz galaxies. For example, a review by \citet{Forster2020} notably shows that by using the resolved kinematics, it is now possible to obtain the rotation curves (RCs) of galaxies up to $z\sim 2.5$. These RCs allow one to probe the baryon and \dm\ content on galactic scales as well as their distribution and physical properties. \citet{Genzel2017} and \citet{Lang2017} were the first to analyze the RCs of star-forming galaxies (SFGs) at high-$z$  ($0.6\leq z\leq 2.6 $), and they found them to be declining; such behavior is only seen in local massive (very high surface brightness) SFGs, while the RCs of most normal SFGs are remarkably flat and rarely decline (e.g., \citealt[][]{rubin1980} and \citealt{PS1996}). Both studies (\citealt{Genzel2017} and \citealt{Lang2017}) suggest that the declining behavior of \RCs\ can be explained by a combination of high baryon fraction and pressure support in the inner regions. Some other high-$z$ studies of late-type and early-type galaxies also report similar low \dm\ fractions within the effective radii \citep[e.g.,][]{Burkert2016, Wuyts2016, Price2016, Ubler2018}. Conversely, \citet{AT2019} studied the shape of \RCs\ in $\approx 1500$ SFGs at $0.6\lesssim z \lesssim 2.2$ and reported flat \RCs\ similar to local SFGs. Moreover, \citet{AT2019} reported a more than $50\%$ \dm\ fraction within $3.5\ R_{\rm e}$, which is similar to local star-forming disk galaxies \citep{PS1996, Martinsson2013, Courteau2015}.
			
			In another follow-up study, \citet{GS21a} studied K-band Multi-Object Spectrograph (KMOS; \citealt{sharples_2014}) Redshift One Spectroscopic Survey (KROSS) data (\citealt{stott2016}) and derived the observed rotation and dispersion curves using 3D forward modeling. These observed \RCs\ were then converted into intrinsic RCs by correcting issues related to \hz measurements, such as beam smearing and pressure gradient. This provided them with a large sample of more than 200 flat \RCs\ of disk-like galaxies at $z\sim 1$ (i.e., 6.5 Gyr look-back time).  In \citet{GS21b}, the authors employed these intrinsic \RCs\ to estimate the invisible mass fraction needed to recover observed kinematics without invoking any DM halo model (such as the cuspy NFW from \citealt{NFW} or the cored \citealt{Burkert}). In this technique, stellar masses ($M_{\rm star}$) are estimated by fitting the spectral energy distribution (SED) of the galaxies and the gas (molecular and atomic) masses by means of the scaling relations by \citet{Tacconi2018} and \citet{Lagos2011}. Assuming an exponential thin disk distribution, they estimated the contribution of baryons to total mass at different scale radii (disk radius: $0.59R_{\rm e}$, optical radius: $1.89R_{\rm e}$, and outer radius: $2.95R_{\rm e}$), where the total mass is the dynamical mass ($M_{\rm dyn} \propto G^{-1} V^2 \ R$) computed directly from the intrinsic rotation curves.\footnote{In general, the scale length (or radius) is associated with various quantities that decrease exponentially, such as the surface brightness. The disk edge is defined as $3.2 \ R_{\rm D}$ ( $=$ $1.89 \ R_{\rm e}$), where the stellar surface luminosity $\propto \exp(-r/R_{\rm D})$. Therefore, scale lengths in terms of the effective radius can be written as: $R_{\rm D} = 0.59 \ R_{\rm e}$; $R_{\rm opt} = 1.89 \ R_{\rm e}$; $R_{\rm out}=2.95 \ R_{\rm e}$. } This work showed that the majority ($> 72 \%$) of \sfgs\ in the KROSS sample at $z\sim 1$ have DM-dominated ($>60\%$) outer disks ($\sim 5-10$  kpc), which agrees well with local SFGs. 
			
			Recently, various other studies have investigated the DM fraction within $R_{\rm e}$, including works by \citet{Genzel2020, Price2021, Bouche2022, Genzel2022}, and \citet{KURVS}. In this article, our focus is on studying the DM fraction within different galactic scales, ranging from $R_{\rm e}$ up to approximately $3 R_{\rm e}$ ($\approx R_{\rm out}$).
			In particular, we expand our understanding of the DM fraction by incorporating a larger sample and extending the redshift range beyond that of \citet{GS21b}. We present a comprehensive investigation using data from the KMOS$^{3D}$ survey \citep{W19} and the KMOS Galaxy Evolution Survey (KGES; \citealp{AT2021}) and previously analyzed KROSS data, covering a redshift range of $0.6< z< 2.5$. 
			
			Our primary objectives are to test the rotation curve analysis method established in \citet{GS21a, GS21b}, examine the redshift evolution of DM fractions across galactic scales, and explore questions related to the assembly history of galaxies. Most importantly, we demonstrate how current constraints on accurately accounting for baryonic content result in substantial uncertainties, both statistical and systematic, in estimating the DM mass of galaxies. By reducing these uncertainties, we as a community have the potential to make a significant leap forward in our understanding of galaxy evolution. The structure of this article is as follows: In Section~\ref{sec:dataset}, we provide an overview of the datasets utilized in this study. Section~\ref{sec:k-model} describes the employed kinematic modeling techniques, the analysis of its outputs, and the establishment of a robust final sample used throughout this work. Moving to Section~\ref{sec:analysis}, we investigate the data for potential discrepancies, present the results of the DM fraction at different galactic scales and cosmic time, and discuss the correlations between \dm\ fraction, baryonic surface density, and circular velocity. Section~\ref{sec:caveats} discusses potential caveats associated with this study. Subsequently, in Section~\ref{sec:discussion}, we delve into a detailed discussion of the main results. Finally, we summarize our findings in Section~\ref{sec:summary}. Throughout the analysis we assumed a flat $\Lambda$CDM cosmology with $\Omega_{m,0} =0.27$, $\Omega_{\Lambda,0}=0.73$, and $H_0=70 \ \mathrm{km \ s^{-1}}$.

			\begin{figure*}
				\begin{center}
					\includegraphics[angle=0,height=10.5truecm,width=17.5truecm]{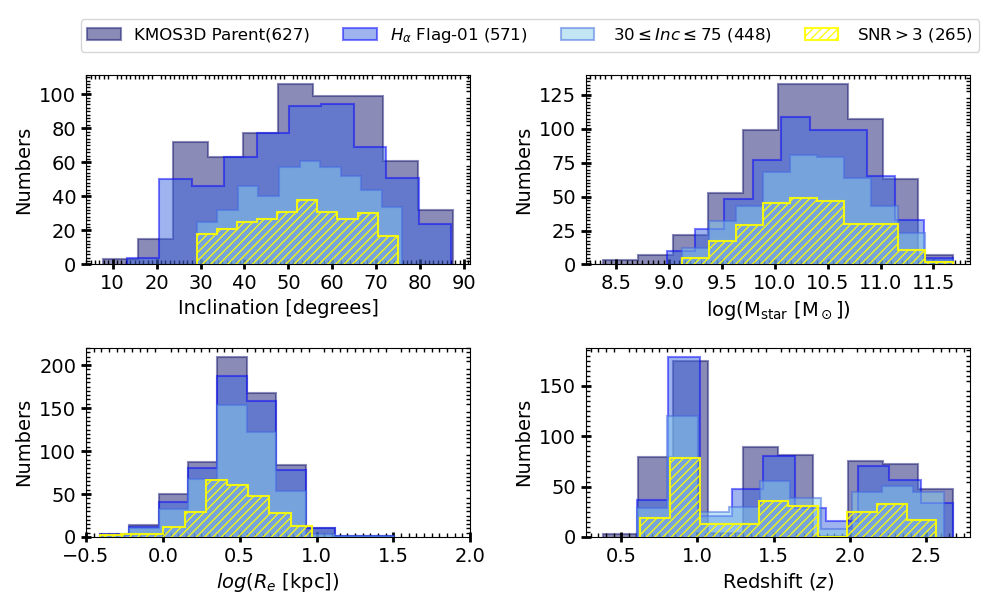}           
					\caption{Distributions of the physical properties of the KMOS$^{\rm 3D}$ parent and selected sample, arranged from top-left to bottom-right: inclinations, stellar masses, effective radii, and redshifts. The dark blue histogram represents all $H\alpha$ datacubes with confirmed spectroscopic redshift. The blue histogram shows $H\alpha$-detected galaxies with Flag-0 or Flag-1 in Table-6 of \citet{W19}. We further select galaxies with inclination angles between $25^{\circ} \leq  \theta_{i}  \leq 75^{\circ}$, shown in the sky-blue histogram. Lastly, we narrow our focus to galaxies with a signal-to-noise ratio greater than 3, depicted by the yellow hatched histogram. This selection criteria allowed us to perform the robust kinematic modeling of these high-$z$ galaxies.  }
					\label{fig:KMOS3D-P}
				\end{center}
			\end{figure*}
			
			
			\section{Datasets} \label{sec:dataset}
			In this study, we utilize publicly available high-$z$ galaxy surveys conducted with the KMOS instrument \citep{sharples_2014}, namely the KMOS$^{\rm 3D}$ survey \citep{W19}, KGES \citep{AT2021}\footnote{https://astro.dur.ac.uk/KROSS/data.html}, and KROSS \citep{stott2016}. KMOS is a second-generation spectrograph located at the Very Large Telescope (VLT), which indeed offers unique advantages for constraining the \dm\ fraction in high-$z$ galaxies. One key feature of KMOS is its capability for integral field spectroscopy, which enables efficient and rapid surveying of a large number of galaxies. Moreover, KMOS operating capability in wavelength range $0.778 - 2.481 \ \mu m$ facilitates the detection of rest-frame optical emission from high-$z$, such as $H\alpha$ ($\lambda_{\rm rest} = 0.6562 \ {\rm \mu m}$). Additional benefit of this broad wavelength coverage is its ability to provide crucial information regarding the stellar populations, star-formation rates, kinematics and dynamics of distant galaxies. In this work, we specifically investigate the dynamics of distant galaxies and analyze the results in the light of prevalent issues, such as, observational uncertainties (e.g., noise), kinematic modeling, and dynamical models.
			
			\subsection{KMOS$^{\mathrm{3D}}$}\label{sec:KMOS3D}
			All objects of the KMOS$^{\rm 3D}$ survey are drawn from the 3D-HST survey \citep{Barmmer2012, Skelton2014, Momcheva2016} within three extra-galactic  fields (COSMOS, \citealt{COSMOS}; GOODS-S, \citealt{GOODS}; and UDS, \citealt{UKIDSS}) covering a wide redshift range ($0.7<z<2.7$). The KMOS$^{\rm 3D}$ observations in the $YJ-$, $H-$, and $K-$bands cover the $H\alpha$ emission line at redshift $0.7< z <1.1$, $1.2 < z< 1.8$, and $1.9<z<2.7$, respectively. In all three filters average  seeing conditions --expressed as the full width half maximum (FWHM) of the point spread function (PSF) -- are less than or equal to $0.55 \arcsec$. We begin with analyzing all $H\alpha$ cubes, and adopt the catalogs released in \citet{W19}. From the physical property catalog (\citealt{W19}, Table~5), we mainly work with quantities such as galaxy-IDs, sky-coordinates, redshift, magnitude, seeing (FWHM), HST-axis ratios, effective radius, stellar masses, and star-formation rates. The aperture size of the $H_\alpha$ datacubes is 1.5$\arcsec$ in radius. The other details, such as IDs, detection and non-detection flags, of $H\alpha$ datacubes are given in the catalog associated with Table~6 of \citet{W19}.   
			\begin{figure*}
				\begin{center}
					\includegraphics[angle=0,height=10.5truecm,width=17.5truecm]{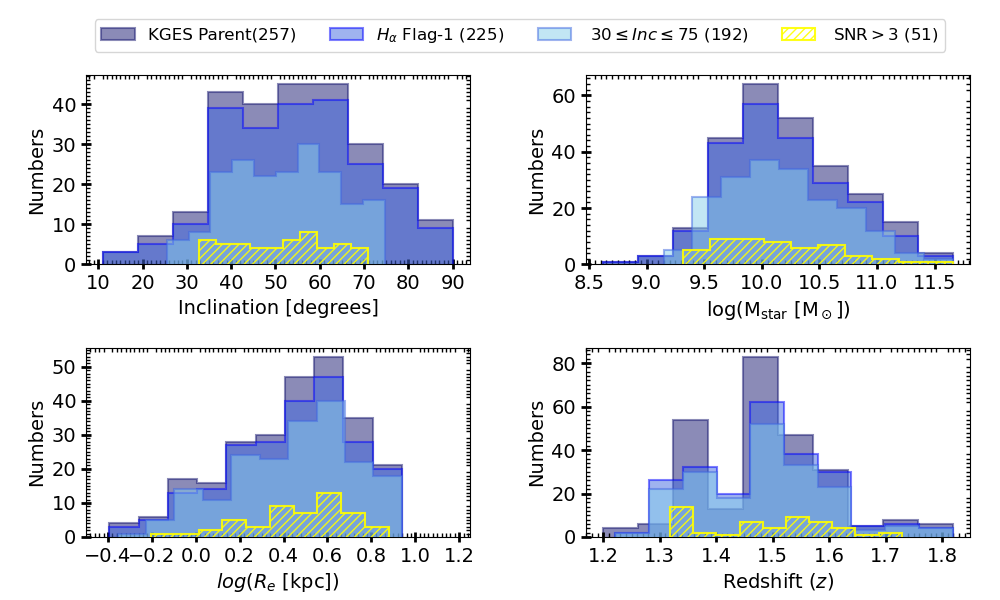}           
					\caption{Distributions of the physical properties of the KGES parent and selected samples, arranged from top-left to bottom-right: inclinations, stellar masses, effective radii, and redshifts. The dark blue histogram represents all $H\alpha$ datacubes with confirmed spectroscopic redshift. The blue histogram shows $H\alpha$-detected galaxies \citet{AT2021}. We further select galaxies with inclination angles between $25^{\circ} \leq  \theta_{i}  \leq 75^{\circ}$, which are shown in the sky-blue histogram. Lastly, we narrow our focus to galaxies with a signal-to-noise ratio greater than 3, depicted by the yellow hatched histogram. This selection criteria allowed us to perform the robust kinematic modeling of these high-$z$ galaxies.  }
					\label{fig:KGES-P}
				\end{center}
			\end{figure*}

			\paragraph{Physical properties of galaxies:} 
			Stellar masses in the KMOS$^{\rm 3D}$ survey \citep{W19} were derived from SED modeling using the FAST fitting code \citep{Kriek2009}, assuming exponentially declining star-formation histories ($\tau > 300$ Myr), solar metallicity, a \citet{Chabrier_2003} initial mass function (IMF), and stellar population synthesis models derived from \cite{Bruzual2003}. The dust attenuation was modeled using the \citet{Calzetti2000} reddening law, with visual extinctions in the range $0 < A_v < 4$. The full KMOS$^{\rm 3D}$ sample covers the stellar mass range $8.3 \leq \log (M_{\rm star}  \ \mathrm{[M_{\odot}]} ) \leq 11.7$.  The effective radius of galaxies are computed from high-resolution CANDELS H-band photometry \citep{Skelton2014}, covering $-0.6 \leq \log (R_{\rm e} \ \mathrm{[kpc]}) \leq 1.5$. The star-formation rate is derived using a cross-calibrated ladder of SFR indicators \citep{Wuyts2011b}. The statistical uncertainty in stellar masses and star-formation rates is assumed to be 10\%. Additionally, following \citet{Pacifici2023}, we incorporate a systematic uncertainty of 0.14 dex in the stellar mass and 0.25 dex in the star formation rate (SFR). To obtain the inclination angle, we use the axis-ratio $a/b$ from \citet{W19} and convert it into inclination $\theta_i$ according to 
			\begin{equation}
				\centering
				cos^2 \theta_{i} = \frac{(b/a)^2 - q_0^2}{1-q^2_0},
			\end{equation}
			where $q_0$ is the intrinsic axial ratio of an edge-on galaxy (e.g., \citealt{TFR1977}), which could in principle have values in the range $0.1$ – $0.65$ (e.g., \citealt{Weijmans2014}); here, we use the commonly assumed value $q_0 = 0.2$, which is applicable for thick discs commonly found at \hz \citep{H17}. The resulting inclinations cover a broad range between $\approx 10^{\circ} - 85^{\circ}$. The kinematic/photometric position angles (PA) are not in the public release of KMOS3D data and obtained through kinematic modeling of the datacubes as discussed in Appendix~\ref{sec: BBoptimization}. The central positions of the sources are similarly re-derived during the kinematic modeling.

			\paragraph{Primary selection criteria:} 
			The \citet{W19} catalog contains 627 galaxies with spectroscopically confirmed redshift (see their  Table~6), which we refer to as the parent sample. From those, we selected H$\alpha$-detected galaxies (i.e., with FLAG 0,1 in the catalog) with high inclination angle ($25^{\circ} \leq \theta_i \leq 75^{\circ}$). \footnote{The line-of-sight velocity is defined as $V_{\rm LOS} = V_{\rm rot} \ \sin\theta_i$, where $V_{\rm rot}$ is the intrinsic rotation and $\theta_i$ is the inclination angle of a galaxy. If a galaxy is face-on (i.e., $\theta_i \approx 0$), the intrinsic rotation results are divergent. Additionally, intrinsic rotation in low-inclination ($0 < \theta_i \leq 25$) galaxies is often obscured by the superposition of radial and tangential motions, leading to a flattened appearance of the velocity profile. This consideration has been fundamental in RC studies since the beginning \citep[see for instance][]{PS1996, Sofue2001}. On the other hand, if a galaxy is edge-on, extinction is prominent due to the increased opacity of the galaxy, which dramatically flatten the inner slope of the RCs above $80^{\circ}$, as demonstrated in various works such as  \citet{Valotto2004} and \citet{Rhee2004}. This is why, we select the galaxies with inclination angle: $25^{\circ} \leq \theta_i \leq 75^{\circ}$.} 
			This left us with 448 objects. Before performing the kinematic modeling, we inspected their signal-to-noise ratio (S/N) and H$\alpha$ images as discussed in Appendix~\ref{sec:SNR} and demonstrated in Figure~\ref{fig:1D-int-spc-SN} and \ref{fig:SN-Haimages}. The signal-to-noise ratio is defined as signal/noise, and it is computed from integrated spectra of flux and noise cubes. The ${\rm S/N}>3$ threshold is the requirement of kinematic modeling tool 3DBarolo (\citealt{ETD15, ETD16}) to produce accurate kinematics. Based on this quantitative and qualitative assessment of dataset, we divide the 448 objects of the sample in three categories: \textbf{Q1:}  ${\rm S/N}>3$ and sharp H$\alpha$ image; \textbf{Q2: } ${\rm S/N}\geq 3$ and moderately visible source; \textbf{Q3: }either ${\rm S/N}<3$ or no appearance of the source in the H$\alpha$ image. We discard all the Q3 galaxies, which yields a final sample of 265 sources for kinematic modeling. The distributions of inclination angle, stellar mass, effective radius, and redshift of the parent and selected KMOS$^{\rm 3D}$ samples are shown in Figure~\ref{fig:KMOS3D-P}. 
			
			\subsection{KGES} \label{sec:KGES}
			The KGES encompasses a sample of 285 galaxies located in the COSMOS, CDFS, and UDS fields, with redshifts ranging from $1.2$ to $1.9$ \citet{AT2021}. The survey primarily targets the $H\alpha$, [NII]6548, and [NII]6583 emission lines, which are redshifted to the H-band wavelength range (approximately $1.46-1.85 \ \rm \mu$m). For the current study, we focus on the subset of 225 KGES star-forming galaxies with confirmed spectroscopic redshifts, $H\alpha$ detection, and not flagged as AGNs. Additionally, we restrict our analysis to galaxies with a K-band magnitude of K$<22.5$, similar to \citet{Gillman2020}. The median redshift of our sample is $z=1.49\pm 0.07$, and the median seeing (FWHM) in the H-band observations is $0.7\arcsec$. The aperture size of the $H\alpha$ cubes is 1.2$\arcsec$ in radius, same as the KROSS $H\alpha$ cubes discussed below in Section~\ref{sec:KROSS}. 
			
			\paragraph{Physical properties of galaxies:}
			The physical properties of the KGES sample can be found in \citet{AT2019, Gillman2020} and \citet{AT2021}. In particular, stellar masses were estimated by fitting the SED of galaxies using a routine called Multi-wavelength Analysis of Galaxy Physical Properties (MAGPHYS; \citealt{MAGPHYS}). The SED of each galaxy was constructed using multi-wavelength photometry ranging from the ultraviolet to the mid-infrared. The MAGPHYS routine compares the observed SEDs with the SEDs from the spectral libraries of \citet{Bruzual2003}, includes the dust attenuation model of \citet{Charlot_Fall_2000}, continuous star-formation histories, and \citet{Chabrier_2003} IMF. The resulting stellar mass range of the $H\alpha$-detected sample is $8.62 \leq \log(M_{\text{star}} \ \mathrm{[M_\odot]}) \leq 11.66$, as reported in \citet{AT2021}. Star formation rates and its uncertainties were estimated from the $H\alpha$ flux, corrected for dust attenuation, assuming a \citet{Calzetti1994} extinction law. The estimated star formation rates range from $0.0 \leq \log(\text{SFR} \  \mathrm{[M_\odot \ yr^{-1}]}) \leq 2.2$. Similar to KMOS3D, the statistical uncertainty on stellar masses and star-formation rates is assumed to be 10\%. Additionally, we add systematic uncertainty of 0.14 dex on stellar mass and 0.25 dex on star-formation rate. Geometrical parameters such as the effective radius, inclination, position angle, and central x-y coordinates were derived using the GALFIT \citep{GALFIT}. For detailed calculations of all physical quantities, we refer the reader to \citet{Gillman2020}.

			\paragraph{Primary selection criteria:} To ensure the quality of datacubes and robustness of our kinematic modeling, we employ the identical selection criteria that were chosen for KMOS$^{\mathrm{3D}}$ (as established in \citealt{GS21a}). Firstly, we select galaxies with confirmed spectroscopic redshifts detected in $H\alpha$. Subsequently, we narrow down the sample based on two additional criteria: a) high inclination angle ($25^{\circ} \leq \theta_i \leq 75^{\circ}$) and b) $\rm S/N > 3$. By applying these selection criteria, we are left with a set of 51 galaxies with sufficiently high S/N, which enables us to conduct accurate kinematic modeling. Figure~\ref{fig:KGES-P} presents the distributions of physical properties of KGES selected sample, same as in Figure~\ref{fig:KMOS3D-P} which is for the KMOS$^{\mathrm{3D}}$ sample.

			\subsection{\textbf{KROSS}} \label{sec:KROSS}
			Additionally, we use the KROSS dataset \citep{stott2016}, which was previously studied in \citet{GS21a, GS21b} and \citet{GS22} (and also by \citealt{H17, HLJ17, AT2019b}). The KROSS targets are selected from extragalactic deep fields covered by multi-wavelength photometric and spectroscopic data: 1) Extended Chandra Deep Field Survey (E-CDFS: \citealt{ECDFS1, ECDFS2}), 2) Cosmic Evolution Survey (COSMOS: \citealt{COSMOS}),  3) Ultra-Deep Survey (UKIDSS: \citealt{UKIDSS}), and 4) SA22 field \citep{SA22}. Some of the targets were also selected from the CF-HiZELS survey \citep{CFHIZELS}. The targets were selected such that the $H_\alpha$ emission is shifted into J-band with a median seeing of $0.7\arcsec$. The aperture size of the $H\alpha$ cubes is 1.2$\arcsec$ in radius. The KROSS sample studied in Sharma et al. contains 225 galaxies with redshift $0.7\leq z\leq 1.04$, inclination range $25^{\circ} < \theta_i \leq 75^{\circ}$, effective radius $0.08 \leq \log (R_{\rm e} \ \mathrm{[kpc]}) \leq 0.89$, stellar mass $9.0 \leq \log (M_{*} \ \mathrm{[M_{\odot}]} ) < 11.0$, and circular velocity $1.45 \leq \log (V_{\rm out} \ \mathrm{[km \ s^{-1}]}) \leq 2.83$, where $V_{\rm out}$ is calculated at $R_{\rm out}=2.95 R_{\rm e}$.
			
			\begin{figure}
				\begin{center}
					\includegraphics[angle=0,height=6.7truecm,width=9.0truecm]{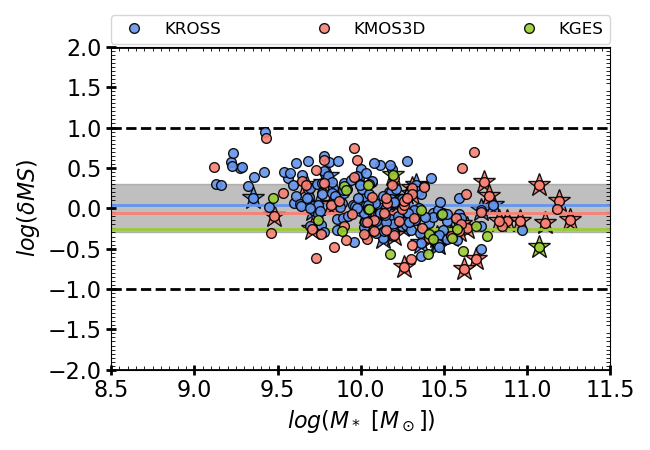}  
					\caption{Position of galaxies (final sample) with respect to the star-forming main sequence (MS). We plot offset from the MS, $\delta {\rm MS} \ (= {\rm SFR}/{\rm SFR}({\rm MS}; z,{M_{\rm star}})$) as a function of the stellar mass ($M_{\rm star}$) with the reference ${\rm SFR}({\rm MS}; z,{M_{\rm star}})$  from \citet{speagle14}. The gray shaded area represents the typical limit (0.3 dex) of star-forming MS  \citep[e.g., see ][]{Rodighiero2011, Genzel2015, Tacconi2018, Freundlich2019}. The KMOS$^{\rm 3D}$, KGES, and KROSS data are denoted by red, green, and blue filled circles, respectively. The solid lines in corresponding colors depict the average offset from the main sequence for each dataset. All galaxies are within 2$\sigma$ scatter of the MS line. The galaxies marked with stars are the galaxies that have $R_e> {\rm PSF}$.}
					\label{fig:Mstar-sfr}
				\end{center}
			\end{figure}
			
			\subsection{Estimating gas masses}\label{sec:Mbar}
			Observations show that typical star-forming galaxies lie on a relatively tight, almost linear, redshift-dependent relation between their stellar mass and star formation rate, the so-called main sequence of star formation \citep[MS; e.g., ][]{Noeske2007, Whitaker2012, speagle14}. Most stars since $z\sim 2.5$ were formed on and around this MS \citep[e.g., ][]{Rodighiero2011}, and galaxies that constitute it, usually exhibit a rotating disk morphology \citep[e.g., ][]{ForsterSchreiber2006, Daddi2010b, Wuyts2011b}. 
			Figure~\ref{fig:Mstar-sfr} shows the position of the final sample (detailed in Section~\ref{sec:final-sample}) with respect to the main sequence of typical star-forming galaxies (MS), i.e., their offset from the main sequence:
			\begin{equation}
				\delta {\rm MS} = {\rm SFR}/{\rm SFR}({\rm MS}; z,{M_{\rm star}}),
			\end{equation} 
			where ${\rm SFR}({\rm MS}; z,{M_{\rm star}})$ is the analytical prescription for the center of the MS as a function of redshift and stellar mass proposed in the compilation by \citet{speagle14}, as a function of stellar mass. This figure shows that 64\% of the galaxies in the total sample are within $1\sigma$ of the main sequence scatter, whereas the remaining 36\% are within the $2-3\sigma$ range.
			This enables us to estimate their molecular gas masses ($M_{\rm H2}$) using the \citet{Tacconi2018} scaling relations, which provide a parameterization of the molecular gas mass as a function of redshift, stellar mass, and offset from the MS stemming from a large sample of about 1400 sources on and around the MS in the range $z=0-4.5$ (see also \citealt{Genzel2015} and \citealt{Freundlich2019}). The scatter around these molecular gas scaling relations and the stellar mass induces a 0.3 dex uncertainty in the molecular gas mass estimates, which is accounted in the error estimates.  The H2 mass of our sample is $9.48 \leq log(M_{\rm H2} \ [M_\odot]) \leq 11.03$, with an average molecular gas fractions ($f_{_{\rm H2}}$) of $0.21\pm 0.09$, $0.14\pm 0.05$, and $0.22\pm 0.06$ for the KMOS3D, KGES, and KROSS sub-samples, respectively.\footnote{Mass fraction of $x$ component of $M_{bar}$ is computed as $f_{x} = M_{x}/M_{bar}$} 
			
			To calculate the atomic mass ($M_{\rm HI}$) content of galaxies within the redshift range $0.6 \leq z \leq 1.04$, we use the HI scaling relation presented by \citet{Chowdhury2022}, which provides the first $M_{\rm star}-M_{\rm HI}$ relation at $z\approx 1$, encompassing 11,419 star-forming galaxies. The relation was derived using a stacking analysis across three stellar mass bins, each bin with a $4\sigma$ detection and an average uncertainty of $\sim 0.3$ dex. This uncertainty in the gas scaling relation is additionally accounted in the error of HI mass estimates. To compute the HI mass at $z> 1.04$, we employ the $M_{\rm star}-M_{\rm HI}$ scaling relation derived from a galaxy formation model under the $\Lambda \mathrm{CDM}$ framework \citep[for details, see][]{Lagos2011}. This scaling relation successfully reproduces both the HI mass functions \citep{Zwaan2005, Martin2010} and the $^{12}CO$ luminosity functions \citep{Boselli2002, Keres2003} at $z\approx 0$ with an uncertainty of around $0.25$ dex, as well as follows the observations of quasars  from $z=0-6.4$ (see \citealt{Lagos2011}, Fig.~12).  The HI Mass range of our sample is $9.90 \leq log(M_{\rm HI} \ [M_\odot]) \leq 11.42$, with an average atomic gas fractions ($f_{_{\rm HI}}$) of $0.42\pm 0.16$, $0.48\pm 0.17$, and $0.37\pm 0.06$ for the KMOS3D, KGES, and KROSS sub-samples, respectively. 
			%

			
			\begin{figure*}
				\begin{center}
					\includegraphics[angle=0,height=9.5truecm,width=15.5truecm]{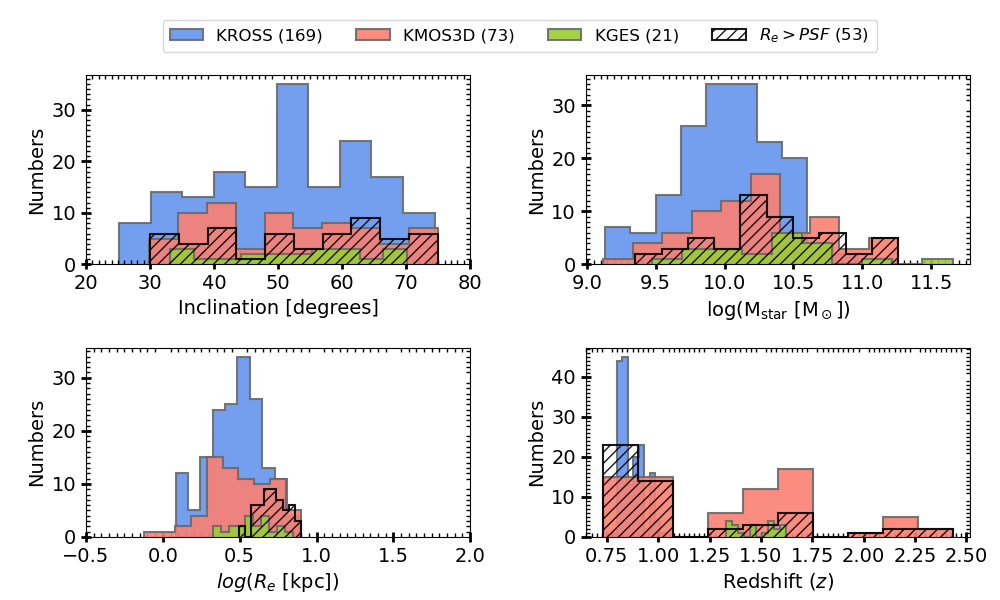}           
					\caption{Distributions of the physical properties of galaxies after kinematic modeling, arranged from top-left to bottom-right: inclinations, stellar masses, effective radii, and redshifts. These distributions are presented for the KMOS$^{3D}$ (in red), KGES (in green), and KROSS (in blue) samples. In total, we use 73 KMOS$^{\rm 3D}$, 21 KGES, and 169 KROSS galaxies to estimate the DM fraction at high-$z$. The distribution of $R_e > PSF$ galaxies (within final sample) is shown by black hatched histograms.  For detailed information on sample selection following kinematic modeling, we refer the reader to Section~\ref{sec:km-inspect}.
					}
					\label{fig:final-hist}
				\end{center}
			\end{figure*}
			
			\begin{figure*}
				\begin{center}

					\includegraphics[angle=0,height=5.5truecm,width=16.5truecm]{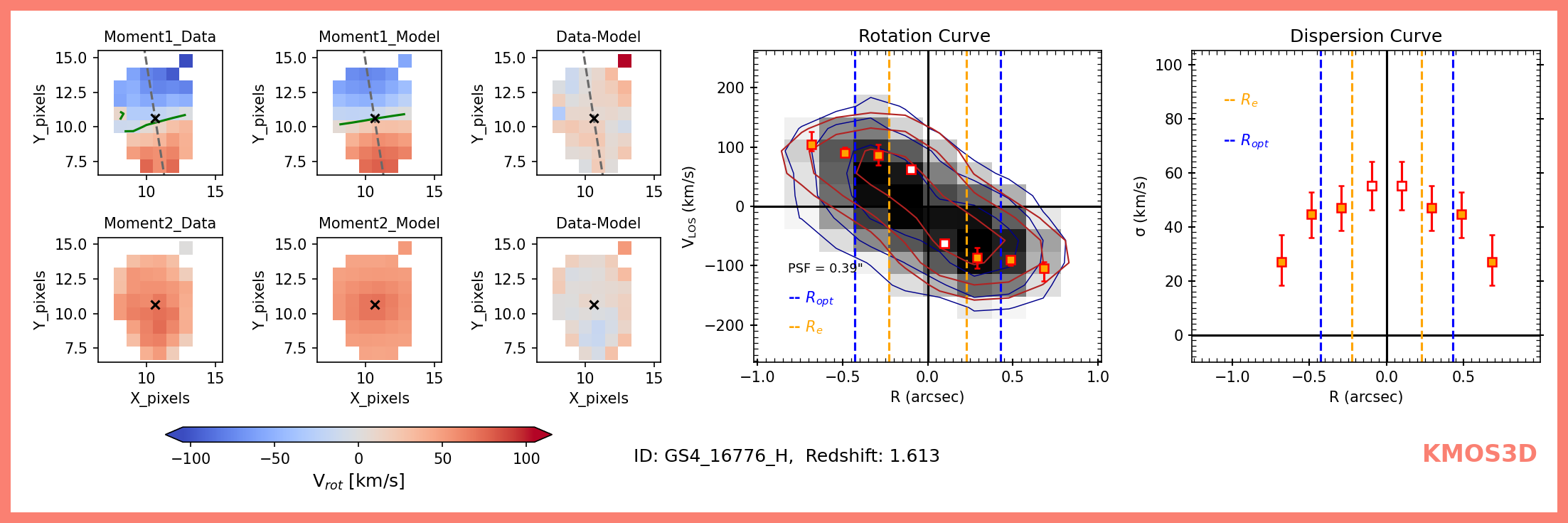} 
					\includegraphics[angle=0,height=5.5truecm,width=16.5truecm]{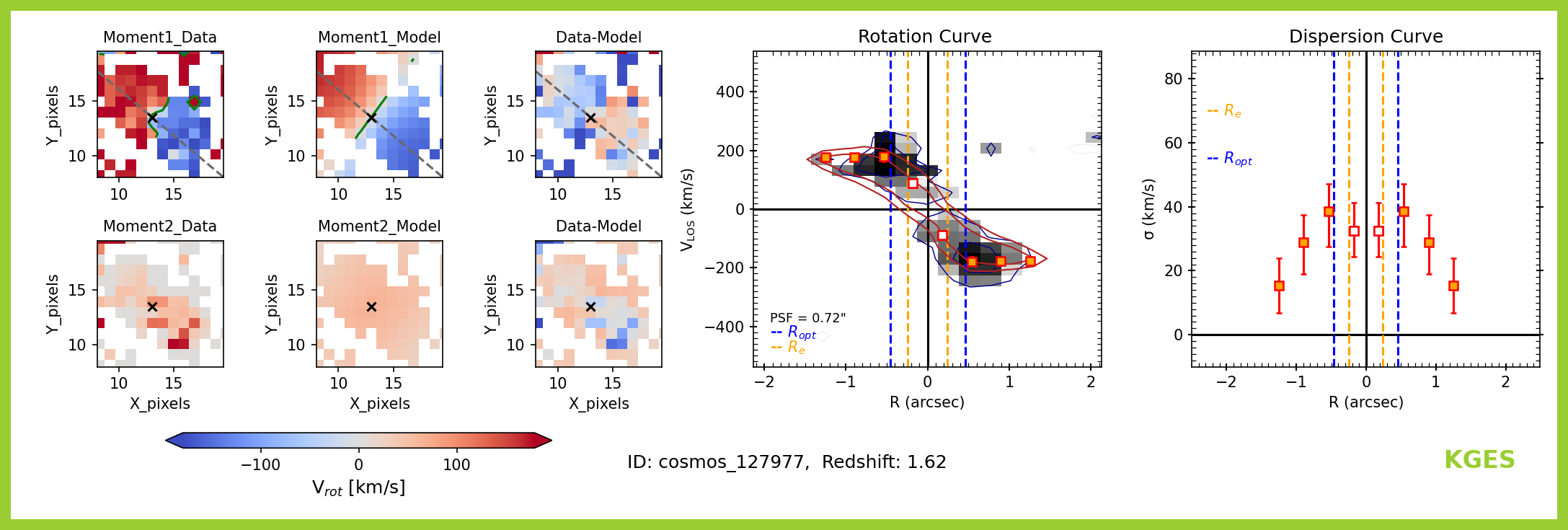} 		
					\includegraphics[angle=0,height=5.5truecm,width=16.5truecm]{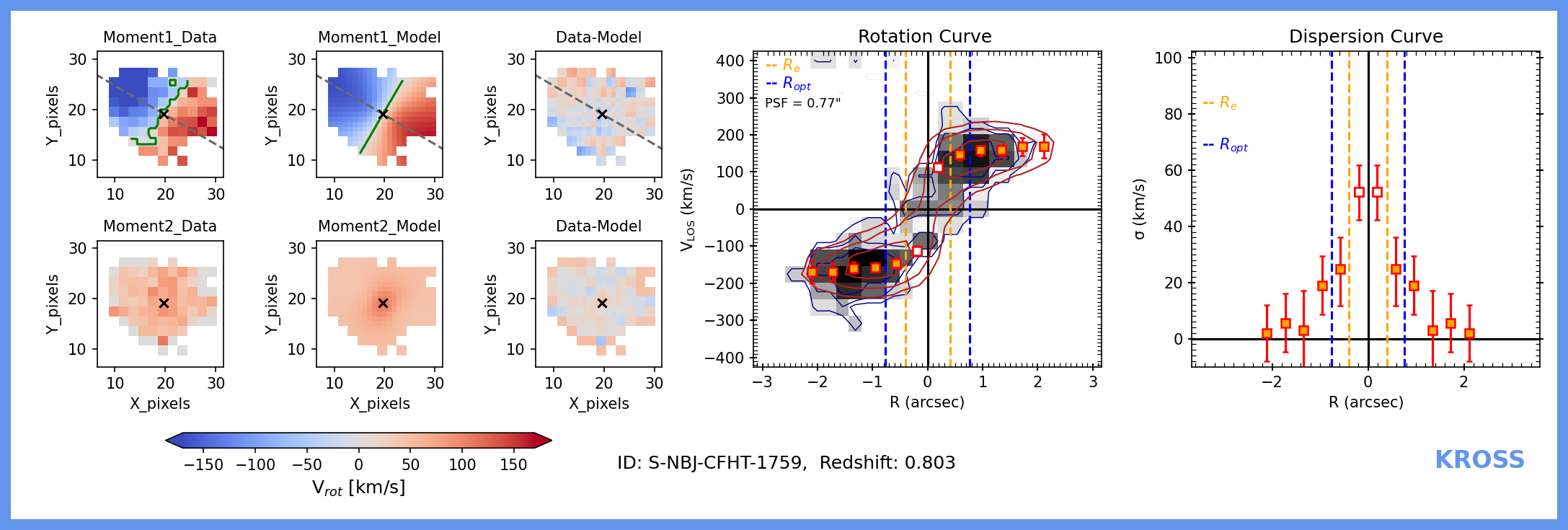} 
					
					\caption{Outputs of the kinematic modeling of selected KMOS$^{\rm 3D}$ (red box), KGES (green), and KROSS (blue) sources obtained using $^{3D}$BAROLO. {\em Row-1, Columns 1-3:} Rotation velocity map data, model, and residuals, respectively. The gray dashed line shows the position angle, the black cross the central x-y position, and the green line marks the plane of rotation. {\em Row-2, Columns 1-3:} Velocity dispersion map data, model, and residuals, respectively.
						{\em Column 4:} Major axis PV diagram, where the black shaded area with blue contour represents the data while the red contour shows the model, and the orange squares with error bars are the best-fit line-of-sight rotation velocity inferred by $^{3D}$BAROLO. The yellow and blue vertical dashed lines represent the effective radius ($R_{\rm e}$), and the optical radius ($R_{\rm opt}=1.89 \ R_{\rm e}$), respectively. {\em Column 5:} Corresponding velocity dispersion curve. The first point in the rotation and dispersion curves is represented by an empty (or white) marker, as it falls under the resolution limit and is therefore excluded from the data analysis.
					}
					\label{fig:BBKMKGKR}
				\end{center}
			\end{figure*}
			
			\section{Forward modeling of the datacubes}
			\label{sec:k-model}
			We conduct a comprehensive reanalysis of the entire KMOS$^{\rm 3D}$, KGES, and KROSS datasets, using the 3D forward modeling approach implemented by $^{3D}$BAROLO. In order to obtain precise kinematics, we used an optimization function in conjunction with $^{3D}$BAROLO to more accurately constrain the essential gas geometrical parameters (see Section~\ref{sec:3dFM}). We found that galaxies with low S/N and a large PSF present a challenge in terms of accurate kinematic modeling. The former is related to the intrinsic brightness of the source, its distance, and the integration time of the observations; the latter to the intrinsic size and atmospheric conditions during observations, which can significantly degrade the resolution. 
			Consequently, we had to discard the low S/N ($\sim 3-5$) and a large PSF  ($PSF \geq R_{\rm max}$ of rotation curve) galaxies from the final analysis (see Section~\ref{sec:km-inspect}). Our final sample consists of 73 KMOS$^{\rm 3D}$, 21 KGES, and 169 KROSS galaxies, i.e., a total of 263 objects (see the following subsections). The distributions of the physical properties of the final sample is shown in Figure~\ref{fig:final-hist}.
			
			\subsection{Kinematic modeling} \label{sec:3dFM}
			We modeled the kinematics of the galaxies in our samples using the $^{3D}$BAROLO code \citep{ETD15}. The main advantages of modeling datacubes with $^{3D}$BAROLO (hereafter 3DBarolo) are that 
			(1) it allows us to reconstruct the intrinsic kinematics in three spatial and three velocity components for given initial guesses that define the kinematics and geometry of a galaxy; 
			(2) the 3D projected modeled datacubes are compared to the observed datacubes in 3D space; 
			(3) it simultaneously incorporates instrumental and observational uncertainties (e.g., spectral smearing and beam smearing) in 3D space.\footnote{In IFU-based spectroscopy and imaging, the PSF describes how a point source (like a distant star or galaxies) appears in an observation due to the effects of the imaging system and atmospheric conditions. Similarly, the line spread function describes how a spectral line from a point source is broadened by the spectrograph and other instrumental effects. That is, beam smearing is associated with a poor PSF, while spectral smearing is due to a poor line spread function.} For details, we refer the reader to \citet[][]{ETD15} and \citet{ETD16}. This three-fold approach of deriving kinematics is designed to overcome the observational and instrumental effects and hence allowed us to stay close to the realistic conditions of the galaxy. Therefore, it provides somewhat improved results compared to the 2D approach,\footnote{The 2D kinematic modeling is a technique that uses datacubes to create velocity-maps and where the rotation curve is derived along the projected major axis of the galaxy.} specifically in the case of small angular sizes and the moderate S/N of high-$z$ galaxies (see \citealt{ETD16}). Basic assumptions under 3DBarolo, its basic requirement, and limitations are detailed in \citet[][section-3.1]{GS21a} and briefly mentioned below. 
			
			
			The kinematic modeling with 3DBarolo requires three geometrical parameters --~namely the galaxy's central position ($x_c,y_c$), the inclination angle ($\theta_i$), and the position angle (PA)~--, and three kinematic parameters --~namely the redshift ($z$), the rotation velocity ($V_{\rm rot}$), and the velocity dispersion of the ionized gas ($V_{\rm disp}$). In our modeling, we set the geometrical parameters and redshift, while kinematic parameters are left free. 3DBarolo comes with several useful features particularly useful for high-$z$ low S/N data\footnote{See its documentation, \url{https://bbarolo.readthedocs.io/en/latest/}}. We used \texttt{3DFIT TASK} for performing the kinematic modeling. 3DBarolo produces mock observations given the input parameters in the 3D observational space ($x,y,\lambda$), where ($x,y$) stands for the spatial axes and $\lambda$ is the spectral axis coordinate, resulting in a datacube: $f(x,y,\lambda)$. These models were fit ring by ring to the observed datacube in the same 3D space, accounting for beam smearing. A successful run of  3DBarolo delivers the beam smearing corrected velocity (or moment) maps, the stellar surface brightness profile, the RC, and the dispersion curve along with the kinematic models. We note, the RC (or PV diagrams) are not derived from the velocity maps, but instead calculated directly from the datacubes by minimizing the difference in $V_{\rm LOS} = V_{\rm rot} \sin \theta_i$ of model and data in each ring.
			
			When we employed 3DBarolo to model the galaxies, we discovered that the photometric geometrical parameters ($x_{c}$, $y_{c}$, and PA) were inadequate for accurately representing (or extracting) their kinematics. This inadequacy arises because gas kinematics differ from stellar kinematics due to their distinct morphologies (i.e., geometry) and alignments. In particular, as we go higher in redshift, gas morphological parameters (PA, $x_c$, and $y_c$) start differing from their photometric measurements, i.e., stellar morphology \citep[as reported in ][]{W15, H17, GS21a}. Moreover, from our previous work with 3DBarolo \citep{GS21a}, we also learned that \texttt{3DFIT TASK} is incapable to constrain simultaneously multiple (more than 2) parameters, most likely due to low quality data at high-$z$.\footnote{For local galaxies, where data-quality is high (very good S/N), 3DBarolo is capable of fitting multiple parameters simultaneously. } 
			Therefore, in this work, we estimated the gas geometrical parameters using an optimization function that runs atop 3DBarolo, namely minimizing the following loss function:
			\begin{equation}
				\label{eq:loss_func}
				\ell  = \sqrt{  \overline{(X_D - X_M)^2}   } +  \frac{N[D]}{N[D: \neq 0, \neq \pm \infty]} + \frac{N[M]}{N[M: \neq 0, \neq \pm  \infty]}, 
			\end{equation}
			where $X_{D/M}$ is an array of data or a model, N represents the length of the array, and D and M stand for data and model, respectively. In the equation, the first term corresponds to the root mean square error, the other two terms are the weights of the data and the model. In the denominators, $N[D: \neq 0, \neq \pm \infty]$ gives the length of the datacube which includes only non-zeros and finite elements;  $N[M: \neq 0, \neq \pm \infty]$ gives the same for the model. That is, weights are higher if the datacube contains more zero or infinite elements; in which case, the loss function ($\ell$) is higher. We use a \texttt{Nelder-Mead} minimization method\footnote{\texttt{Nelder-Mead} is a gradient free optimization method that finds the minimum of function by iteratively updating the vertices of polytope in n-dimensional space. It is particularly used for optimization problem with non-liner or multi-model objective functions, which is the case of galaxy kinematics.}, which is available in the \texttt{scipy.optimize} library. This optimization function enables to fit multiple parameters with 3DBarolo. We ran 3DBarolo on each object as we did in \citet{GS21a}, i.e., the free parameters in 3DBarolo are only $V_{\rm rot}, V_{\rm disp}$ and the extra parameters are constrained by the optimizer. The details of the logical flow of optimization with 3DBarolo are described in Appendix~\ref{sec: BBoptimization}.

			\subsection{Inspection of kinematic modeling outputs} \label{sec:km-inspect}
			We modeled the kinematics of 541 \sfgs: 265 from KMOS$^{\rm 3D}$, 51 from KGES, and 225 from KROSS. For quality assessment and assurance, we inspected the outputs of 3DBarolo+optimization for each individual galaxy. Firstly, we scrutinized the optimization log of all KMOS$^{\rm 3D}$ and KGES galaxies. In KMOS$^{\rm 3D}$, we notice that, 57 of them experienced optimization failure due to quality of observations or inaccurate parameters such as PA, $x_c$, $y_c$ (see discussion in Appendix~\ref{sec: BBoptimization}). Secondly, 3DBarolo could not perform the modeling on 52 galaxies owing to their large PSF (i.e., $R_{max} \leq $ PSF, where $R_{\rm max}$ is the maximum radius of the \RC). Furthermore, we observed that 78 galaxies had a maximum radius comparable to the PSF, thereby allowing 3DBarolo to form only two rings, the first of which being unreliable (see \citealt{ETD16} and \citealt{GS21a}). Consequently, \RCs\ obtained from these galaxies cannot be used in dark matter fraction studies. Lastly, we inspected the velocity maps and high-resolution photometric images, and we noticed that five galaxies have disturbed kinematics due to nearby neighbors, and therefore we discarded them. After kinematic modeling, in total, we had to exclude 192 objects. It is not surprising for us to lose a lot of high-$z$ data, as these observations are often noisy and the angular size of the objects very small. This is consistent with the findings of previous studies on KMOS$^{\rm 3D}$ data, which were often conducted on a relatively small sub-sample \citep[see,][ and references therein]{Genzel2017, Genzel2020}. The final KMOS$^{\rm 3D}$ sample contains 73 galaxies, which is still large enough to perform a statistical study. 
			
			Within the KGES dataset, three galaxies encountered optimization failures, while 11 galaxies exhibited a larger PSF compared to the actual galaxy size. Additionally, 16 galaxies displayed a maximum radius that was comparable to the PSF,  thereby allowing 3DBarolo to form only two rings, the first of which being unreliable. Consequently, these galaxies were deemed unsuitable and were excluded from further analysis. As a result, our final KGES sample contains only 21 galaxies. Furthermore, for the details of comparison sample, i.e., KROSS objects, see Appendix~\ref{sec: BBoptimization}. In the end, we retain 263 galaxies, 73 from KMOS3D, 21 from KGES, and 169 from KROSS.
			
			\begin{figure}
				\begin{center}
					\includegraphics[angle=0,height=6.0truecm,width=9.0truecm]{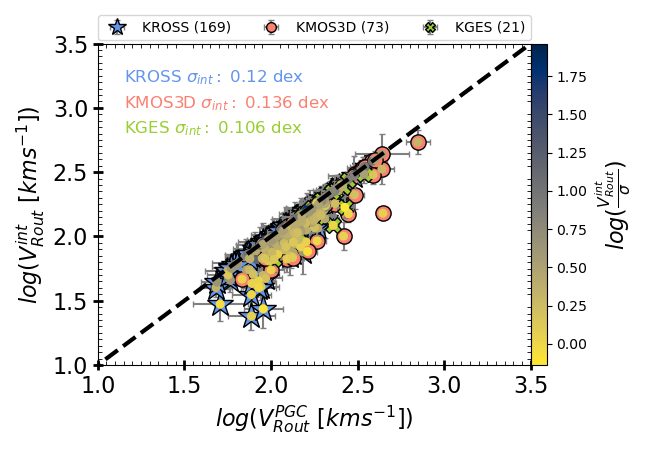}  
					\caption{Impact of the pressure gradient on the circular velocity of galaxies computed at $R_{\rm out}$. The KMOS$^{\rm 3D}$, KGES, and KROSS samples are denoted by red circles, green cross, and blue stars, respectively. The interior of each data point is color-coded for rotation-to-dispersion ratio computed before pressure support correction ($V^{int}_{{Rout}} / \sigma$). The black dashed line represents the one-to-one relation between the two measurements. We observe that galaxies with higher velocity dispersion require large pressure corrections.
					}
					\label{fig:PGCorr}
				\end{center}
			\end{figure}
			
			\begin{figure}
				\begin{center}
					\includegraphics[angle=0,height=6.0truecm,width=9.0truecm]{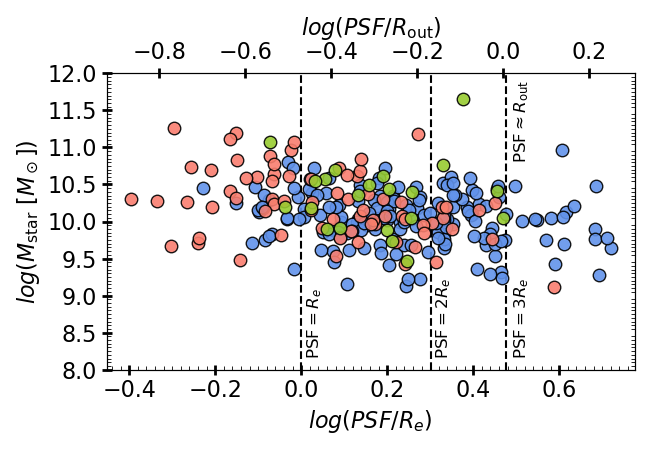}  
					
					\caption{Stellar masses as a function of ratio between ${\rm PSF} / R_{\rm e}$. The KMOS$^{\rm 3D}$, KGES, and KROSS samples are represented by red circles, green crosses, and blue stars, respectively, with a vertical dashed line indicating ${\rm PSF}= x R_{\rm e}$. This figure indicates that only a limited number (= 53) of galaxies are applicable for determining the dark matter fraction within $R_{\rm e}$. However, it is important to note that our analysis is not restricted within $R_{\rm e}$. In fact, we evaluate the dark matter fraction at outer radii (i.e., $R > R_{\rm e}$).  }
					\label{fig:ReRpsf}
				\end{center}
			\end{figure}
			
			\subsection{Kinematic modeling results}
			\label{sec:kmresults}
			Figure~\ref{fig:BBKMKGKR} shows a few examples of the kinematic modeling results. In this figure, the first three columns in each row show the moment maps \footnote{For example, the first moment (or moment-1) is the intensity-weighted velocity map, which shows the mean velocity of the emission cube at each point.}, data, model, and residuals, respectively, from left to right. As we can see, model closely matches the data, and residuals lie around (below) 5\% of the total velocity. However, we do notice slight structures in the residual maps that could potentially be associated with irregular gas motions or turbulent flows, which were not accounted in 3DBarolo, one of the caveats (along with other ones) highlighted in Section~\ref{sec:caveats}. Nevertheless, these results represent the most optimal outcome achievable through the utilization of both 3DBarolo and the optimization function. The fourth column displays the major axis position-velocity diagram (PV diagram), which is the line-of-sight (LOS) rotation velocity of the galaxy at each spatial bin.\footnote{A spatial bin is based on the size of the spherical ring in which the line-of-sight velocity is computed.} The red contour represents the model, and the black shaded area with blue contour represents the data. The orange squares with error bars indicate the best-fit LOS rotation velocity. The yellow and blue vertical dashed lines represent the effective radius ($R_{\rm e}$) and optical radius ($R_{\rm opt}=1.89 \ R_{\rm e}$) of the galaxy, respectively. The last column presents the velocity dispersion curve. The last column presents the velocity dispersion curve. We note that the first (inner) data point in the rotation curves and velocity dispersion profiles occasionally exhibits unexpectedly high or low values, a known issue with the 3DBarolo code \citep{ETD16, GS21a}, which most-likely arises due to limited resolution in the data. In our calculations, we discard this inner data point as it is irrelevant for science. However, we discuss (and present) it here to indicate the limitations of our kinematic modeling technique.
			
			To estimate the DM fraction, we first correct the rotation curves for pressure support, as the turbulent interstellar medium in high-$z$ galaxies makes them pressure-supported systems \citep{Burkert2010, Ubler2019}. In \citet{GS21a}, it was demonstrated that high-z galaxies exhibit a non-uniform and non-isotropic velocity dispersion, inducing a pressure gradient. This pressure gradient significantly hampers the motion of gas, leading to a decrease in the rotation velocity of the gas in the inner region of galaxies by  $\sim 50\%$ of its original value. In some cases, it also affects the outer rotation curves, causing them to decline. 
			
			To address this issue, \citet{GS21a} introduced a method known as the `pressure gradient correction' (PGC), which effectively corrects for gas pressure. This approach is analogous to the `asymmetric drift correction,' which addresses stellar pressure, as discussed in \citet[][Sec.~3.2]{GS21a}. In this study, we applied the PGC method to all datasets and investigated the impact of velocity dispersion on the circular velocity of galaxies. In Figure~\ref{fig:PGCorr}, we present the intrinsic velocities ($V^{\rm int}$) and pressure-corrected circular velocities ($V^{\rm PGC}$) within $R_{\rm out}$, color coded for intrinsic rotation-to-dispersion ratio ($V^{\rm int}_{\rm Rout}/\sigma$), where $V^{\rm int}$ is rotation velocity without pressure support corrections. As shown, systems primarily supported by rotation ($V^{\rm int}_{\rm Rout}/\sigma >1$) exhibit minimal or no pressure correction, whereas dispersion-dominated systems induce significant correction. Additionally, there are substantial pressure corrections at the lower end of the velocity range ($\log(V_{R_{\rm out}} [km/s]) < 2.3$), that  gradually decrease toward higher velocities and approaches zero.  We note to the readers that, prior to the implementation of PGC, there were only 9 dispersion dominated galaxies (3 KMOS$^{\rm 3D}$, 1  KGES, and 5 KROSS). However, after applying PGC, none of these galaxies have $V^{^{\rm PGC}}_{\rm Rout}/\sigma < 1$, as depicted in Figure~\ref{fig:voversigma}. Therefore, we do not exclude these galaxies from our analysis. Hence, the full sample is a good representative of rotation supported system.
			
			\begin{figure}
				\begin{center}
					\includegraphics[angle=0,height=6.0truecm,width=8.5truecm]{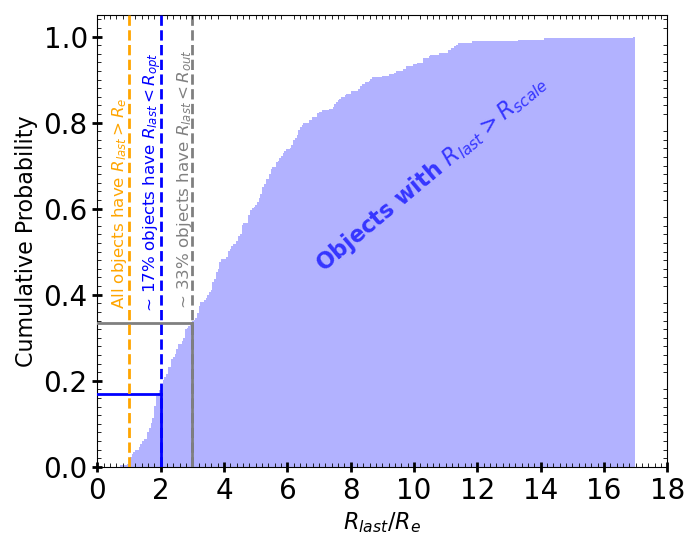}  
					
					\caption{Cumulative distribution of the ratio between the last observed radius in the rotation curves ($R_{\rm last}$) and the effective radius ($R_e$). The yellow, blue, and gray dashed lines indicate the scale radii $R_e$, $R_{\rm opt}$, and $R_{\rm out}$, respectively. This plot shows that for the entire sample, $R_{\rm last} > R_e $, while approximately 83\% and 67\% of objects have $R_{\rm last}$ greater than $R_{\rm opt}$ and $R_{\rm out}$, respectively.}
					\label{fig:ReRlast}
				\end{center}
			\end{figure}
			
			Finally, we examined the relationship between the size of the PSF and the effective radius and found that 59\% (43) of KMOS$^{\rm 3D}$, 90\% (19) of KGES, and 88\% (148) of KROSS galaxies possess a PSF larger than their effective radius, i.e., only 53 galaxies have $R_{\rm e} > {\rm PSF}$ as shown in Figure~\ref{fig:ReRpsf}. This suggests that the majority of the sample cannot be used for studying the dark matter fraction within $R_{\rm e}$, as doing so would result in highly uncertain outcomes. Hence, we only used the 53 galaxies with a large enough effective radius compared to the PSF to characterize the DM fraction within that effective radius. Additionally, 16\% (12) of KMOS$^{\rm 3D}$, 19\% (4) of KGES, and 42\% (71) of KROSS galaxies have a PSF larger than their optical radius. However, only 1.4\% (1) of KMOS$^{\rm 3D}$, 0.0\% (0) of KGES, and 12\% (20) of KROSS galaxies have a PSF larger than their outer radius. Therefore, the most reliable measurement of the dark matter fraction is obtained within the outer radius ($R_{\rm out}$). Consequently, in this work we mostly focus on interpreting the results that are computed  within (or at) $R_{\rm out}$. 
			
			It is important to note that about 83\% and 67\% of the galaxy rotation curves extend to $R_{\rm opt}$ and $R_{\rm out}$, respectively. As shown in Figure~\ref{fig:ReRlast}, only 17\% and 33\% of galaxies have $R_{\rm opt}$ and $R_{\rm out}$ beyond the last observed radius ($R_{\rm last}$) in the rotation curves. For cases where the rotation curve does not reach the reference radius, we interpolate the velocity estimates. We did not assume any specific functional form for the rotation curve; instead, we used the \texttt{numpy.interp} routine, which applies linear interpolation. This approach ensures that if the rotation curve is declining, it will continue to decline, and vice versa. Our approach is as follows: if $R_{\rm opt}$ or $R_{\rm out}$ exceeds $R_{\rm last}$, we calculate $f_{\rm DM}(<R_{\rm scale})$ at the nearest observed point. This method ensures that our analysis remains within the observed region of each galaxy. 
			Moreover, when the characteristic radii (\( R_{\rm out}, R_{\rm opt}, R_e \)) are not covered in the observed rotation curve, we impose additional uncertainties on the circular velocity (\( V_c \)) measurements at those radii to account for potential errors introduced by interpolation or extrapolation. Specifically, we increase the uncertainties by 10\% when \( V_c \) is estimated at \( R_{\rm out} \) or \( R_{\rm opt} \), and by 25\% when evaluated at \( R_{\rm e} \). The smaller penalty at \( R_{\rm out} \) and \( R_{\rm opt} \) is motivated by the fact that they lie in the outskirts where rotation curves are relatively flat; hence, interpolation is less prone to large deviations. In contrast, the inner regions within \( R_{\rm e} \) often exhibit steeper velocity gradients, justifying a more conservative error adjustments. 
			%

			
			\subsection{Final sample} \label{sec:final-sample}
			To summarize the sample selection, it is crucial to note that the initial selection of galaxies for kinematic modeling was based on the following criteria: (1) confirmed H$\alpha$ detection and spectroscopic redshift, (2) inclination angles within the range of $25^{\circ} \leq \theta_i \leq 75^{\circ}$, and (3) S/N $ > 3$. This primary selection criteria is detailed in Section~\ref{sec:dataset} for the KMOS3D and KGES datasets and outlined in Table~\ref{Tab:sample-selection}. Following the primary selection criteria, our chosen sample comprises 265 KMOS3D galaxies and 51 KGES objects, as depicted in Figure~\ref{fig:KMOS3D-P} and Figure~\ref{fig:KGES-P}. For the KROSS dataset, we utilized the complete set of 256 galaxies analyzed in \citet{GS21a}.
			
			Following the kinematic modeling process (Section~\ref{sec:k-model}), we implemented secondary selection criteria as detailed in Section~\ref{sec:km-inspect} and outlined in Table~\ref{Tab:sample-selection}. Under this criteria, galaxies were excluded if they met the following conditions: (1) 3DBarolo+ optimization did not succeed, indicating unreliable optimized parameters such as PA or central coordinates; (2) No mask was created, implying 3DBarolo's failure to mask true emission due to moderate signal-to-noise; (3) $\mathrm{R_{\rm max}< PSF}$, indicating 3DBarolo's inability to create rings and hence fails to produce kinematic models; (4) $\mathrm{R_{\rm max}=PSF}$, in this case resulting kinematic models provide only two measurements in \RCs, which were insufficient for dynamical modeling. This secondary selection criteria resulted in a final sample of 263 galaxies, comprising 169 from KROSS, 73 from KMOS3D, and 21 from KGES. The distribution of relevant physical quantities for the final sample across all datasets is presented in Figure~\ref{fig:final-hist}, and their location with respect to star-forming galaxies is shown in Figure~\ref{fig:Mstar-sfr}. 
			
			\begin{table*}
				\centering
				\caption{Primary and secondary sample selection criteria.}
				\begin{tabular}
					{|p{0.5cm}|p{8cm}||p{1.5cm}|p{1.5cm}|p{1.5cm}|}
					\hline
					\multicolumn{5}{|c|}{\textbf{Primary Selection Criteria}} \\
					\hline
					No. & Criteria & KMOS3D & KGES & KROSS\\
					\hline
					1 & Confirm H$\alpha$ detection and spectroscopic redshift   & 571    & 225    & -   \\
					2 & $25^{\circ} \leq \theta_i \leq 75^{\circ}$ & 448     &  192   & -  \\
					3 & S/N $>3$ & 265 & 51  & - \\
					\hline
					\multicolumn{5}{|c|}{\textbf{Secondary Selection Criteria}} \\
					\hline
					1  &  3DBarolo successfully runs &  213    &  40   & 225   \\
					2 & 3DBarolo+Optimization success = True   &  156  & 37   & 169  \\ 
					3 & $R_{\rm max} > {\rm PSF}$ & 78   &  21  & -  \\
					4 & Non-interacting system in HST images &  73 & - & -  \\
					\hline
					\hline
					* & Final Sample & 73 & 21 & 169 \\
					\hline
				\end{tabular}
				\tablefoot{Primary and secondary selection criteria applied before and after kinematic modeling, respectively. Table gives the  sample size of all three dataset after imposing each criterion.  We note that the KROSS sample has been previously studied in \citet{GS21a}; thus, in this work, we do not apply primary selection criteria to this specific sample. Lastly, $R_{\rm max}$ is the maximum radius of rotation curve.}
				\label{Tab:sample-selection}
				
			\end{table*}
			
			In addition, we emphasize to the reader that the computation of the dark matter fraction within $R_{\rm e}$ was exclusively performed on 53 galaxies having $R_{\rm e} > {\rm PSF}$, as depicted in Figure~\ref{fig:ReRpsf} and discussed in Section~\ref{sec:kmresults}. These 53 objects are also highlighted separately in the Figure~\ref{fig:Mstar-sfr} and Figure~\ref{fig:final-hist}. However, the goal of this work is to go beyond the specific case of the `dark matter fraction within $R_{\rm e}$'. As elaborated in Section~\ref{sec:analysis}, we present the dark matter fraction and scaling relations at outer radii ($R > R_{\rm e}$, where $R$ is the running radius of rotation curve) using the full sample of 263 galaxies. Furthermore, we note that the maximum radius ($R_{\rm max}$ or $R_{\rm last}$) of galaxies in the final sample is larger than the PSF, see Figure~\ref{fig:Rmax-Rpsf}, i.e., this sample is suitable for estimating dark matter at outer radii.
			
			
			\section{Results}\label{sec:analysis}
			Our primary goal is to measure the dark matter fraction across various galactic scales. Recent studies at \hz suggest that galaxies contain baryon-dominant inner regions (within $R_{\rm e}$), where dark matter constitutes less than 20\% of the total mass \citep{Genzel2020, Genzel2022}. Adding to this perspective, a study by \citet{Lelli2023} found that the kinematics of two main-sequence galaxies, from the cosmic dawn, could be entirely explained by baryon only dynamical models, eliminating the need for dark matter. These findings amplify the long-standing disk-halo degeneracy issue, a challenge that persists even in local galaxies \citep{VanAlbada1985}, and remains unresolved, particularly when mass-to-light ratios are not entirely reliable \citep{Sofue2001, Bullock2017}. Therefore, in this work, we adopt a conservative approach, beginning with the mass-modeling of rotation curves under the conditions of maximal baryonic disk, using the Bayesian inference technique established in \citet[][Sec. 3.1]{GS22}. We assumed that stars and gas follow an exponential distribution \citep{Freeman} such that the mass-modeled rotation curve of a galaxy is given by
			\begin{equation}
				\label{eq:Vd}
				V_{\_}^2(R)= \frac{1}{2} \Big( \frac{GM_{\_}}{R_{\rm scale}} \Big) \  \Big( \alpha \frac{R}{R_{\rm scale}} \Big)^2 \  [I_0K_0 - I_1K_1],
			\end{equation}
			where $M_{\_}$ and $R_{\rm scale}$ are the total mass and the scale length of the different components (stars, H2, and HI), respectively, and $I_n$ and $K_n$ are modified Bessel functions computed at $\alpha=1.6$ for stars and $\alpha = 0.53$ for gas \citep[c.f.][]{PS1996, Karukes2017}.
			
\begin{figure}
				\begin{center}
					\includegraphics[angle=0,height=6.0truecm,width=8.5truecm]{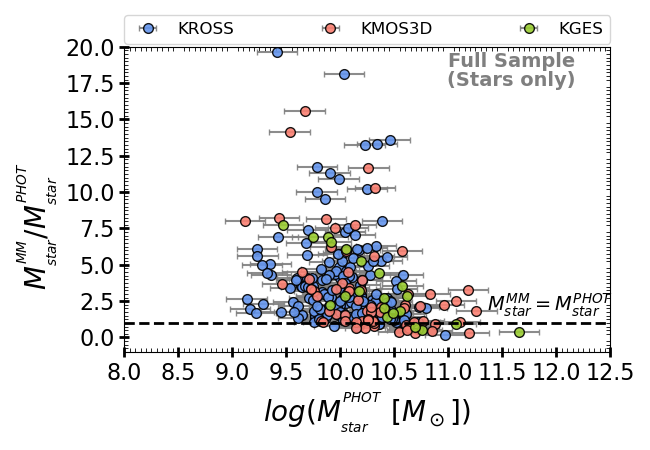}  
					
					\includegraphics[angle=0,height=6.0truecm,width=8.5truecm]{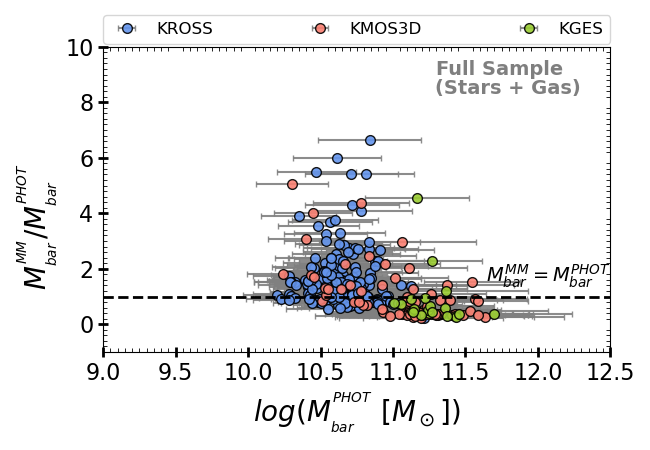}  
					
					\caption{Comparison of photometric (PHOT) stellar and baryonic (stars+gas) masses with mass-modeled stellar and baryonic masses in the case of the maximum stellar disk and maximum stellar+gas disk scenarios, upper and lower panels, respectively. The KMOS$^{\rm 3D}$, KGES, and KROSS datasets are depicted in red, green, and blue colors, respectively. The dashed black line indicates when mass-modeled masses are equivalent to their fiducial values (i.e., photometric masses), suggesting that objects lying above this line require the presence of a dark matter halo.}
					\label{fig:MM-best-param}
				\end{center}
			\end{figure}

			We mass-modeled the \RCs\ under two scenarios that consider the maximum contribution from (1) the stellar disk, i.e., $M_{\_} = M_{\rm star}$, and (2) baryonic (stars+gas) disk, i.e., $M_{\_} = M_{\rm star}+(1.33*M_{\rm HI})+M_{\rm H2}$, details are compiled in Appendix~\ref{sec:mass-models}. Figures~\ref{fig:Mass-models-1} and \ref{fig:Mass-models-2} provide a few examples of mass-modeled RCs. We compare these \RCs\ with their fiducial (fid) values, i.e., velocity profile derived directly from the photometric stellar and gas masses discussed in Section~\ref{sec:Mbar}.  In Figure~\ref{fig:MM-best-param}, we present a comparison between the fiducial (i.e., masses from photometry) and the mass-modeled stellar and baryonic masses in the case of maximum stellar disk and maximum baryonic disk scenarios (upper and lower panels, respectively). When considering the maximum stellar disk, we observed that fitting the \RCs\ requires stellar masses that are twice the values obtained from photometric measurements for the majority of the sample. This finding strongly suggests the requirement of an additional gas component at \hz, which we incorporate in the maximum baryonic disk scenario. However, as illustrated in the bottom panel of Figure~\ref{fig:MM-best-param}, even after including the gas component, more than 50\% of the sample still demands baryonic masses that are a factor of two higher than their fiducial values. Such values are unrealistic to obtain within the range of observed uncertainties. That is, with a large sample of 263 galaxies, we are unable to rule out the presence of dark matter halos at \hz.  
			
			Moreover, recent observations of \hz galaxies suggest that longer integration time is crucial for accurately mapping the complete kinematics. In particular, a work by \citet[][]{KURVS} on KMOS Ultra-deep Rotation Velocity Survey (KURVS) has demonstrated that deep observations enhance both the amplitude and radial extent of the \RCs. Consequently, deep observations of our current sample will further demand the inclusion of an additional halo component, as anticipated while analyzing the mass-modeled \RCs\ (see Figure~\ref{fig:Mass-models-2}). Additionally, the Freeman model assumes a razor-thin disk, which is an extreme case. Allowing for a finite thickness of the disk would further decrease the circular velocity of the baryonic component and will provide the more room for the dark matter. Henceforth, we proceed the investigation of the rotation curves in the presence of a dark matter halo component.

			\begin{figure}
				\begin{center}
					
					\includegraphics[angle=0,height=6.2truecm,width=8.5truecm]{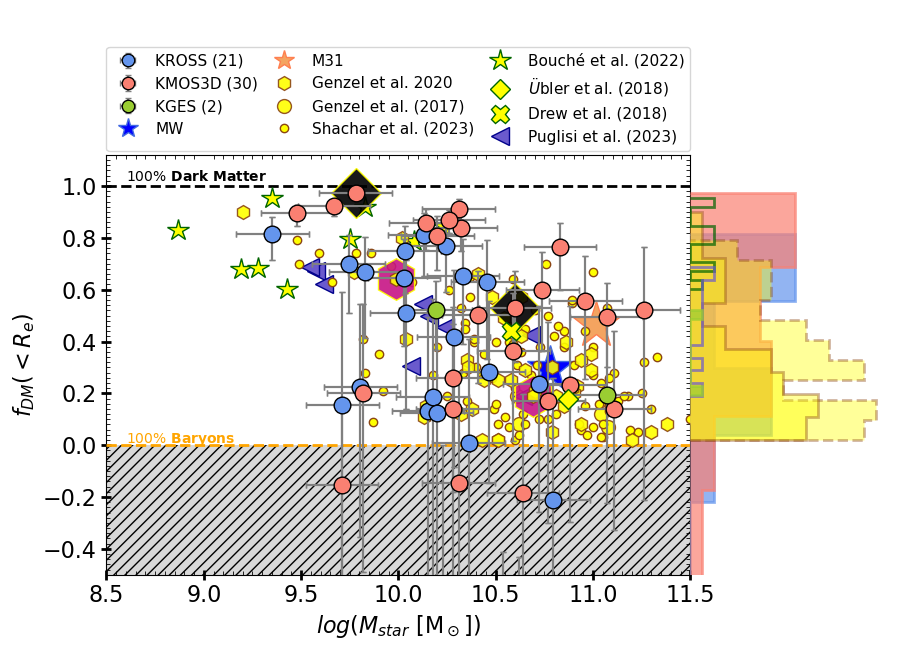}  
					
					\caption{Dark matter fraction within $R_{\rm e}$ as a function of stellar masses for galaxies having $R_{\rm e} > PSF$. The KMOS$^{\rm 3D}$, KGES, and KROSS datasets are depicted in red, green, and blue colors, respectively. The errors on the datasets are 68\% confidence intervals. For comparison, we include two local massive disk galaxies: the Milky Way (MW) and Andromeda (M31), represented by the blue and the orange star, respectively. Additionally, we compare these results with previous studies at high redshift, such as \citet[][yellow circles]{Genzel2017}, \citet[][yellow hexagons]{Genzel2020}, \citet[][yellow dots]{Genzel2022}, \citet[][yellow star]{Bouche2022}, \citet[][yellow diamond]{Ubler2018}, \citet[][yellow cross]{Drew2018}, and \citet[][purple triangle]{KURVS}. The black square represents the two galaxies that are in-common with \citet{Genzel2020} drawn from this work, while pink hexagons shows the measurements of same objects from \citet{Genzel2020}. The black and yellow horizontal dashed lines represent the regimes of 100\% dark matter dominance and baryon dominance, respectively. The gray shaded area indicates the `forbidden' region, where galaxies with $M_{\mathrm{dyn}} < M_{\mathrm{bar}}$ are located.  Horizontal histograms (aligned to the right y-axis) compare datasets: KMOS\(^{\rm 3D} \), KGES, and KROSS (filled red, green, and blue hists), \citet{Genzel2017, Genzel2020} (yellow fill with solid brown line), \citet{Genzel2022} (yellow fill with dashed brown line ), and \citet{Bouche2022} and \citet{KURVS} (open histograms in green and purple, respectively).
					} 
					\label{fig:FdmRe}
				\end{center}
			\end{figure}
			
			\begin{figure}
				\begin{center}
					\includegraphics[angle=0,height=6.5truecm,width=8.5truecm]{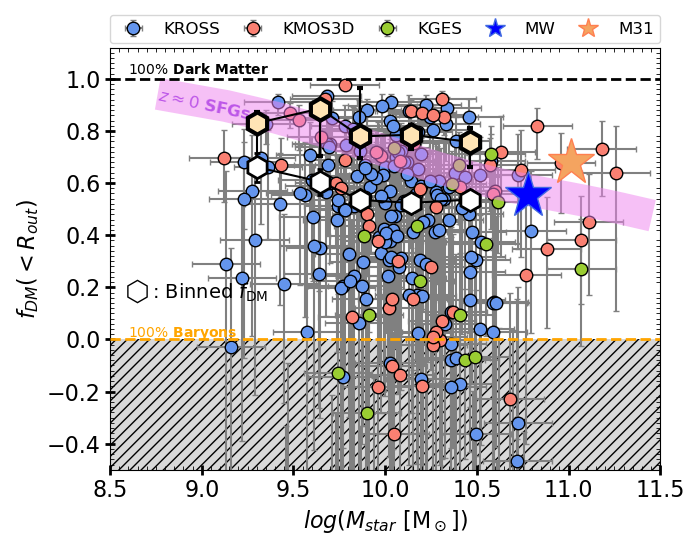} 

					\caption{Dark matter fraction within $R_{\rm out}$.  The KMOS$^{\rm 3D}$, KGES, and KROSS datasets are represented in red, green, and blue colors, respectively. The binned dark matter fraction, calculated using weighted mean and root-mean-square statistics discussed in Section~\ref{sec:fdm-R}, is represented by off-white and white color hexagons, respectively, connected by black lines. The uncertainties on individual and binned data points represent the 68\% confidence interval. For comparison, we include two local massive disk galaxies: the Milky Way (MW) and Andromeda (M31), represented by the blue and the orange star, respectively. The pink shaded area represents local star-forming disks \citep{PS1996}. The solid black dashed line represents the 100\% dark matter regime, while the yellow dashed line represents the baryon-dominated regime. The gray shaded area indicates the `forbidden' region, where galaxies with $M_{dyn}< M_{\rm bar}$ are located. These color codes are same throughout the text.}
					\label{fig:FdmRout}
				\end{center}
			\end{figure}
			
			The total dark matter mass can be calculated by subtracting the baryonic mass contribution from observed rotation curve, \textbf{($V^2(R) - V_{\rm bar}^2(R)= GM_{\rm DM}/R$)}
			as a function of radius, as explained in Appendix~\ref{sec:mass-models}. That is, dark matter halo modeling is not necessary to study the amount of dark matter. In order to estimate the dark matter fraction of our datasets, we used Equation~\ref{eq:fdm}, which is a halo model independent approach (previously postulated in \citealt{GS21b}), yet it allowed us to estimate the dark matter at different galactic scales.  We estimated the uncertainties in the dark matter fraction using Monte Carlo sampling, propagating the errors in velocity and mass estimates throughout the analysis. We begin by inspecting dark matter fraction within $R_{\rm e}$, which is a rather controversial issue for high-$z$ galaxies. For example, \citet{Genzel2020} and \citet{Genzel2022} reports dark matter-deficiency within $R_{\rm e}$ at \hz, on the other hand, \citet{GS21b} and \citet{Bouche2022} reported a similar amount ($>45\%$) of dark matter fraction as seen in local star-forming disk galaxies. 
			
			\subsection{Dark matter fraction within $R_{\rm e}$} \label{sec:fdm-Re}
			Here, we present the \dm\ fraction only for the galaxies that have $R_{\rm e} > {\rm PSF}$. In total, we have only 20\% (53) galaxies that abide this criteria, and cover the redshift range of $0.73 \leq z \leq 2.43$. \footnote{Redshift range of samples plotted for dark matter fraction within $R_{\rm e}$; KROSS: $0.79 \leq z \leq 0.99$, KMOS3D: $0.73 \leq z \leq 2.43$, and: KGES $1.35 \leq z \leq 1.55$.} The results are shown in Figure~\ref{fig:FdmRe}. We observe that \textbf{six} galaxies fall in the `forbidden' region, where $M_{dyn}< M_{\rm bar}$, possibly indicating inaccurate photometric stellar mass estimates or incomplete sampling of stellar and gas motion in the observed \RCs. However, we do expect some data points with low dark matter fraction to fall within this region given the uncertainties on the measurements of stellar masses and SFRs. Apart from forbidden region galaxies, $\sim 8$ galaxies exhibit dark matter deficiency within $R_{\rm e}$ ($f_{_{\mathrm{DM}}} < 20\%$). Notably, the dark matter fraction of very massive ($> 10^{10.5} \ \mathrm{M_\odot}$) galaxies in our sample surpasses that of local massive galaxies, such as the Milky Way \citep[][and ref. therein]{MilkyWay} and Andromeda galaxies \citep[][and ref. therein]{Andromeda}, represented by blue and orange stars in Figure~\ref{fig:FdmRe}, respectively.
			
			We compared our findings with those of previous studies, such as \citet{Genzel2017, Ubler2018, Drew2018, Genzel2020, Genzel2022, Bouche2022}, and \citet{KURVS}, as shown by the different markers in Figure~\ref{fig:FdmRe}. Our results are in full agreement with \citet{Bouche2022, Drew2018}, and a few galaxies of \citet{Genzel2020, Genzel2022} and \citet{KURVS}. However, we observe that the majority of \citet{Genzel2022}, \citet{Genzel2020}, and \citet{Genzel2017} are baryon-dominated, including a galaxy of \citet{Ubler2018}. Specifically, the very massive galaxies studied in \citet{Genzel2020} and \citet{Genzel2022} are dark matter deficient, which contrasts with our findings, where majority of galaxies in the same mass range have $\sim 30\% - 58\%$ dark matter, except two. However, we cannot draw any concrete conclusions due to the limited number of data points and their large uncertainties. Nevertheless, to aid the reader and facilitate a visual comparison between studies, we include horizontal histograms corresponding to each dataset along the y-axis of Figure~\ref{fig:FdmRe}, illustrating the degree of agreement or discrepancy.
			
			Next, we cross-matched KMOS$^{\rm 3D}$ sample and the galaxies studied in \cite{Genzel2020}, resulting in the identification of only two overlapping systems: GS4\_05881 ($z=0.99$) and GS4\_43501 ($z=1.61$). These systems are represented by black squares in Figure~\ref{fig:FdmRe}. According to \cite{Genzel2020}, the reported dark matter fractions within $R_{\rm e}$ for these two systems are $f_{_{\rm DM}} (<R_{\rm e}) = 0.64$ and $0.19$, respectively, shown by pink hexagons. While our estimates are about $0.98$ and $0.55$, respectively, i.e. $\sim$ 1.5 and 3 times higher than the estimates reported in \cite{Genzel2020}. This difference is most likely attributed to the distinct kinematic modeling and pressure support corrections implemented in this work, quantified and reported previously in \citet{GS21a, GS21b}. Moreover, it is noticeable that the stellar masses of these objects reported in \citet{Genzel2020} are marginally higher than those in the present study. This discrepancy arises because we employ photometric stellar masses, whereas \citet{Genzel2020} utilizes dynamically mass-modeled (best-fit) stellar masses.

			Furthermore, we had the opportunity to refine the measurement of $R_{\rm e}$ for a subset of the KROSS sample through the latest observations from the James Webb Space Telescope (JWST). This particular subset of galaxies is situated within the COSMOS field \citep{Skelton2014}, which has recently undergone observation by the COSMOS-WEB team \citep{COSMOS-Web}. Estimates of $R_{\rm e}$ for these galaxies were derived using GALFITM \citep{GALFITM2013} setup equal to that presented in  \citet{Martorano2023}. As discussed in Appendix~\ref{sec:extra1} and illustrated in Figure~\ref{fig:FdmRe-JWST}, we observe that the new $R_{\rm e}$ estimates shift the galaxies of low dark matter fractions ($<20\%$) toward higher values. In this specific sub-sample, majority (~95\%) of the galaxies have a dark matter fraction above $20\%$. Consequently, we suggest that a low ($<0.2$) dark matter fraction within $R_{\rm e}$ is possible for massive disk-like galaxies at high redshifts ($0.65 \leq z \leq 2.2$), but it is very unlikely for all mass ranges ($8.5 \leq log(M_* \ [M_\odot]) \leq 10.5$). 

			\begin{figure*}
				\begin{center}
					\includegraphics[angle=0,height=6.5truecm,width=8.5truecm]{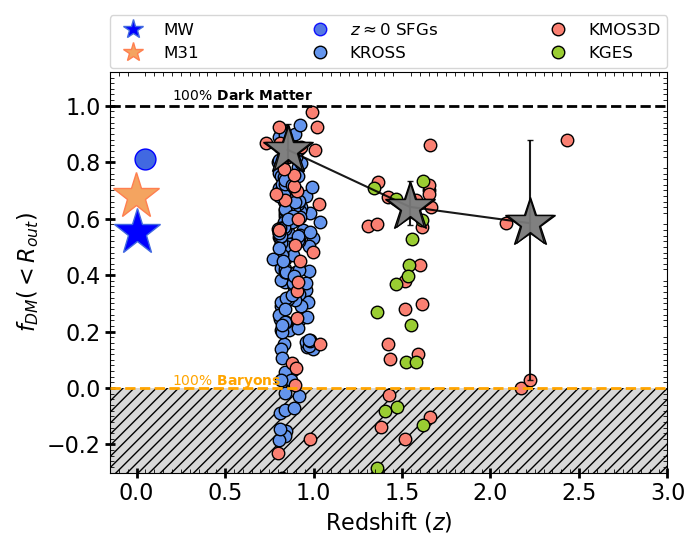}  
					\includegraphics[angle=0,height=6.5truecm,width=8.5truecm]{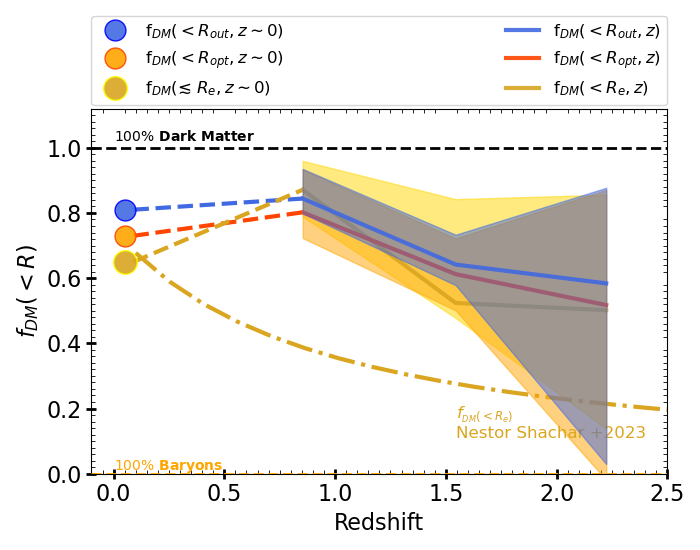} 
					\caption{Dark matter fraction across the cosmic time. {\em Left panel:} Dark matter fraction within $R_{\rm out}$ as a function of redshift. The color codes of datasets are same as Figure~\ref{fig:FdmRout}, and given in the legend of the plot. The gray stars represents the binned data points with 1$\sigma$ uncertainty shown by error-bars.  {\em Right panel:} Averaged dark matter fraction within $R_{\rm e}$, $R_{\rm opt}$, and $R_{\rm out}$ as a function of redshift, represented by yellow and blue lines, respectively. The bootstrap-errors representing 68\% confidence interval are indicated by shaded areas color-coded with the same color as the averaged line. Dark matter fraction of local \sfgs\ within $R_{\rm opt}$ and $R_{\rm out}$  are from \citet{PS1996}, and within $R_{\rm e}$ they are taken from disk galaxy survey of \citet{Courteau2015}. The uncertainties on these measurements are, on average, $\pm 0.25$. The color code of these estimates are same as high-$z$ galaxies. The dark-yellow dotted-dashed lines shows the dark matter fraction within $R_{\rm e}$ from \citet{Genzel2022}. 
					}
					\label{fig:Fdm-z}
				\end{center}
			\end{figure*}
			
			\subsection{Dark matter fraction within $R_{\rm out}$}\label{sec:fdm-R}
			In Figure~\ref{fig:FdmRout}, we plot the dark matter fraction within $R_{\rm out}$ ($= 2.95 \times R_{\rm e}$), as a function of stellar mass. Firstly, we observe that the 47\% of the sample exhibits dark matter-dominated outer disks, with 23\% of objects showing $0.2 \leq f_{\mathrm{_{DM}}} (<R_{\rm out}) <0.5$, and  28\% of objects having $f_{\mathrm{_{DM}}} (<R_{\rm out})<0.2$ (including objects in the forbidden region). Secondly, we notice a slightly decreasing trend of $f_{_{\rm DM}}$ as a function of stellar mass, excluding a few outliers. These outliers are six massive galaxies in the KMOS$^{\rm 3D}$ sample that display prominent disks in HST images and exhibit rising-flat rotation curves, as shown in Figure~\ref{fig:massive_gal}.\footnote{The following massive galaxies have very high dark matter fractions: COS4\_23890 ($z = 0.85$, $M_*=10^{10.83}$, $f_{_{\rm DM}} = 0.83$), U4\_19708 ($z= 1.66$, $M_*=10^{11.26}$, $f_{_{\rm DM}} = 0.64$), U4\_31577 ($z= 1.52$, $M_*=10^{11.07}$, $f_{_{\rm DM}} = 0.37$), and U4\_26875 ($z= 1.36$,  $M_*=10^{11.18}$, $f_{_{\rm DM}} = 0.73$), U4\_28358 ($z= 1.42$, $M_*=10^{10.98}$, $f_{_{\rm DM}} = 0.68$), U4\_21533 ($z= 0.92$, $M_*=10^{11.11}$, $f_{_{\rm DM}} = 0.46$).} Notably, their dark matter halos are as massive as those of the most massive systems in the present-day Universe,  such as the Milky Way (MW) and Andromeda (M31), shown by blue and orange stars, respectively, in Figure~\ref{fig:FdmRout}. 
			
			To compare these measurements with local star-forming galaxies (SFGs) \citep{PS1996}, we average the dark matter fraction, \( f_{\mathrm{_{DM}}} (<R_{\rm out}) \), across five stellar mass bins, excluding the massive galaxies (\(\log(M_{\rm star}/M_\odot) > 10.5\)). Recalling \citet{GS21b}, authors estimated the dark matter fraction using the same method as in this work. However, the uncertainties were propagated using the \texttt{python-uncertainty} package without accounting for systematics in stellar masses, star formation rates, and gas masses, resulting in relatively small uncertainties. Therefore, simple root-mean-square statistics was used to compute the average values and errors (for details, see \citet{GS21b}). The resulted dark matter fractions agreed with local studies. In this work, applying the same statistics results in 20\% low dark matter fraction compared to local studies as shown by white hexagons in Figure~\ref{fig:FdmRout}.
			
			We remark that in this work, we estimated the uncertainties by accounting for both statistical and systematic uncertainties in stellar masses and star formation rates (SFRs), as well as the scatter in gas mass scaling relations (see Section~\ref{sec:KMOS3D} \& \ref{sec:KGES} ). These uncertainties are propagated throughout the analysis using Monte Carlo sampling with 10,000 samples. This procedure leads to more reliable results, albeit at the cost of significantly larger uncertainties on individual data points. Consequently, the conventional binning method (as employed in \citealt{GS21b}) used to average out the dark matter fraction likely underestimates its values. Therefore, we also applied weighted-mean statistics, given by $\bar{X} = \frac{\Sigma_{i=1}^{n}x_i \times w_i}{\Sigma_{i=1}^{n}w_i}$, where, $ x_i=\mathrm{data \ in \ specific \ stellar \ mass \ bin}, \ w_i=1/error^2, \ \mathrm{and} \ \bar{X}=\mathrm{averaged \ data}$. This binning technique assigns greater influence to data points with higher precision by weighting the errors during averaging. The uncertainties on the binned data points are estimated using bootstrap resampling. As shown by off-white hexagons in Figure~\ref{fig:FdmRout}, the binned dark matter fraction within $R_{\rm out}$ matches the local studies. Additionally, we observe a slightly declining trend as a function of stellar mass.

			In Appendix~\ref{sec:extra1}, Figure~\ref{fig:FdmRopt} shows the results of dark matter fraction within $R_{\rm opt}$ ($= 1.89 \times R_{\rm e}$). We notice that dark matter fraction is on average 10\% less within $R_{\rm opt}$ than $R_{\rm out}$, suggesting that dark matter dominates the outer-disks at high-$z$, which is very similar to local disk galaxies \citep{PS1996}. Finally, we investigate galaxies with $f_{\mathrm{_{DM}}} (<R_{\rm out})<20\%$ in relation to their PSF and S/N, but no dependencies are found.

			To further investigate, we estimated the stellar masses (within scale radius, e.g., $R_{\rm out}$) of these objects using the Sérsic profile. It is important to note that the stellar masses derived under the Freeman disk assumption are, on average, 1.02 times higher than those obtained using the Sérsic profile (see Appendix~\ref{sec:method-fdmrac} and Figure~\ref{fig:Mstar-sersic}). In ideal cases where dark matter dominates, the dark matter fraction estimated using the Sérsic profile—assuming a Sérsic index of \( n = 1 - 1.35 \)—is only $\sim 2\%$ higher. However, significant changes arise when the stellar or gas mass constitutes a substantially larger fraction of the total mass, typically by 0.7–1 dex (i.e., a factor of five to ten), as observed in galaxies with low dark matter fractions in our sample. Furthermore, assuming a Sérsic index of \( n > 1.5 \) implies that galaxies are more dark matter dominated in the outskirts while being baryon-dominated in the inner regions. Consequently, modifying the assumed stellar mass profile can significantly increase the estimated dark matter fraction for systems that lie in the `forbidden' region or exhibiting low \( f_{\mathrm{DM}} (<20\%) \) within \( R_{\rm out} \).  Since Sérsic indices for the full sample are not available and computing them is beyond the scope of this study, we adopt the Freeman disk profile throughout. This choice is motivated by the fact that galaxies in our sample are predominantly disky morphology seen in the high-resolution images. Nevertheless, we acknowledge the importance of further investigating the role of galaxy morphology and structural parameters in shaping the inferred dark matter fraction.

			\subsection{Dark matter fraction across cosmic time}\label{sec:fdm-z}
			To gain a deeper understanding of the evolution of dark matter with cosmic time, we initially divided the dark matter fraction, $f_{\mathrm{_{DM}}} (<R_{\rm out})$, into three redshift bins: $0.5<z\leq1.1$, $1.1<z\leq1.8$, and $1.8<z\leq2.5$. We employed weighted mean statistics to determine the binned values, and the errors were estimated using a bootstrap method as explained in Section~\ref{sec:fdm-R}. We note that during binning, we use only resolved galaxies with $f_{\rm_{DM}} (<R_{\rm out}) > 0$. The results are presented in Figure~\ref{fig:Fdm-z} (left panel), where the binned data points are denoted by large gray stars connected by a solid black line. Upon inspecting the figure, it becomes evident that the dark matter fraction exhibits a decreasing trend within the redshift range of $0.85 \leq z \leq 1.8$. To confidently establish this trend beyond $z>1.8$, additional data is required. Consequently, we refrain from further discussing the higher-redshift bin ($z>1.8$) in our analysis.
			
			In the right panel of Figure~\ref{fig:Fdm-z}, we present the binned dark matter fraction within $R_{\rm e}$, $R_{\rm out}$, and $R_{\rm out}$, as a function of redshift. We observe that the dark matter fraction increases on galactic scales as we move outward, from \( R_{\rm opt} \) to \( R_{\rm out} \). We had anticipated that the dark matter fraction within \( R_e \) would be lower than the \( R_{\rm opt} \) and \( R_{\rm out} \). However, our findings show the contrary: \( f_{\mathrm{DM}}(<R_{\rm e}) > f_{\mathrm{DM}}(<R_{\rm out}) \) at \( z \sim 1 \). This discrepancy could be real or an artifact caused by the very low number of resolved galaxies within \( R_e \). Nevertheless, at $z\sim 1$ the difference in dark matter fraction at different radii is not significant, but it is certainly not declining rapidly. On an average, the dark matter fraction within the effective radius, at $z=1-1.8$, does not go below 50\%. These findings, especially, $f_{_{\mathrm{{DM}}}} (<R_{\rm e})$ stand in contrast to those of \citet{Genzel2022}, who reported a rapid decrease and low dark matter fraction within $R_e$ as a function of redshift.  However, it is worth noting that in order to further refine our understanding on dark matter content within $R_{\rm e}$, more resolved (high-quality) observations are required.

			\begin{figure*}
				\begin{center}
					\includegraphics[angle=0,height=6.0truecm,width=8.5truecm]{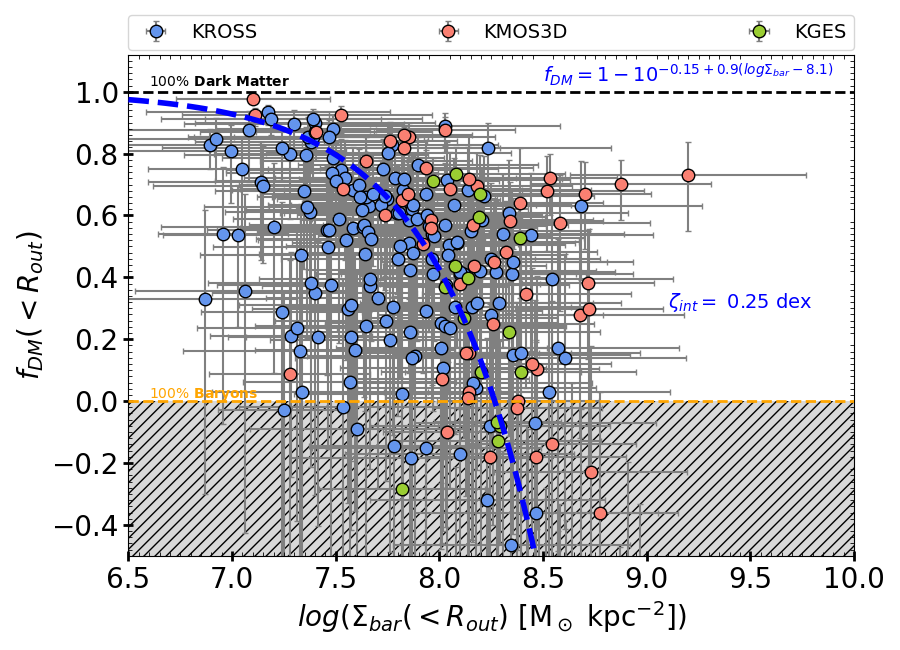}  
					\includegraphics[angle=0,height=6.0truecm,width=8.5truecm]{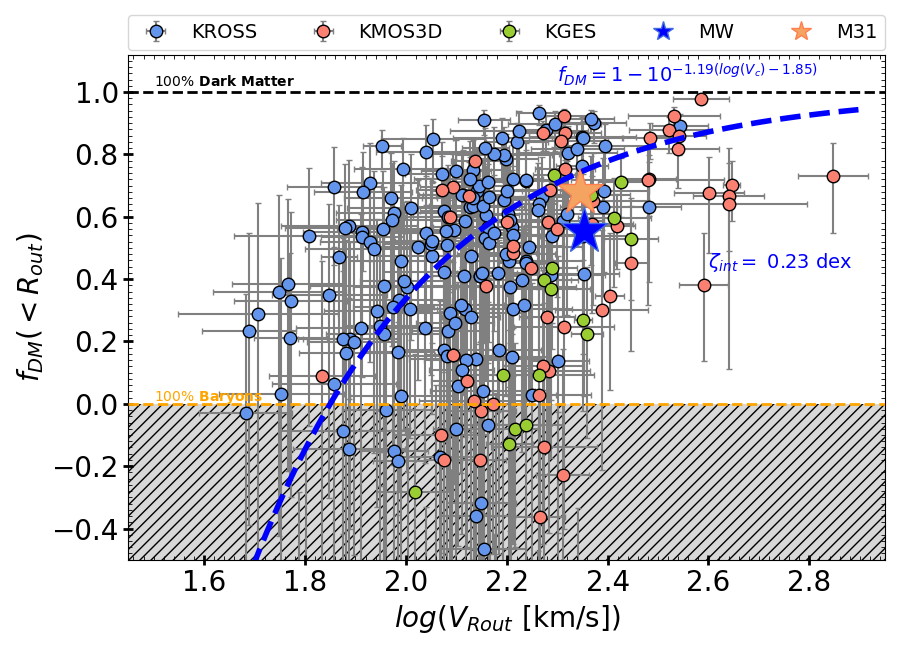} 
					
					\caption{Scaling relations of dark matter, $f_{_{\rm DM}} - \Sigma_{\rm bar}$ and $f_{\mathrm{_{DM}}}  - V_{\rm c}$, shown in the left and right panels, respectively. The blue dashed line represents the best fit to the data in both relations, with the associated intrinsic scatter ($\zeta_{int}$) indicated on the plot. We note that the $f_{\mathrm{_{DM}}}  - V_{\rm c}$ relation can be seen as the result of a combination of the mass-velocity and mass-size relations. The color codes used for data in both panels are consistent with Figure~\ref{fig:FdmRout}. We note that the uncertainties on data points represent the 68\% confidence interval.}
					\label{fig:Fdm-SD-Vc}
				\end{center}
			\end{figure*}

			\begin{figure*}
				\begin{center}
					\includegraphics[angle=0,height=4.0truecm,width=6.0truecm]{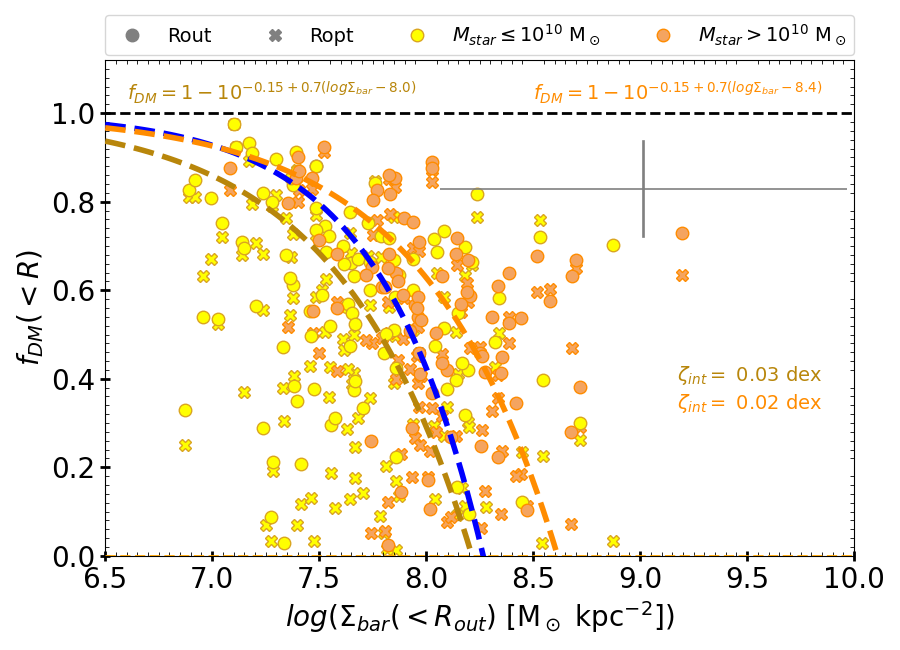}  
					\includegraphics[angle=0,height=4.5truecm,width=12.0truecm]{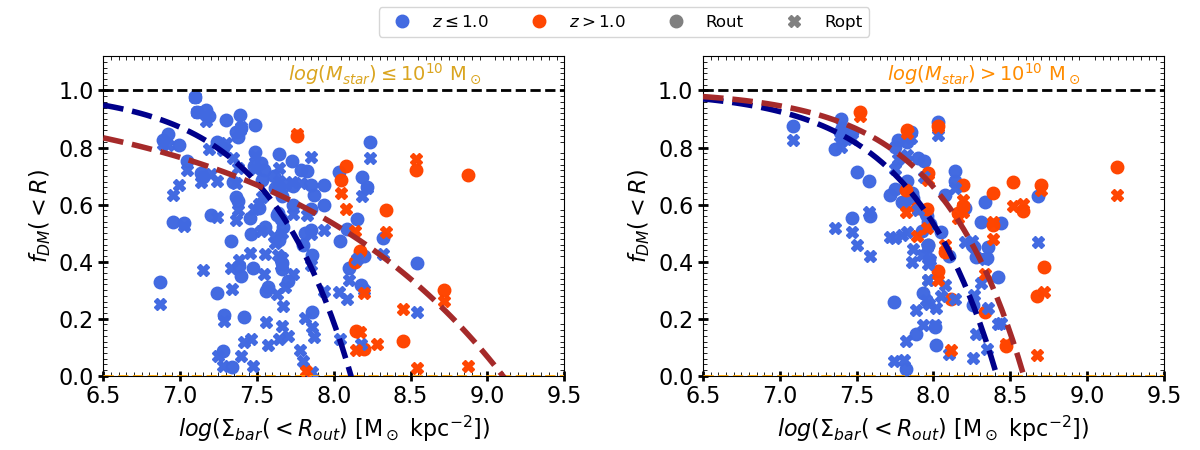} 
					
					\caption{Exploration of the $f_{_{\rm DM}} - \Sigma_{\rm bar}$ relation across different stellar masses and redshift intervals. {\em First panel:} $f_{_{\rm DM}} - \Sigma_{\rm bar}$ relation for the low-stellar mass bin ($M_{\text{star}} \leq 10^{10} \ M_\odot$) in yellow and the high-stellar mass bin ($M_{\text{star}} > 10^{10}\ M_\odot$) in orange. The original best-fit to $f_{_{\rm DM}} - \Sigma_{\rm bar}$ relation is shown blue dashed line, while the best fit of low- and high-stellar mass bins are shown by brown and orange dashed lines, respectively. {\em Second-third panel:} $f_{_{\rm DM}} - \Sigma_{\rm bar}$ relation separated for low- and high-stellar masses, left and right panels respectively. In both panels, low-$z$ ($z\leq 1$) and high-$z$ ($z>1$) galaxies are shown in blue and red, respectively, and corresponding best fits are shown in dark blue and red dashed-lines. In the first panel, median error on the data points is indicated by gray cross. The filled crosses \& circles represent the measurements taken within $R_{\rm opt}$ and $R_{\rm out}$, respectively. We note to the reader that these relations are plotted exclusively for objects with  $f_{_{\rm DM}}>0$, while the full sample is shown in Figure~\ref{fig:Fdm-SD-Vc}. }
					\label{fig:Fdm-SD-mass-z}
				\end{center}
			\end{figure*}
			
			\subsection{Dark matter scaling relations}\label{sec:fdm-Srelations}
			Here, we present the dark matter fraction correlation with the baryon surface density $\Sigma_{\text{bar}} (<R_{\text{out}})$ and the circular velocity $V_{\text{Rout}}$ of galaxies, as shown in the left and right panels of Figure~\ref{fig:Fdm-SD-Vc}, respectively. We fit both correlations using an exponential power law, represented mathematically as follows:
			\begin{eqnarray}\label{eq:fdm-Srel-1}
				f_{_{\rm DM}} (<R_{\text{out}}) & = & 1- 10^{-0.15\pm 0.2 + 0.9\pm0.3(\log(\Sigma_{\text{bar}}) - 8.1\pm0.5)} 
			\end{eqnarray}
			\begin{eqnarray}\label{eq:fdm-Srel-2}
				f_{_{\rm DM}} (<R_{\text{out}}) & = & 1- 10^{-1.19\pm0.2(\log(V_{\rm c}) - 1.85\pm0.2)}
			\end{eqnarray}
			These relations were fit by minimizing the intrinsic scatter. As shown in Figure~\ref{fig:Fdm-SD-Vc}, the dark matter fraction displays a negative correlation with baryon surface density, which is expected and observed at lower redshifts (see \citealt{McGaugh2010}). On the contrary, it exhibits a positive correlation with circular velocity. Together these relations imply that, although the dynamics of galaxies are dominated by dark matter, the baryons still play an important role in hampering the presence of dark matter, i.e., the evolutionary stages of baryonic matter most likely seem to strongly impact the distribution of dark matter within galaxies. This has long been known at low redshift, but it is highly interesting that the trend seems to continue at higher redshift, too.
			
			To thoroughly examine the scenario postulated above, we divided our full sample into two stellar mass bins: a low-mass bin ($M_{\text{star}} \leq 10^{10} \ M_\odot$) and a high-mass bin ($M_{\text{star}} > 10^{10} \ M_\odot$). We plotted the $\Sigma_{\text{bar}}(<R_{\text{out}}) - f_{_{\text{DM}}}(<R_{\text{out}})$ correlations in the first panel of Figure~\ref{fig:Fdm-SD-mass-z}. We note to the reader that this relation is plotted exclusively for objects with  $f_{_{\rm DM}}>0$. The low stellar mass objects are represented by bright yellow, while the high mass objects are depicted in orange. We observed that the $\Sigma_{\text{bar}}(<R_{\text{out}}) - f_{_{\text{DM}}}(<R_{\text{out}})$ relation, as given in Equation~\ref{eq:fdm-Srel-1} remains roughly the same for both mass ranges, indicating no change in the slope. Moreover, this relation remains consistent when observed for $f_{_{\rm DM}} (<R_{\rm opt})$, represented by cross marks, i.e., it is valid at different radii within galaxies. 
			
\begin{figure*}
				\begin{center}
					\includegraphics[angle=0,height=4.0truecm,width=6.0truecm]{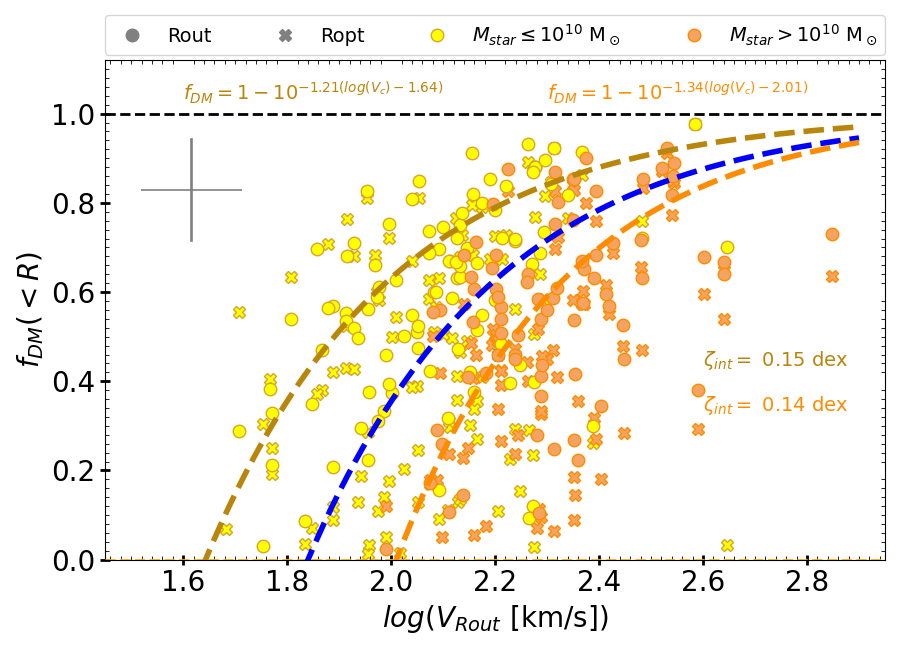}  
					\includegraphics[angle=0,height=4.5truecm,width=12.0truecm]{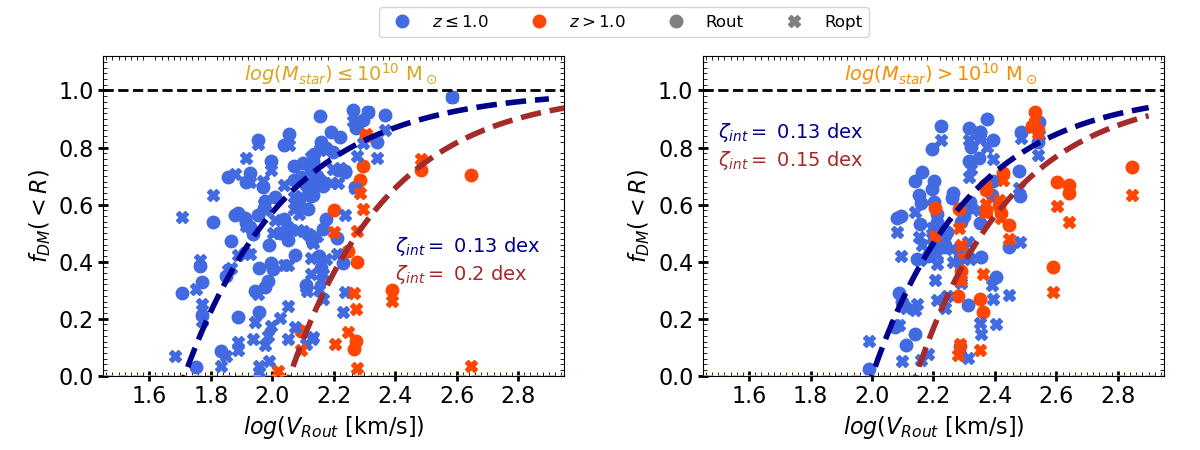} 
					
					\caption{Exploration of the $f_{_{\rm DM}} - V_{\rm c}$ relation across different stellar masses and redshift intervals. {\em First panel:} $f_{_{\rm DM}} - V_{\rm c}$ relation for the low-stellar mass bin ($M_{\text{star}} \leq 10^{10} \ M_\odot$) in yellow and the high-stellar mass bin ($M_{\text{star}} > 10^{10}\ M_\odot$) in orange. The best-fit to original $f_{_{\rm DM}} - V_{\rm c}$ relation is shown blue dashed line given by Equation~\ref{eq:fdm-Srel-2}. The Equations corresponds to best fit of low- and high-stellar mass bins is written on the plot, shown by brown and orange dashed lines, respectively. The intrinsic scatter ($\zeta_{int}$) associated with fits are printed on plot using same color code. {\em Second-third panel:} $f_{_{\rm DM}} - V_{\rm c}$ relation separated for low- and high-stellar masses, left and right panel respectively. In both panels, low-$z$ ($z\leq 1$) and high-$z$ ($z>1$) galaxies are shown in blue and red, respectively, and corresponding best-fits are shown by dashed blue and red lines. In all panels, filled crosses and circles represent the measurements taken within $R_{\rm opt}$ and $R_{\rm out}$, respectively, and median error on the data points is indicated by gray crosses.  Low surface brightness and low velocity low mass galaxies at high redshift might be missed from the sample. On the other hand, it is clear that DM fractions are lower at fixed rotational velocity at higher redshift, i.e. the plots are not biased on the vertical axis. We note to the reader that these relations are plotted exclusively for objects with  $f_{_{\rm DM}}>0$ for the clarity on the matter, while the full sample is shown in Figure~\ref{fig:Fdm-SD-Vc}. }
					\label{fig:Fdm-Vc-mass-z}
				\end{center}
			\end{figure*}

			Conversely, when we plot $V_{\text{Rout}} - f_{_{\text{DM}}}$ (at both radial scales: $R_{\text{opt}}$ and $R_{\text{out}}$) for different mass bins we observed a distinct offset in the relation, as shown in the first panel of Figure~\ref{fig:Fdm-Vc-mass-z}. This suggests that galaxies with high stellar mass are fast-rotating systems with a relatively low dark matter fraction at the outer radius for a given rotational velocity, while the opposite trend is observed for low stellar mass systems. This likely suggests an evolution in the distribution of dark matter due to baryonic processes that take place in massive galaxies. 
			
			Next, we segregated low and high stellar mass galaxies into low and high redshift ($z\leq 1$ and $z>1$, respectively) as shown in the second and third panel of Figure~\ref{fig:Fdm-Vc-mass-z}. We observed very clearly distinct behaviors within the low mass galaxy population. Specifically, at $z>1$ in this population, galaxies exhibited lower dark matter fractions at a given rotational velocity compared to the same mass range at $z\leq 1$. In contrast, high mass systems demonstrate a more similar behavior for the two redshift bins, with only a slight variation in circular velocities and dark matter fractions.\footnote{We note to the reader that correlations in Figure~\ref{fig:Fdm-SD-mass-z} \& \ref{fig:Fdm-Vc-mass-z} are plotted exclusively for objects with  $f_{_{\rm DM}}(<R_{\rm opt})>0$.} 
			
			We also segregated low and high stellar mass galaxies into low and high redshift ($z\leq 1$ and $z>1$, respectively) for the baryon surface density relation, as shown in the second and third panel of Figure~\ref{fig:Fdm-SD-mass-z}. We notice that in the high stellar mass bin, both low and high-$z$ galaxies follows a similar trend with the same shape as given by Equation~\ref{eq:fdm-Srel-1}. 
			The same is true for the low-$z$ galaxies of the low stellar mass bin. However, we noted that the high-$z$ galaxies of the low stellar mass bin display a sharp cutoff at surface densities lower than $log(\Sigma_{\text{bar}}(<R_{\text{out}})) \approx 8.0 $. This is most likely hinting to an observational bias, i.e., low baryon density galaxies seem to be missing at $z>1$. This causes the fit to not follow the Equation~\ref{eq:fdm-Srel-1}, but the discrepancy is not as clear visually as for the $V_{\text{Rout}} - f_{_{\text{DM}}}(<R_{\text{out}})$ relation where the redshift evolution is obvious.

			\section{Caveats}
			\label{sec:caveats}	
			We estimated the dark matter fraction using a halo-model independent approach, assuming an exponential thin disk distribution for stars and gas. Our analysis provides reliable measurements of the dark matter fraction within different radii: $R_{\rm e}$, $R_{\rm opt}$, and $R_{\rm out}$. However, it is important to note that only 20\% (53 objects) of the total sample (263 objects) satisfy the criterion $R_{\rm e} > {\rm PSF}$. Nevertheless, we report that galaxies are not dark matter deficient at high-$z$, especially, within $1-3R_{\rm e}$. Although we report a high dark matter fraction within $R_{\rm e}$ for the \hz galaxies in a larger sample, we advise the reader to interpret these measurements with caution. For instance, the JWST photometry will have the capability to resolve the inner stellar disk and bulge components of galaxies (see also Section~\ref{sec:extra1} and Figure~\ref{fig:FdmRe-JWST}). Moreover, deep spectroscopic observations will further improve the shape of the rotation curves, as reported in \citet[][their Fig.~2]{KURVS}. Both the former and the latter will impose even tighter constraints on the dark matter distribution in the inner regions of galaxies, which may differ from the results presented here. In particular, the geometry of the disk, especially the gaseous disk, could be different \citep{Romeo2020, Renaud2021, Romeo2023}, whilst many systematic uncertainties in the actual stellar mass and gas mass could change the results, although we have shown in Section~\ref{sec:analysis} that it would take a very serious mismatch to erase the dark matter signature from the rotation curves entirely.
			
			Our analysis yields unphysical dark matter fractions for $\sim 22 \%$ of the sample due to $M_{\mathrm{dyn}} < M_{\mathrm{bar}}$, and an additional $\sim 15\%$ exhibits low dark matter fraction (0-0.2) within $R_{\rm opt}$ and $R_{\rm out}$, see Figure~\ref{fig:FdmRout} and \ref{fig:FdmRopt}. In order to understand these objects, we carefully examined their rotation curves, moment maps, and high-resolution images, yet no apparent anomalies were found. Consequently, our guess is that the photometric stellar masses of these galaxies are likely overestimated. The reason is, SED fitting involves modeling of observed SED by comparing it to a grid of model spectra. The accuracy of the derived stellar mass depends on the assumptions made in these models, such as the choice of star formation histories, stellar population synthesis models, initial mass function (IMF), dust attenuation laws etc. If these assumptions do not accurately represent the true physical conditions of the galaxy (e.g., fraction of binary stars), it can introduce systematic errors \citep[see, e.g.,][]{Pacifici2023, Leja2019}, which may impact the dark matter estimates and its trend as a function of stellar mass or redshift, for example. Moreover, uncertainties in the gas scaling relations (reported in Section~\ref{sec:Mbar}) could also impact the total baryonic mass and consequently the dark matter estimates.

			To comprehensively assess the impact of overall uncertainty in stellar mass, star-formation rate, and gas scaling relations, we estimated the dark matter fraction under two scenarios: one accounting for statistical and systematic uncertainties (referred to as Case-1) and with only statistical  uncertainties (referred to as Case-2). We conducted Monte Carlo sampling with 10,000 realizations to propagate errors on all parameters. In Case-1, statistical and systematic uncertainties on stellar masses are 0.04 and 0.14 dex, respectively, while for star formation rates, they are 0.04 and 0.25 dex, respectively \citep[c.f.][]{Pacifici2023}. During the computation of molecular and atomic gas masses, as defined in Section~\ref{sec:Mbar}, we incorporated an additional uncertainty of 0.3 dex (error in gas scaling relations) for each component individually (H2 and HI masses).  Conversely, in Case-2, only statistical uncertainties on stellar mass and star-formation rates were considered for the analysis. Upon comparison, we observed only an increase of 1\% in the dark matter in Case-2 than Case-1.  In both cases, the overall trends of the dark matter fraction as a function of stellar mass, galactic scales, and redshift remained consistent, albeit the large errors in case-1. In this work we present the results of Case-1, i.e., modeling statistical as well systematic uncertainties.
			
			Furthermore, the issues regarding  $M_{\mathrm{dyn}} < M_{\mathrm{star}}$ have been previously reported in studies concerning \hz galaxies, including works by \citet{W15, Forster2020}, and \citet{GS21a}. We propose several conditions under which the dynamical mass could appear smaller than the baryonic mass:
			
			\begin{itemize}
				\item Incomplete sampling of the velocity distribution: If the velocity measurements used to estimate the dynamical mass are limited to only a small fraction of the stars or gas in a galaxy, the resulting estimate of the dynamical mass could be lower than the actual value, as hinted in \citet{KURVS}. In this analysis, we extracted the best possible kinematics information available in our datasets, but with better dataset these estimates will improve further. 
				
				\item Non-equilibrium conditions: If a galaxy has recently undergone a major disturbance, such as a merger or a violent epoch of star formation, the velocity distribution of its stars and gas may not yet have settled into a stable equilibrium \citep[][and reference therein]{Riechers2014, Lemaux2017, Forster2020}. In this case, the dynamical masses based on rotation velocity could be lower than the actual value.  In our sample, we visually inspect the galaxy \RCs\ moment maps, and high resolution photometry, and discard all the potential mergers. However,  it is worth noting that our understanding of the merging history of galaxies at high-$z$ is still limited.
				
				\item Non-circular orbits: If the stars and gas in a galaxy are on non-circular orbits, the velocity measurements used to estimate the dynamical mass will only provide a lower limit on the actual mass \citep[][references therein]{GalDynBook}. This can only be understood with future high-resolution spectroscopy. That is, there is room for improvement in our dynamical masses.
			\end{itemize}
			On the contrary, the most massive galaxies in our sample with $\log(M_{\mathrm{star}} \ [\mathrm{M_\odot}]) > 10.7$  exhibit a significant amount of dark matter across the galactic scales, surpassing the average dark matter fraction observed in the overall sample. Remarkably, the dark matter halos of these galaxies are more massive than those observed in massive systems in the local Universe, such as the Milky Way and Andromeda (see Figure~\ref{fig:FdmRe} and Figure~\ref{fig:FdmRout}), the two plausible scenarios are:
			
			\begin{enumerate}
				\item  These massive galaxies may have followed distinct evolutionary pathways, resulting in the formation of exceptionally massive dark matter halos. Or,
				\item The baryonic masses of these objects are underestimated, for instance due to the presence of massive compact objects. Indeed, if a galaxy contains a substantial number of compact objects, such as black holes or neutron stars, their gravitational influence could dominate the motion of stars and gas, leading to an overestimate of the dynamical mass \citep[see references therein]{Naab2017}
			\end{enumerate} 
			
			Finally, it is important to highlight that galaxies with low baryon surface densities ($ \log(\Sigma_{\text{bar}}(<R_{\text{out}})) < 8.0 $), often categorized as low surface-brightness galaxies, are noticeably absent from our sample at $ z > 1 $, as depicted in Figure~\ref{fig:Fdm-SD-mass-z} (second panel). This absence is potentially attributable to Tolman Dimming, a relativistic phenomenon in which the observed surface brightness of a celestial object diminishes as $ (1 + z)^2$ \citep{Tolman1930, Tolman_diming_1996, Tolman_diming_2010}. Consequently, low surface-brightness galaxies could go missing in high-redshift observations. It should also be noted that these galaxies present challenges for observation even in the local Universe. To acquire resolved rotation curves for such systems, observatories with substantial aperture sizes, such as the forthcoming Extremely Large Telescope, would be required.

			With these caveats in mind, it is noteworthy that the majority ($\sim$ 70\%) of galaxies in our sample contain significant amount of dark matter,  consistent with the findings of a previous study conducted by \citet{GS21b}, and not very different from local galaxies. Consequently, the reliability of our sample, techniques, and measurements instills confidence and provides a firm foundation for further discussion on: 1) the evolution of dark matter in galaxies across cosmic time, and 2) challenges in constraining the dark matter at \hz.
			
			\section{Discussion}\label{sec:discussion}
			In this section, we discuss the main findings of this work.

			\paragraph{Dark matter fraction:}
			In local star-forming galaxies, where dark matter is believed to constitute the majority of the mass in most galaxies, dark matter fraction (within $R_{\rm opt}$) estimates typically range from 70\% to 90\% of the total mass. The  dark matter fraction in these galaxies is relatively higher in the outskirts compared to the inner regions, indicating that the inner regions are dominated by baryonic processes while the dynamics of the outskirts is governed by dark matter  \citep{rubin1980, PS1996, Martinsson2013, Courteau2015, McGaugh2016}. A similar trend is observed in \hz\ galaxies, as shown in Figure~\ref{fig:FdmRout} \& \ref{fig:Fdm-z}. Based on the datasets presented in this study, we show that the galaxies at high redshifts (0.75 < z < 1.8) are predominantly influenced by dark matter from $R_{\rm opt}$ to $R_{\rm out}$, with fractions ranging from 50\% to 90\%. 

			While we identify dark matter-dominated systems at \hz, we observe a significant scatter in relations, such as $f_{_{\rm DM}}-M_{\rm star}$, $f_{_{\rm DM}}-z$,  $f_{_{\rm DM} }-V_{\rm Rout}$, and  $\Sigma_{\rm bar}-f_{_{\rm DM}}$, see Figure~\ref{fig:FdmRout}, \ref{fig:Fdm-z}, \& \ref{fig:Fdm-SD-Vc}, respectively. A similar scatter is also reported in \citet{GS21b}. The scatters in these relations suggest that these galaxies might still be undergoing the process of building or acclimating their distribution of baryons and \dm. In other words, they are at different stages of `galaxy assembly'.\footnote{Galaxy assembly refers to the process by which galaxies form and evolve over time, involving the accretion of gas and dark matter, enhance star formation, and the merging of smaller galaxies into larger structures.} The scatter in the dark matter fraction itself may stem from various other factors, including differences in formation history of galaxies, the irregular distribution of baryons that can affect the distribution of dark matter, and the environment in which these galaxies reside \citep{Dutton2016,Behroozi2019}. Moreover, it could also be due to the diversity in the \dm\ halo properties, which are closely coupled to the properties of the baryonic matter. However, we note that if any systematic uncertainties in estimating baryonic masses have been neglected, this would naturally increase the scatter, never decrease it.

			\paragraph{Dark matter halo assembly:} According to the current cosmological model, Lambda Cold Dark Matter ($\Lambda$CDM), dark matter plays a fundamental role in the assembly of cosmic structures in the Universe. This assembly process follows a hierarchical pattern, wherein smaller structures form first and subsequently merge to form larger ones, and gradually increase in mass and size \citep{Peebles1993}. This hierarchical merging process is primarily driven by gravity and influenced by the distribution of matter in the Universe, encompassing both dark and baryonic matter. Since dark matter is invisible and cannot be directly observed, studying the distribution and properties of baryons, such as surface density and motion governed by the total gravitational potential, provides a way to gain insights into the distribution of dark matter and its assembly history.
			
			In accordance with the aforementioned concepts, we investigated $\Sigma_{\rm bar}-f_{_{\rm DM}}$ relation, which was previously examined by \citet{Genzel2020} within the effective radius. Expanding upon their work, we extended this relation to encompass the outer radius ($R_{\rm opt}$ and $R_{\rm out}$). Notably, our data well fit the relationship described by Equation~\ref{eq:fdm-Srel-1}, with an intrinsic scatter of $0.25$ dex. 
			Intriguingly, we observed that the $\Sigma_{\rm bar}-f_{_{\rm DM}}$ relation maintains a consistent slope also when examined at the optical radius, and the outer radius in both low and high mass systems, as illustrated in the first panel of Figure~\ref{fig:Fdm-SD-mass-z}. Moreover, this relationship remains unchanged across different redshift ranges, as demonstrated in second-third panel of Figure~\ref{fig:Fdm-SD-mass-z}. Thus, the $\Sigma_{\rm bar}-f_{_{\rm DM}}$ relation exhibits a uniform nature, implying that the influence of baryonic matter on dark matter appears to be similar across various stellar mass ranges, and redshift intervals. However, uncertainty on the individual measurement and intrinsic scatter in this relation is very high. Only high-quality data will allow the exact scatter of this relation to be pinpointed. Taking the current scatter at face value would indeed indicate that the relation is not fundamental \citep{Milgrom, FamaeyMcGaugh} but that galaxies are likely at distinct evolutionary stages, such as being in the process of establishing their respective disks. 
			
			On the other hand, when analyzing the relationship between  $V_{\rm c} - f_{_{\rm DM}}(<R_{\rm out})$, as depicted in the right panel of Figure~\ref{fig:Fdm-SD-Vc},  the correlation was slightly tighter, with an intrinsic scatter of $0.23$ dex. However, this relation,  which is the result of a combination of the mass-velocity and mass-size relations is not at all universal, as further exploration of this relationship for low and high stellar mass galaxies, as illustrated in the first panel of Figure~\ref{fig:Fdm-Vc-mass-z}, revealed the emergence of distinct sequences as a function of both mass and redshift. Galaxies with high stellar mass naturally exhibit a lower fraction of dark matter at fixed rotational velocity, which is related to their baryonic size and dark matter halos being different from lower stellar mass ones. In the second-third panel of Figure~\ref{fig:Fdm-Vc-mass-z}, we divide the  $f_{_{\rm DM}}-V_{\rm c}$ relation of low and high stellar masses in low and high-$z$ ($z\leq 1$ and $z>1$, respectively). Notably, a distinct offset was observed between low-z and high-$z$ objects of the low-mass bin ($\log(M_{\rm star} \ [M_\odot]) \leq 10^{10}$). Specifically, low-mass galaxies at high-$z$  exhibit a lower fraction of dark matter at fixed rotational velocity. This is intriguing, suggesting that low-mass mass galaxies most likely undergo higher degree of evolution in terms of their respective dark matter and baryon distribution. On the other hand, high-mass systems already seems to be settled at higher-$z$, as evidenced in the third panel of Figure~\ref{fig:Fdm-Vc-mass-z}. 
			\paragraph{Uncertainties \& limitations:} 
			As highlighted in caveats, in Section~\ref{sec:caveats}, the study of dark matter fraction at \hz is very challenging and requires significant refinements in both baryonic constraints and kinematic modeling. In this section, we discuss a few key challenges of the field.
			
			For baryons, SED fitting techniques, which are commonly used to estimate stellar masses and SFRs, exhibit substantial offsets from each other \citep{Leja2019, Pacifici2023}. These discrepancies not only lead to larger uncertainties in stellar masses and SFRs, but also impact the gas mass scaling relations. For instance, \citet{Tacconi2018}, molecular gas mass scaling relation, uses the \citet{speagle14} main-sequence relation ($M_*$-SFR plane). At \(z \sim 0.1\), the \citet{speagle14} main sequence differs from \citet{Chang2015} by \(\sim 0.4\) dex, and at \(z \sim 1\), it is \(\sim 0.25\) dex off from the 3DHST study \citep{Nelson2021}. At higher redshifts, this offset is not yet quantified. Therefore, the systematic offset and observed scatter in gas scaling relation, which relay on main-sequence relation is complex to accurately quantify, and beyond of the scope of this study.
			
			Due to the lack of direct measurements of HI gas at high redshifts (\(z > 1\)), accurate gas masses at \(z > 1\) are still unavailable. At \(z < 1\), HI mass scaling relations are derived using stacking analyses, as current observing facilities only allow HI signal detection without resolution \citet{Chowdhury2022}. These HI scaling relations are general and not specifically for star-forming disk-like galaxies. That is, we can use them as a first order approximation, but they are not accurate.  Refining HI mass relations will requires future observations from the Square Kilometer Array.
			
			Moreover, current kinematic modeling techniques have limitations, such as not accounting for non-circular motions, gas inflows, and outflows \citet{Oman2019}. Consequently, accurately modeling the total dynamical mass of the system is  challenging. This issue can only be resolved with very high-resolution large galactic-scale surveys at high redshifts, which are currently not feasible even in optical and near-infrared astronomy. Further progress might only be possible with the construction of Extremely Large Telescope.
			
			Lastly, the `Tolman Dimming' effect, hinders the observation of low to intermediate mass galaxies at higher redshifts. Consequently, our observations at \hz are biased toward more massive systems. Additionally, we know that, in the local Universe and at \hz, dark matter shows a slightly decreasing trend as a function of stellar mass, i.e., massive galaxies show relatively low dark matter fraction. Therefore, the observed low dark matter fraction at \hz, or the decreasing trend of dark matter fraction as a function of redshift, could be an artifact caused by the absence of low to intermediate mass galaxies. To accurately constrain the trend of dark matter fraction across cosmic time, further deep observations of low and intermediate mass galaxies at high redshifts are required.
			
			Nevertheless, given the best possible information available on  baryonic content, observed kinematics, and their uncertainties, we present conservative dark matter fraction estimates to the best of our abilities. However, we acknowledge that these findings require further refinement, which will be achieved as the field progresses.

			
			\section{Summary and conclusions}\label{sec:summary}
			Through this study, our aim was to investigate the fraction of dark matter across different galactic scales and cosmic time. To achieve this objective, we utilized a substantial sample consisting of 263 main-sequence star-forming disk-like galaxies within the redshift range $0.6\leq z \leq 2.2$. This sample encompasses 73 galaxies from the KMOS$^{\rm 3D}$ survey, 21 from the KGES survey, and 169 from the KROSS survey. We performed 3D forward modeling of datacubes using the 3DBarolo code, as described in Section~\ref{sec:k-model}. Figure~\ref{fig:BBKMKGKR} provides an illustrative example of the kinematic modeling results obtained using 3DBarolo, displaying the velocity distribution (moment-1 and moment-2 maps), data, model, and residuals, followed by the major axis PV diagram and the velocity dispersion curve. We applied pressure support corrections to the rotation curves inferred from 3DBarolo, these pressure corrected \RCs\ are referred to as intrinsic rotation curves.
			
			To estimate the dark matter fraction, we subtracted the baryon contribution from the intrinsic rotation curves, assuming that both the star and gas are rotationally supported and confined to an exponential disk. The residual contribution was assumed to originate from the dark matter, which was required to explain the observed kinematics. For a detailed explanation of this methodology, we refer the reader to Section~\ref{sec:mass-models} and \ref{sec:method-fdmrac}. We estimated the dark matter fraction within $R_{\rm e} \ \mathrm{till} \ R_{\rm out}$ (see Sect.~\ref{sec:fdm-Re} \& \ref{sec:fdm-R} and Figure~\ref{fig:FdmRe} \& \ref{fig:FdmRout}). Additionally, we explored the variation of the dark matter fraction across different redshift ranges, presented in Section~\ref{sec:fdm-z} and illustrated in Figure~\ref{fig:Fdm-z}. Furthermore, we investigated the interplay between baryonic matter and dark matter by examining the scaling relations of dark matter with baryon surface density and circular velocity at the outer radius of galaxies in Section~\ref{sec:fdm-Srelations}, as shown in Figure~\ref{fig:Fdm-SD-Vc}, \ref{fig:Fdm-SD-mass-z}, and \ref{fig:Fdm-Vc-mass-z}. The results obtained from these analyses offer valuable insights into the role of dark matter in shaping galaxies. Our key findings are as follows:
			
			\begin{itemize}
				\item  Similar to local disk galaxies, \hz galaxies also exhibit dark matter-dominated outer halos (from $R_{\rm e}$ to $R_{\rm out}$), with a dark matter fraction ($f_{_{\rm DM}}$) ranging between $50\%$ and $90\%$. It is noteworthy that at $z=1-1.8$, the median value of $f_{_{\rm DM}}$ remains above 50\% across all the galactic scales, which is very similar to local disk galaxies.
				\\
				\item Due to the lack of resolved observations within \( R_e \), it remains uncertain whether or not the dark matter fraction in \hz disk-like galaxies gradually increases from \( R_e \) to \( R_{\rm out} \). To establish a clear trend of dark matter fraction as a function of redshift, deeper observations of low- and intermediate-mass galaxies at high redshift are required along with precise measurements of their baryonic and total dynamical masses.
				\\
				\item The $V_{\rm c}-f_{_{\rm DM}}$ relation revealed distinct sequences with stellar mass and redshift intervals. Specifically, low stellar mass ($M_{\rm star}\leq 10^{10} \ {\rm M_\odot}$) galaxies exhibit a higher degree of dynamical evolution with clearly lower dark matter fractions at a given rotational velocity at higher redshift ($z > 1$). This can potentially be attributed to the hierarchical assembly of dark matter halos governing the evolution of galaxies. 
				\\
				\item The $\Sigma_{\rm bar}-f_{_{\rm DM}}$ relation demonstrates a consistent trend, slope, and scatter across different stellar mass ranges and redshift intervals, indicating a universal nature of the influence of baryonic matter on the distribution of dark matter. The relation also holds at different radii within galaxies. This implies that the interplay between baryons and dark matter in galaxies is the same, regardless of their mass, size, and age. 
				\\
				\item Numerous uncertainties related to the baryonic content result in substantial errors in the estimated dark matter fraction of galaxies. By reducing these errors, through a combination of observational and theoretical improvements, we as a community have the potential to make a significant leap forward in our understanding of galaxy evolution and the properties of dark matter.
			\end{itemize}	

			Existing studies (including this work) on dark matter in high-redshift galaxies support the presence of dark matter but fall short on being able to robustly establish further insights, such as its radial distribution.
			Nevertheless, these studies represent an important step forward in our understanding of galaxies and dark matter halo evolution. To establish tighter constraints on our findings, the current kinematic modeling techniques need to be amended, as they rely heavily on the assumptions of axisymmetry and dynamical equilibrium, which might not be the case at high-$z$. More accurate assessment of the baryonic content at \hz is required. Lastly, lower surface-brightness galaxies are underrepresented in high-redshift observations, making high-quality data indispensable for resolving the present conundrums. 
			
			\section*{Data availability}
			\label{sec:DA}
			The catalogs of galaxy physical properties such as stellar mass, gas mass, rotation velocity, dark matter fraction etc. are publicly accessible via Zenodo (\href{https://zenodo.org/records/15458539}{Highz\_DMfrac\_catalogs}). Additionally, position–velocity diagrams and velocity dispersion plots for the complete datasets are available at Zenodo (\href{https://zenodo.org/records/15450208}{Figure~E.4-E.9}). The corresponding datacubes and 3DBarolo model outputs can be provided by the authors upon reasonable request. 

			 \begin{acknowledgements} 
			We thank the anonymous referee for providing valuable comments that have improved the quality of the manuscript. We express our gratitude to Emily Wisnioski and her team, as well as Mark Swinbank and his team, for the public release of KMOS3D and KGES H$\alpha$ datacubes and catalogs, along with their valuable discussions and support. We also thank Benoit Famaey, Jonathan Freundlich, and Florent Renaud for their efforts in contributing to this paper at various stages.
			G.S., acknowledges SARAO postdoctoral fellowship (UId No.: 97882), and thanks Ambica Govind for providing HST images of full sample. G.S. also thanks Mihael Petac for various fruitful discussions. 
			GvdV acknowledges funding from the European Research Council (ERC) under the European Union’s Horizon 2020 research and innovation programme under grant agreement No 724857 (Consolidator GrantArcheoDyn). 
			G.S. acknowledge support from the University of Strasbourg Institute for Advanced Study (USIAS), within the French national programme Investment for the Future (Excellence Initiative) IdEx-Unistra. M.M., thanks financial support of the Flemish Fund for Scientific Research (FWO-Vlaanderen), research project G030319N. 
			\end{acknowledgements}

			
			
			%
			\bibliographystyle{aa} 
			\bibliography{DM_fraction_2023} 
			%
			
			%

			\begin{appendix} 
				
				\section{Signal to noise estimation}
				\label{sec:SNR}
				
\begin{figure*}
					\begin{center}
						\includegraphics[angle=0,height=4.5truecm,width=16.0truecm]{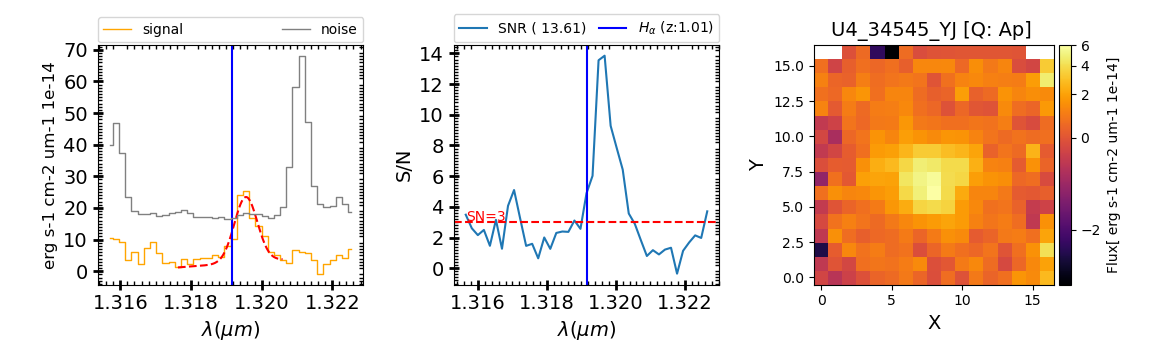} 
						\includegraphics[angle=0,height=7.0truecm,width=10.5truecm]{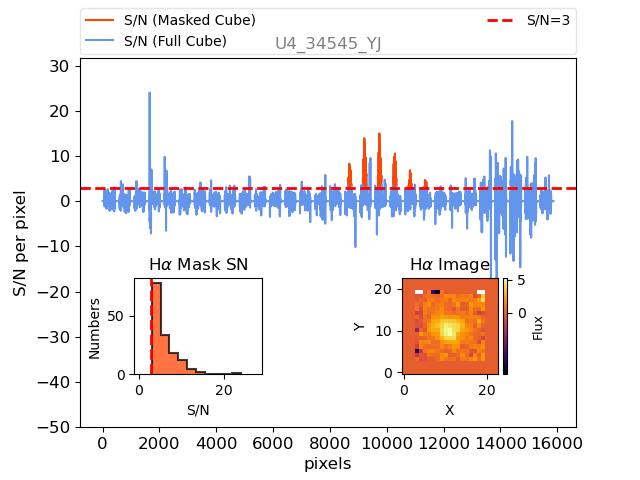}  
						
						\caption{Example showing the S/N in datacubes. {\em Upper Panel:} S/N derived from the integrated spectra. From left to right are presented: integrated spectrum of the flux (yellow) and noise (gray)  as a function of wavelength, the S/N as a function of wavelength, and the collapsed H$\alpha$ image. To compute the continuum and S/N from the integrated spectra of H$\alpha$-emission, we utilized the \texttt{gauss\_fit} module from MPDAF library, which provides the Gaussian-fit, flux at the peak emission, and average continuum. The Gaussian fit is represented by the red dashed line in the spectrum plot (first panel), and the S/N around the H$\alpha$-peak is specified in the legend of the S/N plot (middle panel). Here, we computed the S/N value with in $\lambda_{H\alpha} \pm 0.005 \mu m$, which approximately covers the Gaussian of H$\alpha$ flux. {\em Lower Panel:} S/N derived from the 1D spectra of the datacube, with blue denoting the full H$\alpha$ cube and red indicating the masked region (i.e., identified H$\alpha$ emission). Additionally, in same panel, we show the distribution of S/N in the mask and the collapsed H$\alpha$ images, left and right small windows, respectively. We compare the S/N of the integrated spectra (upper row, middle panel) with the S/N distribution of 1D spectra (lower panel, left zoom-in window). Throughout the analysis, we observe that the average S/N of the 1D spectra is approximately similar to the value obtained from the integrated spectra. This justifies our S/N measurement techniques.
						}
						\label{fig:1D-int-spc-SN}
					\end{center}
				\end{figure*}

\begin{figure*}
					\begin{center}
						\includegraphics[angle=0,height=4.5truecm,width=18.0truecm]{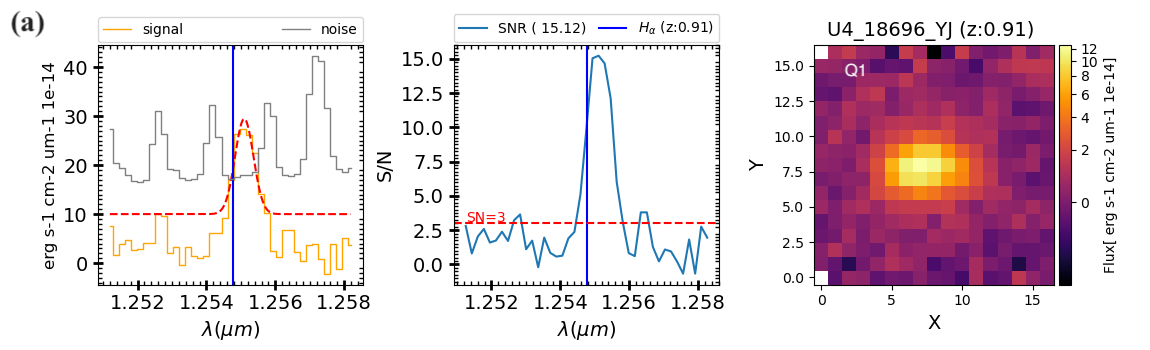}  
						
						\includegraphics[angle=0,height=4.5truecm,width=18.0truecm]{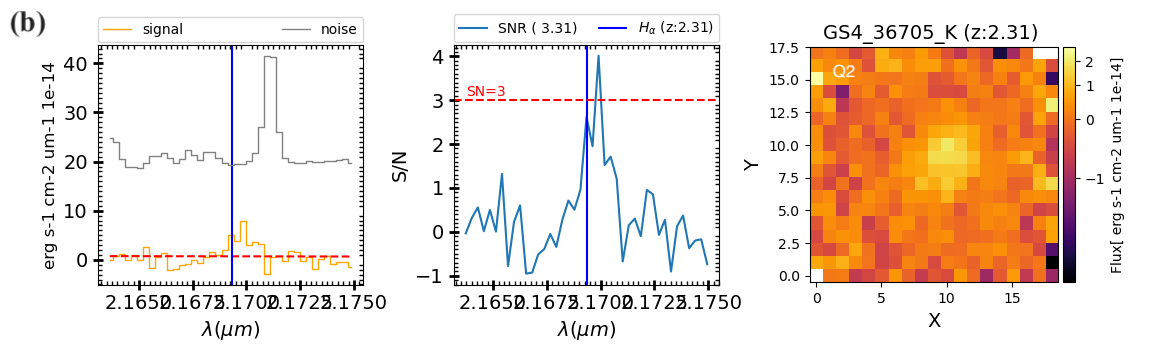} 
						
						\includegraphics[angle=0,height=4.5truecm,width=18.0truecm]{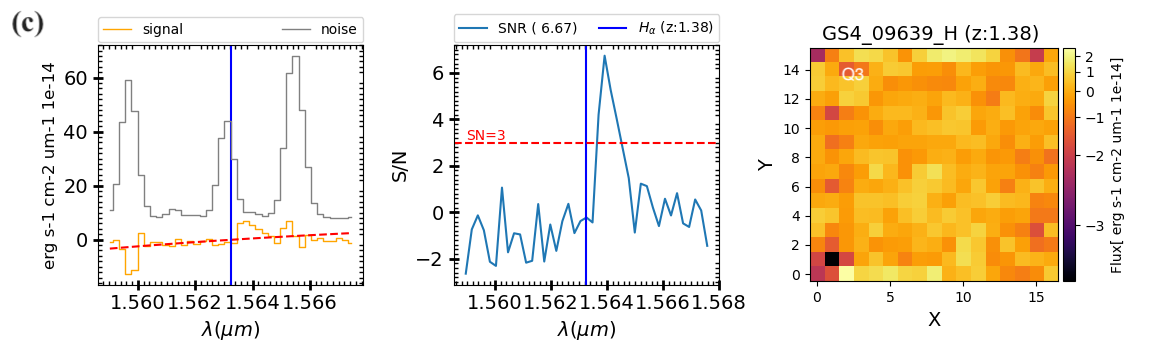} 
						
						\includegraphics[angle=0,height=4.5truecm,width=18.0truecm]{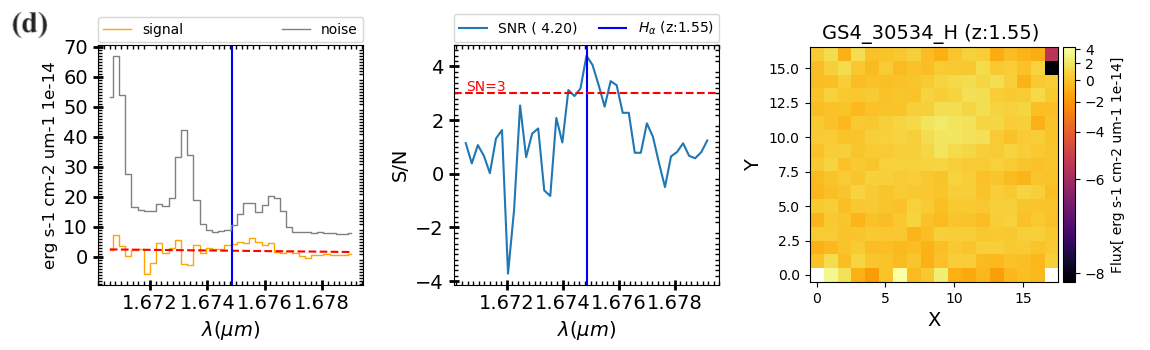}
						
						\caption{Example of quality assessment of KMOS$^{\rm 3D}$ and KGES sample. Row 1 \& 2 shows Q1 and Q2 galaxies, respectively, Row 3 \& 4 are Q3 galaxies. The color codes are given in the legend of the plots.}
						\label{fig:SN-Haimages}
					\end{center}
				\end{figure*}

				Observations of high-$z$ galaxies are often characterized by limited spatial resolution and low S/Ns, posing challenges for kinematic modeling. Therefore, before performing kinematic modeling with \texttt{3DBarolo}, we first examine the integrated spectra of the H$\alpha$ datacubes.\footnote{The integrated spectrum of a datacube is computed by summing the flux over all spatial pixels (spaxels) for each spectral channel, mathematically expressed as $F_{\text{integrated}}(\lambda) = \sum_{x,y} F(\lambda, x, y)$. } 
				Prior to any analysis, we crop the publicly available datacubes into smaller sub-cubes centered on H$\alpha$. Each cropped H$\alpha$ sub-cube consists of 31 spectral channels, spanning 15 channels on either side of the central H$\alpha$ wavelength ($\lambda_{H\alpha}$), while no cropping is applied in the spatial dimensions. These sub-cubes are utilized to compute the integrated spectra, 1D spectra (defined below), and H$\alpha$ images.  
				Furthermore, the publicly available H$\alpha$ datacubes include both flux and noise extensions, enabling us to estimate the ${\rm S/N}(\lambda) = F_{signal}(\lambda)/F_{noise}(\lambda)$, as illustrated in the upper panel of Figure~\ref{fig:1D-int-spc-SN}. Additionally, we evaluate the integrated H$\alpha$ images of each galaxy.\footnote{An integrated image from a datacube is obtained by summing the flux over all wavelength slices, effectively collapsing the spectral dimension. This results in a 2D image representing the total flux across all wavelengths at each spatial position, mathematically given by $I(x,y) = \sum_{\lambda} F(\lambda, x, y)$.}  
				In our analysis, the S/N provides a quantitative measure of the data quality (shown in the first and second columns of Figure~\ref{fig:SN-Haimages}), while qualitative assessments are based on the visual inspection of H$\alpha$ images (shown in the third column of Figure~\ref{fig:SN-Haimages}). Based on this combined quantitative and qualitative inspection, we classify the sample into three categories:  
				\begin{itemize}
					\item \textbf{Q1:} S/N $>3$ with a well-defined H$\alpha$ image.
					\item \textbf{Q2:} S/N $\geq 3$ with a moderately visible source.
					\item \textbf{Q3:} S/N $<3$ or no discernible source.
				\end{itemize}  
				The S/N threshold of three is a fundamental requirement of \texttt{3DBarolo}; results below this threshold are considered unreliable \citep{ETD15, ETD16}. Examples of this classification are presented in Figure~\ref{fig:SN-Haimages}. We exclude all Q3 galaxies from further analysis, restricting kinematic modeling to Q1 and Q2 galaxies, in accordance with the primary selection criteria discussed in Section~\ref{sec:dataset}.  
				
				After performing kinematic modeling, we estimate the S/N from the unfolded cubes detailed in \cite{GS21a} and shown in the bottom panel of Figure~\ref{fig:1D-int-spc-SN}. Briefly, \texttt{3DBarolo}'s \href{https://bbarolo.readthedocs.io/en/latest/tasks/fit3d.html}{3DFIT-Task} provides an option to utilize the \href{https://www.atnf.csiro.au/people/Matthew.Whiting/Duchamp/}{Duchamp} three-dimensional source-finding algorithm via the \texttt{Mask=Search} option. This feature generates a robust mask that distinguishes true emission within the datacube from background noise. 
				More specifically, the mask produced by \texttt{3DBarolo} is a cube of the same dimensions as the H$\alpha$ datacube, where detected emission pixels are assigned a value of 1, while all other pixels are set to 0. In the unfolded cube method, pixels where \texttt{mask} $= 1$ are considered as signal, while all other pixels are treated as noise. In other words, when flux is plotted as a function of pixels, we refer to it as the 1D spectrum, where masked pixels (\texttt{mask} $= 1$) correspond to the signal, and the remaining pixels contain only noise. To quantify the noise level, we compute the root-mean-square (rms) of the flux from noise-designated pixels. This root mean square noise is then used to compute and plot the signal-to-noise ratio per pixel from the 1D spectrum, i.e., ${\rm S/N \ per \ pixel} = F(pixel)/\sigma_{noise}$. We use this information and conduct a thorough cross-verification of the S/N in the integrated spectra with the 1D spectra as shown in Figure~\ref{fig:1D-int-spc-SN}. We note that if \texttt{SEARCH task} is unsuccessful in finding source (i.e., at providing a mask), we do not study these galaxies, and mark them Quality-3 because S/N is not sufficient for 3DBarolo. As shown in Figure~\ref{fig:1D-int-spc-SN}, it is remarkable that the S/N of the integrated spectra matches very well with that of the 1D spectra, despite the fact that the two methods are independent.
				
				Furthermore, as discussed in \citet{GS21a}, for the KROSS dataset noise information is not provided, which compelled us to first run 3DBarolo and subsequently assign the Quality of the objects based on S/N from 1D-spectrum . This process is particularly challenging due to the extensive dataset ($> 500$ objects per survey), and has proven to be significantly time-consuming. This is one of the primary reasons why studying the S/N from integrated spectra is crucial. This information can be used in primary selection criteria, thereby expediting the analysis process. Lastly, we observe that the catalogs used in this work provide somewhat round-off values for spectroscopic redshifts. Consequently, in some cases the peak of H$\alpha$ appears slightly shifted ($\sim 10^{-4} \mu m$), as indicated by the vertical blue line in Figures~\ref{fig:1D-int-spc-SN} and \ref{fig:SN-Haimages}.

				\section{Constraining gas geometrical parameters with 3DBarolo}
				\label{sec: BBoptimization}
			
				As we go higher in redshift the information on morphological gas central x-y position ($x_c, y_c$) and position angle (PA) starts differentiating from their photometric measurements \citep[as reported in ][]{W15, H17, GS21a}. On the other hand, from our previous work with 3DBarolo \citep{GS21a}, we learned that  \texttt{3DFIT TASK} is not feasible in constraining the multiple parameters.  Therefore, in this work we estimate gas geometrical parameters using an optimization function that runs atop of 3DBarolo run. This optimization function takes following inputs: (1) loss function given in Eq.~\ref{eq:loss_func}, (2) Initial guess and bounds on parameter to be fit, (3) minimization method, and (4) `par file' to execute 3DBarolo run.  In particular, we optimize PA, $x_c$, and $y_c$ . Initial guess on these parameters comes from the photometric information or from H$\alpha$ datacubes (discussed in Sec.~\ref{sec:dataset}). The bounds are as following: $[x_c = x_{phot} \pm 4, y_c = y_{phot} \pm 4]$ and  PA[0, 360]. To be accurate, we also optimize systemic velocity ($V_{\rm sys}$), the initial gas on it is computed from `Doppler shift' of H$\alpha$ line, and bounds are [-200, 200]. However, choice of fitting $V_{\rm sys}$ is tricky, for example, when we observe high residuals in moment maps, we fix it to a median value given in 3DBarolo output files.
				The optimize-minimize function works in three steps:
				\begin{enumerate}
					\item Run 3DBarolo (free: $V_{\rm rot}$, $V_{\rm disp}$, and in some cases $V_{\rm sys}$), which gives moment maps, PV- diagrams, and estimates on $V_{\rm rot}$ and $V_{\rm disp}$.
					
					\item Compute the loss function (Eq.~\ref{eq:loss_func}). The data and model inputs in loss function depend on the choice of parameter to be fit, which we have discussed below.
					
					\item Save the loss, and optimization log. The optimization log gives us the best estimate of parameters and success rate (True or False).
				\end{enumerate}

				\begin{figure}
						\includegraphics[angle=0,height=6.5truecm,width=8.5truecm]{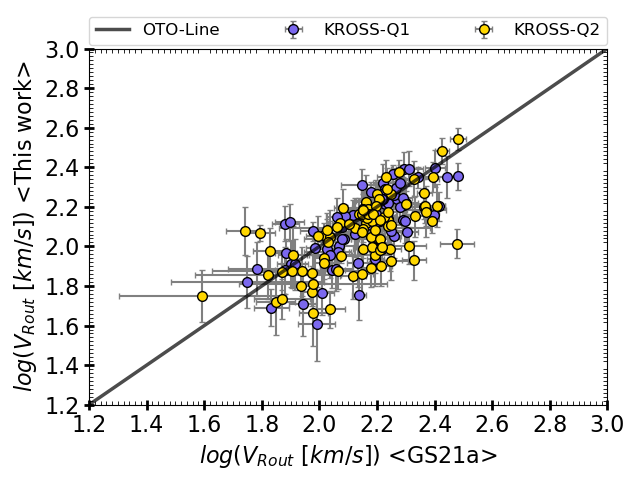}           
						\caption{This figure compares the circular velocity of KROSS galaxies as computed in this study (y-axis) with that of \citet[][x-axis]{GS21b}. The Quality-1 and Quality-2 galaxies, as classified in \citet{GS21a}, are represented by yellow and purple circles, respectively, and the black line shows one-to-one relationship between two studies.}
						\label{fig:Vnew-old}
				\end{figure}
				
				\begin{figure}
					\begin{center}
						\includegraphics[angle=0,height=6.0truecm,width=8.5truecm]{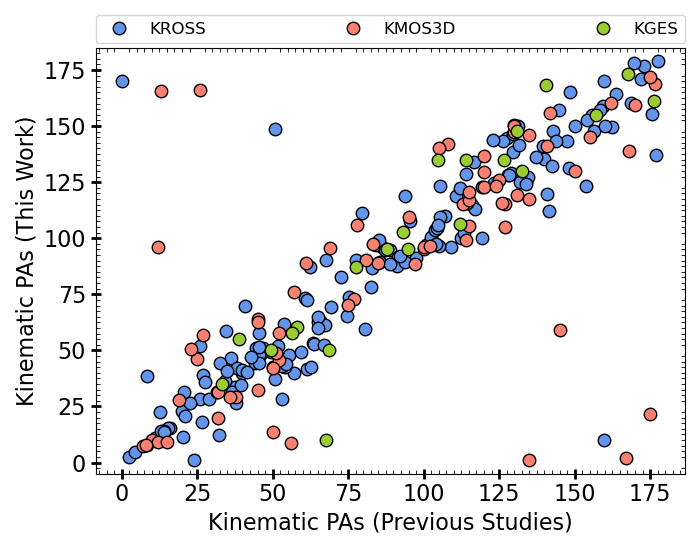}  
						
						\caption{Comparison of kinematic PAs computed in this work with respect to previous studies. KROSS daste compared with \citet{GS21a}, KMOS3D with \citet{W19}, and KGES with \citet{AT2021}. The color codes are given in the legend of the plot.}
						\label{fig:W19-PAs}
					\end{center}
				\end{figure}		
				
				The optimization procedure follows a logical flow consists of four stages. Firstly, the $x_c$ and $y_c$ parameters are constrained using moment-0 maps, which provide information on the gas distribution. In the second stage, the previously optimized $x_c$ and $y_c$ positions are fixed to optimize the position angle (PA) parameter using minor axis PV diagrams. As noted in \cite[][appendix-A]{GS21a}, the symmetry of the minor axis PV diagram around the axis indicates the correctness of the PA parameter; an asymmetric PV diagram implies an incorrect PA. In the third stage, the previously constrained $x_c$, $y_c$, and PA parameters are fixed to estimate the systemic velocity ($V_{\rm sys}$) parameter. The loss function for $V_{\rm sys}$ uses major axis PV diagrams since $V_{\rm sys}$ directly affects the rotation velocity, which is determined by the major axis PV diagram. In the final stage, all optimized parameters are used to compute the final loss using moment-1 maps since our goal is to constrain the rotation velocity of the system. However, for cross-check, the final loss is also computed using moment-2 maps, although the difference is negligible ($< 0.001$) and does not appear in the residual maps. Lastly, the output corresponding to the minimum loss is saved.
				
				We noted that this optimization procedure does not work for random guesses on the parameter values and requires reasonably close initial estimates, such as those provided by photometric estimates. Additionally, the procedure is computationally intensive, with each galaxy taking approximately 20-30 minutes to optimize when running on 8-cores. Nevertheless, this procedure is worthwhile for studying the nature of dark matter.
				
				We evaluated the effectiveness of 3DBarolo+optimizer on the KROSS data, which we previously analyzed and discussed in \citet{GS21a}. As expected, the optimization process failed for galaxies with a low S/N ($<5$), specifically those with Quality2. Consequently, we were compelled to exclude 56 out of 225 galaxies. In Figure~\ref{fig:Vnew-old}, we compare the circular velocities (computed at $R_{\rm out}$) of KROSS sample, obtained with old and new kinematic modeling techniques. It is noteworthy that the previous and current circular velocity measurements are in reasonable agreement, with an intrinsic scatter of 0.03 dex. Both Quality-1 and Quality-2 galaxies in the GS21 sample exhibit variations in their circular velocities. Nonetheless, we used the new measurements because assessment of  data-to-model fits in the PV diagram are better in the new analysis, as shown Figure~E.4-E.9 uploaded at Zenodo (see section data availability). Moreover, moment-map residuals stays close to zero.
				
				When we applied 3DBarolo and the optimizer on the KMOS3D and KGES dataset, we noticed that in some galaxies having moderate S/N levels ($ \sim 3-5$), 3DBarolo was unable to identify the source mask, most-likely due to very noisy pixels. Additionally, for certain galaxies, the optimization of the position angle was unsuccessful, although the exact cause remains unknown. These instances were categorized under a batch of poor S/N (low-quality) galaxies for simplification purposes. In contrast, we notice that some galaxies with moderate signal-to-noise levels produced satisfactory results. This outcome may be attributed to the initial guess of the position angle closely aligning with the angle of the galaxy or relatively low noise in the unmasked pixels. It is important to acknowledge that these occurrences highlight the limitations of kinematic modeling. Therefore, our approach involves selecting results that are both available and reliable, i.e., where \texttt{3DBarolo} yields reliable kinematic outputs and \texttt{optimization} demonstrate success.

				To assess the reliability of the optimized parameters, specifically PAs, we compare the best-fit PAs with previously studied kinematic PAs: KROSS with \citet{GS21a}, KMOS3D with \citet[][PA received in private correspondance]{W19}, and KGES with \citet{AT2021}. As shown in Figure~\ref{fig:W19-PAs}, our estimates aligns very well with previous studies. Therefore, we concluded that  3DBarolo+optimizer works well on high-$z$ moderate signal-to-noise data and hence can be used to constrain extra parameters.
				
\begin{figure}
					\begin{center}
						\includegraphics[angle=0,height=6.0truecm,width=8.5truecm]{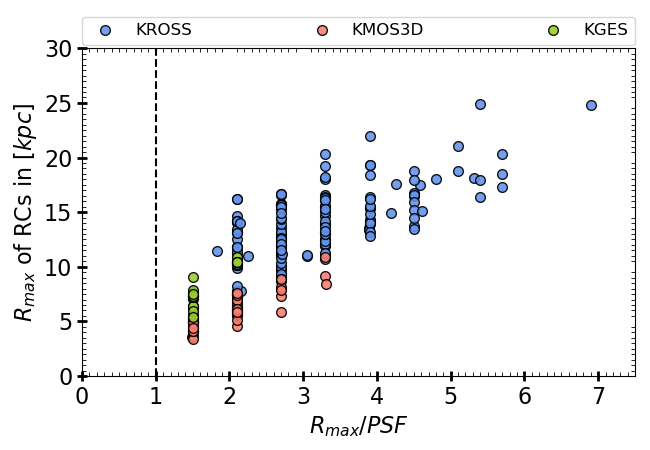}  
						
						\caption{$R_{\rm max}$ of rotation curves as a function of $R_{\rm max}/PSF$.}
						\label{fig:Rmax-Rpsf}
					\end{center}
				\end{figure}

				Furthermore, we examined the maximum radius of rotation curves with respect to the PSF of final sample, see discussion in Section~\ref{sec:final-sample}. As shown in Figure~\ref{fig:Rmax-Rpsf}, all the galaxies in final sample avid the criteria of $R_{\rm max} > PSF$. Therefore, this sample can easily be used for dynamical modeling and dark matter estimates.

				\begin{figure}
					\begin{center}
						\includegraphics[angle=0,height=4.7truecm,width=9.0truecm]{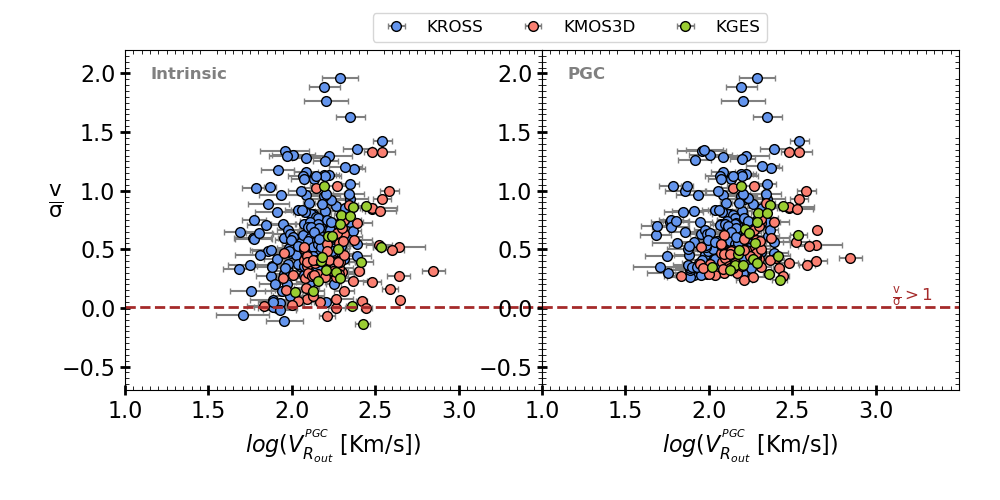}  
						
						\caption{This figure shows the rotation to dispersion ratio ($v/\sigma$) of galaxies as a function of circular velocity computed at $R_{\rm out}$. The left and right panels show $v/\sigma$ before and after PGCs, respectively. The red horizontal dashed line indicates $v/\sigma = 1.1$. This figure demonstrates that, after pressure support corrections, all the galaxies are rotation-dominated. Therefore, we do not exclude any galaxies based on their observed $v/\sigma$ values. }
						\label{fig:voversigma}
					\end{center}
				\end{figure}
				
				\begin{figure*}
					\begin{center}
						\includegraphics[angle=0,height=6.0truecm,width=8.0truecm]{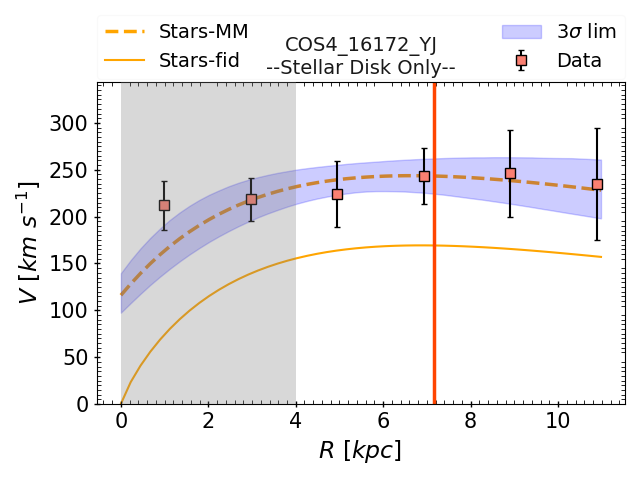}
						\includegraphics[angle=0,height=6.5truecm,width=8.0truecm]{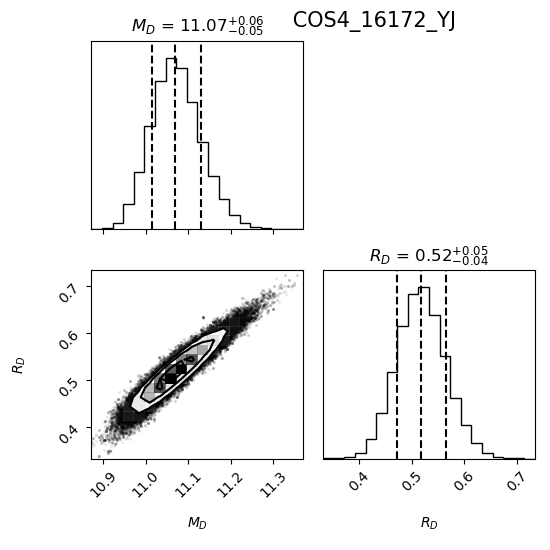}  
						
						\includegraphics[angle=0,height=6.0truecm,width=8.0truecm]{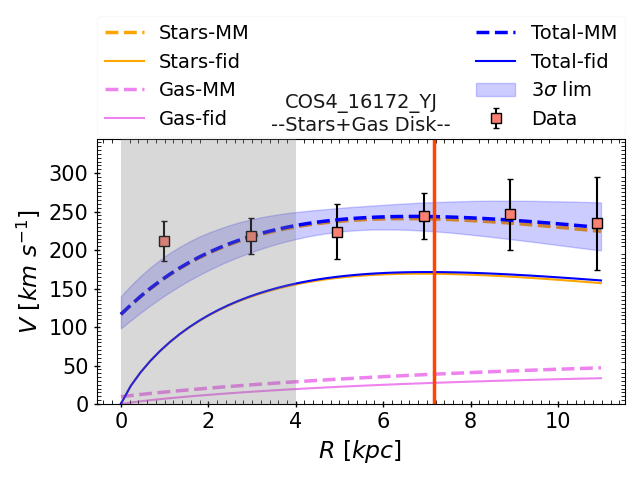}
						\includegraphics[angle=0,height=6.5truecm,width=8.0truecm]{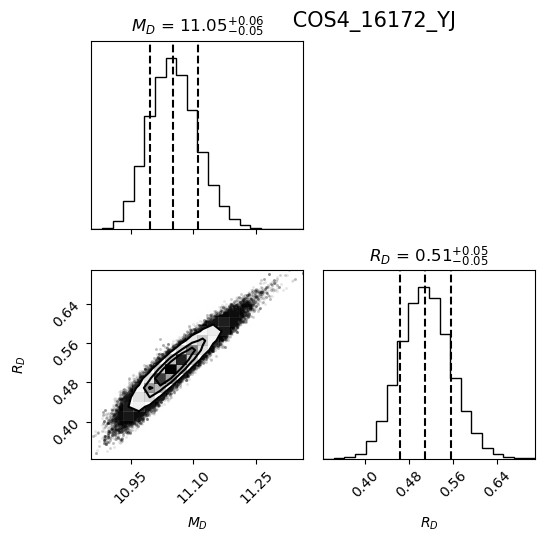}  
						
						\caption{Examples of mass modeled \RCs\ in the case of the maximum stellar disk scenario (upper panel) and maximum stellar+gas disk (lower panel). The best-fit parameters are shown in the right column, and color codes are given in the legend of the plot, where `MM' stands for dynamical mass-model profiles estimated using Bayesian inference , and `fid' stands for fiducial profile derived from given photometric mass. This example favors the maximum stellar and baryonic disk scenario, upper and lower panels, respectively. }
						\label{fig:Mass-models-1}
					\end{center}
				\end{figure*}
				
				\begin{figure*}
					\begin{center}
						\includegraphics[angle=0,height=6.0truecm,width=8.0truecm]{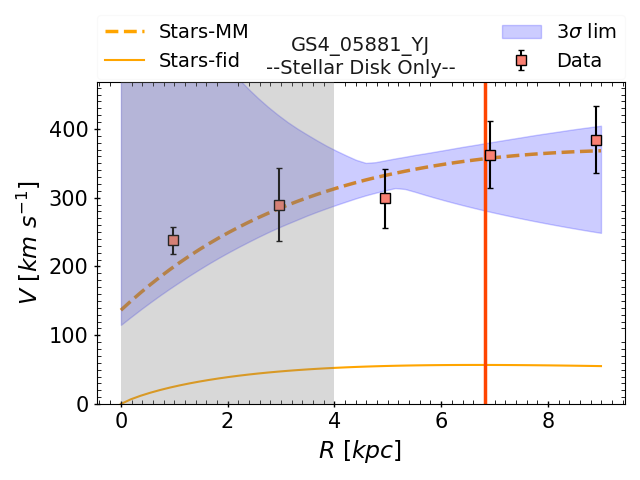}
						\includegraphics[angle=0,height=6.5truecm,width=8.0truecm]{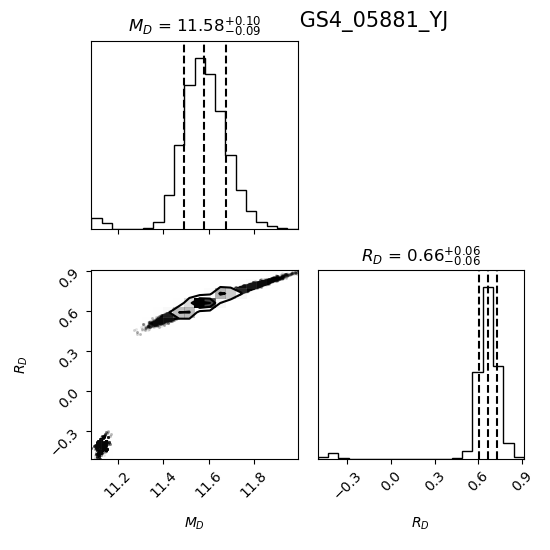}  
						
						\includegraphics[angle=0,height=6.0truecm,width=8.0truecm]{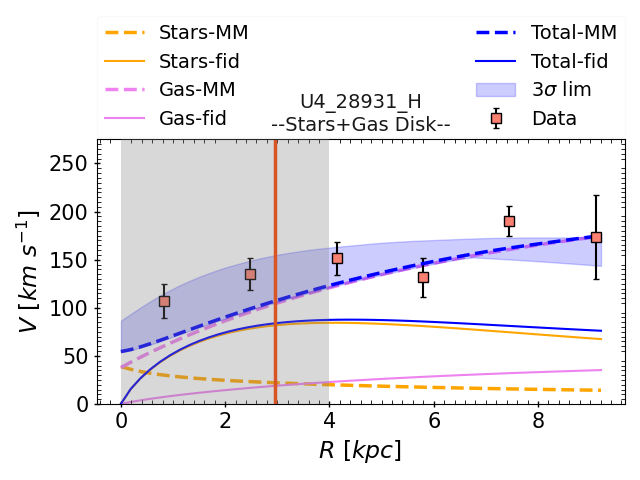}
						\includegraphics[angle=0,height=6.5truecm,width=8.0truecm]{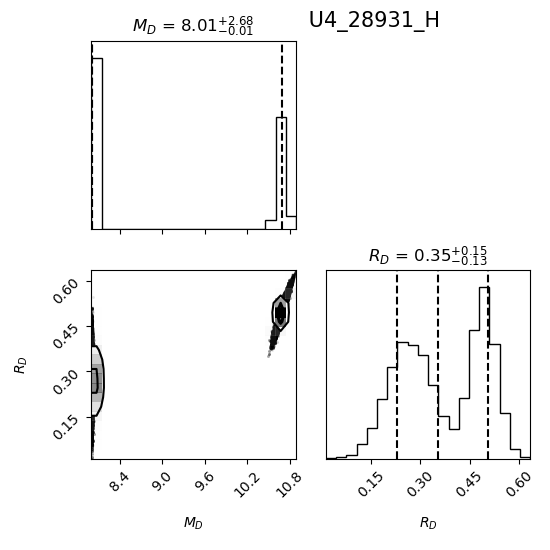}  
						
						\caption{Examples of mass modeled \RCs\ in the case of the maximum stellar disk scenario (upper panel) and maximum stellar+gas disk (lower panel). The best-fit parameters are shown in the right column, and color codes are given in the legend of the plot, where `MM' indicates dynamical mass-model profiles estimated using Bayesian inference and `fid' stands for the fiducial profile derived from a given photometric mass.  This example demonstrates that (1) the model is incapable of fitting the observations using only baryons, and (2) the best-fit values for stellar or gas mass are an order of magnitude higher than their fiducial values, which is unrealistic given the uncertainties in the observations. Therefore, we suggest that the dark matter halo component is required in \hz galaxies.}
						\label{fig:Mass-models-2}
					\end{center}
				\end{figure*}

				\section{Mass modeling of \RCs}\label{sec:mass-models}

				Initially, we assume the absence of dark matter halos around high-$z$ galaxies and dynamically model the baryonic component to fully fit the observed rotation curves. In this case, observed velocity is defined as
				\begin{equation}
					\label{eq:vtot}
					V^2_{\rm c}(R) = V^2_{\mathrm{stars}}(R) + V^2_{\mathrm{gas}}(R),
				\end{equation}
				where $V_{\mathrm{stars}}$ represents the contribution of stars in both the bulge and disk, while $V_{\mathrm{gas}}$ encompasses the combined effect of molecular and atomic gas components. It is important to note that in the maximum stellar-disk scenario, the second term on the right-hand side of the equation becomes null.
				
				We performed the dynamical modeling using a Bayesian inference technique established in \citet{GS22}, which employs flexible models. For Bayesian inference, we employed Markov chain Monte Carlo analysis, assuming a $\chi^2$ test statistic on observed kinematics and modeled kinematics. In our modeling approach, we assumed that the distribution of stars and gas follows an exponential mass profile \citep{Freeman}, while the bulge is treated as a point mass. It is worth noting that the bulge remains unresolved in all datasets. Moreover, we only fit the outer rotation curves; therefore, we fixed the bulge mass at $10^{9} \ M_\odot$ as a first-order approximation. To estimate the molecular and atomic gas masses, we use observed scaling relations, as described in Section~\ref{sec:Mbar}, where we fixed the star-formation rate at its fiducial value. This approach allowed us to obtain realistic gas masses for a given stellar mass and star-formation rate. In the modeling, we kept flat priors on the stellar disk mass ($8.0 \leq log(M_{\rm star} \ \mathrm{M_\odot}) \leq 12.0 $) and kept a Gaussian prior with a 25\% relative uncertainty for the stellar disk radius. A comprehensive discussion of this mass modeling technique can be found in \citet[][sec: 3.1]{GS22}.
				
				In Figures~\ref{fig:Mass-models-1} and~\ref{fig:Mass-models-2}, we provide a few examples of mass-modeled rotation curves (\RCs) to illustrate different scenarios. The upper panel of Figure~\ref{fig:Mass-models-1} depicts the maximum stellar disk scenario, while the lower panel represents the maximum stars+gas disk scenario. The corresponding best-fit parameters are displayed in the right column.  Figures~\ref{fig:Mass-models-1} demonstrates the robustness of our mass modeling technique. However, Figure~\ref{fig:Mass-models-2} reveals the limitations of using the same models, as we are unable to precisely constrain the stellar mass, whether considering the maximum stellar disk or the stars+gas disk scenario. In some cases depicted in the same figure (lower panel), the mass modeling allows for extremely high gas masses while suppressing the stellar masses to values lower than their observed photometric masses. Furthermore, we observe that the majority of the sample exhibits best-fit stellar masses that are an order of magnitude higher than their fiducial values (see Figure~\ref{fig:MM-best-param}). These discrepancies are both alarming and unrealistic, indicating the probable necessity of an additional halo component in these galaxies.

				\section{Halo model independent $f_{\mathrm{_{DM}}}$} \label{sec:method-fdmrac}
				
				In our previous work \citet{GS21b}, we established a halo model independent framework to study the dark matter fraction of high-z. In this framework, first we compute the total dynamical mass of galaxy directly from its rotation curve. The dynamical mass of a galaxy is defined as 
				\begin{equation}
					\label{eq:Mdyn}
					M_{\rm dyn}(<R) = \kappa(\mathrm{{\tiny R} }) \frac{V^2(R) \ R}{G} 
					,\end{equation}
				where $V(R)$ is the circular velocity computed at radius $R$ and $\kappa$({\tiny R}) is the geometric factor accounting for the presence of a stellar  disk alongside  the  spherical bulge and  halo  \citep[see,][]{Persic1990}. For $R_{\rm e}$, $R_{\rm opt}$, and $R_{\rm out}$, the values of $\kappa$ are 1.2, 1.05, and 1.0, respectively (see also \citealt{GS21b}).  The bulge mass contribution within 5 kpc is negligible for local spirals; therefore, we do not model it. However, the geometric parameter $\kappa$({\tiny R}) in the dynamical mass calculation takes into account the distribution of the mass located in the bulge and in the disk.	
				
				Second, we disentangled the contribution of total baryonic mass from dynamical mass.\footnote{Total baryonic mass is $M_{\rm bar} = M_* + \ M_{\rm H_{2}} + 1.33 M_{\rm HI}$, where factor 1.33 accounts the Helium content. } Briefly, we know that our samples are fair representative of main-sequence of star-forming galaxies as shown in Figure~\ref{fig:Mstar-sfr}, which allowed us to compute their molecular and atomic masses using scaling relations \citep[][respectively]{Tacconi2018, Chowdhury2022, Lagos2011}.  Our goal is to calculate the \dm\ fraction of our galaxies within $R_{\rm e}, \ R_{\rm opt}$, and $R_{\rm out}$, and therefore we must first determine the baryonic masses within these radii. In  \citep{GS21a}, we found that \RCs\ of $z\sim 1$ galaxies are  similar to local star-forming disk galaxies. This suggests that the radial distribution of stellar and molecular gas masses within these galaxies can be well approximated by the Freeman disk \citep{Freeman}:
				\begin{equation}
					\label{eq:FMs}
					\begin{array}{r@{}l}
						V_{_{\rm star}}^2(R)= \frac{1}{2} \Big( \frac{GM_{_{\rm star}}}{R_{_{\rm star}}} \Big) \  \Big( \alpha \frac{R}{R_{_{\rm star}}} \Big)^2 \  [I_0K_0 - I_1K_1], \\ \\
						
						V_{_{\rm H2}}^2(R)= \frac{1}{2} \Big( \frac{GM_{_{\rm H2}}}{R_{_{\rm H2}}} \Big) \  \Big( \alpha \frac{R}{R_{_{\rm H2}}} \Big)^2 \  [I_0K_0 - I_1K_1],\\ \\
						
						V_{_{\rm HI}}^2(R)= \frac{1}{2} \Big( \frac{GM_{_{\rm HI}}}{R_{_{\rm HI}}} \Big) \  \Big( \alpha \frac{R}{R_{_{\rm HI}}} \Big)^2 \  [I_0K_0 - I_1K_1],
					\end{array}
				\end{equation}
				where $M_{\_}$ and $R_{\rm scale}$ are the total mass and the scale length of the different components (stars, H2, and HI), respectively, and $I_n$ and $K_n$ are modified Bessel functions computed at $\alpha=1.6$ for stars and $\alpha = 0.53$ for gas \citep[c.f.][]{PS1996, Karukes2017}. In this scenario, stars are assumed to be distributed in the stellar disk ($R_{*} = R_{\rm D}$) known from photometry, discussed in Section [\ref{sec:dataset}]. The molecular gas is generally distributed outward through the stellar disk (up to the length of the ionized gas $R_{\rm gas}$); therefore, we take $R_{\rm H2} \equiv R_{\rm gas}$. Here, we estimate the gas scale length $R_{\rm gas}$ by fitting the $H_\alpha$ surface brightness, see \citet[][see appendix-C]{GS21b}.  Moreover, studies of local disk galaxies have shown that the surface brightness of the HI disk is much more extended than that of the H2 disk \citep[][see their Fig. 5]{Jian2010}; see also \cite{Leroy2008} and \citet{Cormier2016}. Therefore, we assume $R_{\rm HI}=2 \times R_{\rm H2}$, which is a rough estimate, but still reasonable, considering that at high-$z$ no information is available on the $M_{\rm HI }$ (or $M_{\rm H2}$) surface brightness distribution. Thus, the Equation [\ref{eq:FMs}] allowed us to estimate $M_{*}$, $M_{\rm HI}$, and $M_{\rm H2}$ within different radii ($R_{\rm e}, \ R_{\rm opt}$, and $R_{\rm out}$) using spherical symmetry ($M (<R) =  V^2 \times R / G$).
				
				Given the information on bayonic and dynamical masses, the \dm\ fractions within radius $R$ can be computed as
				\begin{equation}
					\label{eq:fdm}
					f_{_{\rm DM}} (<R)= 1- \frac{M_{\rm bar} (<R)}{M_{dyn} (<R)}.
				\end{equation}
				Therefore, using Equation[\ref{eq:fdm}], we computed $f_{_{\rm DM} }$ within $R_{\rm e}, \ R_{\rm opt}$, and $ R_{\rm out}$ for all three datasets. We note that, owing to the limited spatial resolution in our \RCs, the measurements of $f_{_{\rm DM}}$ within $R_{\rm e}$ are less accurate than for $R_{\rm opt}$ and $R_{\rm out}$.

\begin{figure}
					\begin{center}
						\includegraphics[angle=0,height=6.5truecm,width=9.0truecm]{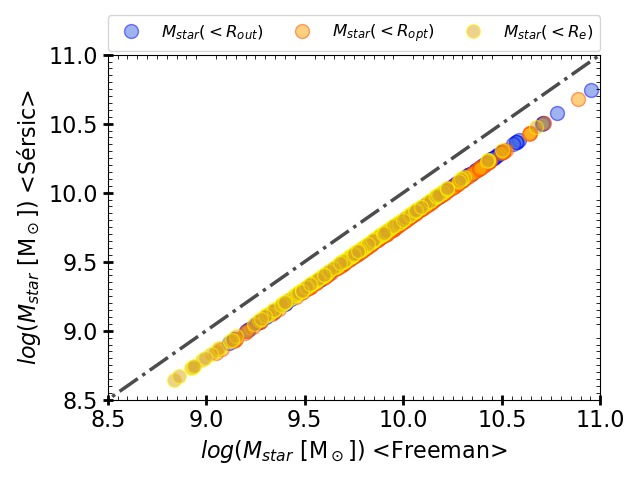} 

						\caption{Comparison of stellar masses derived using S$\acute{e}$rsic profile (y-axis) and Freeman disk (x-axis). Different color code are representing the different scale radii of galaxies within which stellar masses are estimated (blue: $R_{\rm out}$, orange: $R_{\rm opt}$, and yellow: $R_{\rm e}$), and black dotted-dashed line represents the one-to-one relation. 
						}
						\label{fig:Mstar-sersic}
					\end{center}
				\end{figure}

				The stellar disk in \hz\ galaxies is often unresolved, making it difficult to determine whether the disk follows an exponentially thin profile \citep[well described by][]{Freeman} or a thicker structure better modeled by a S$\acute{e}$rsic profile \citep{sersic}. To address this uncertainty, we estimate the stellar mass within a characteristic scale radius (e.g., $R_{\rm out}$) using both the S$\acute{e}$rsic profile \citep{sersic} and the Freeman disk \citep{Freeman}. We adopt a S$\acute{e}$rsic index of $n=1.35$, which represents the average value for KROSS sample. The results are presented in Figure~\ref{fig:Mstar-sersic}. We find that assuming a Freeman disk yields stellar masses that are, on average, 1.02 times ($\approx 0.21$ dex) higher than those obtained using the S$\acute{e}$rsic profile. In ideal cases where dark matter dominates the total mass, this results in only a minor decrease in the estimated dark matter fraction.  
				
				However, a significant change in $f_{\rm DM}$ would occur if the stellar or baryonic mass were increased by a much larger fraction of the total mass, by an order of magnitude (0.7–1 dex, i.e., a factor of five to ten), which is seen in some low dark matter fraction galaxies in our sample. This is most likely due to a high gas mass fraction or other systematic effects discussed in Section~\ref{sec:caveats}. Conversely, adopting a S$\acute{e}$rsic profile with $n>1.5$ would systematically shift the low dark matter fraction galaxies toward higher $f_{\rm DM}$ values, especially, in the outskirts. However, for the majority of KMOS3D and KGES galaxies, published S$\acute{e}$rsic indices are not available. Therefore, following our previous studies \citep{GS21b, GS22} and seen obvious disk morphology in high-resolution images of our sample, we assume that the stars (and gas) are well distributed according to the Freeman disk \citep{Freeman}.

				\section{Extra}
				\label{sec:extra1}
				
				\begin{figure}
					\begin{center}
						\includegraphics[angle=0,height=6.5truecm,width=8.5truecm]{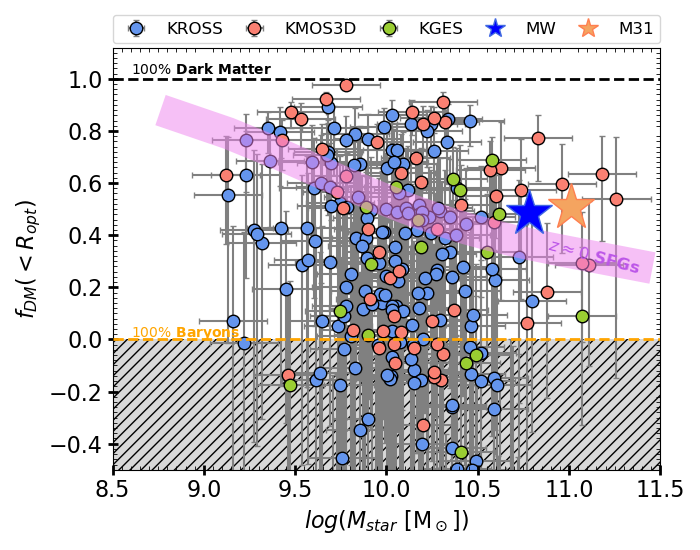}  
						
						\caption{Dark matter fraction within $R_{\rm opt}$.  The KMOS$^{\rm 3D}$, KGES, and KROSS datasets are represented in red, green, and blue colors, respectively. For comparison, we include two local massive disk galaxies: the Milky Way (MW) and Andromeda (M31), represented by the blue and the orange star, respectively. The solid black dashed line represents the 100\% dark matter regime, while the yellow dashed line represents the baryon-dominated regime. The gray shaded area indicates the forbidden region, where galaxies with $M_{dyn}< M_{\rm bar}$ are located.
						}
						\label{fig:FdmRopt}
					\end{center}
				\end{figure}
				
				\paragraph{Dark matter fraction within $R_{\rm opt}$: } In Figure~\ref{fig:FdmRopt}, we plot the dark matter fraction within $R_{\rm opt}$  ($\sim 2 \ R_{\rm e}$) of galaxies as function of stellar mass. Firstly, we observe that the majority of the sample exhibits dark matter-dominated outer disks, with 23\% of objects showing $0.2 \leq f_{_{\rm DM}} (<R_{\rm opt}) <0.5$, and 36\% of objects having $f_{_{\rm DM}} (<R_{\rm opt})<0.2$ (including objects in the forbidden region). Secondly, we notice a slightly decreasing trend of $f_{_{\rm DM}}$ as a function of stellar mass, except for a few outliers. Lastly, we compared the \dm\ fraction within $R_{\rm opt}$ and $R_{\rm out}$. Interestingly, we observed that galaxies generally exhibit slightly lower dark matter fractions in the inner region ($R_{\rm opt}$) compared to outskirts ($R_{\rm out}$). On average,  \dm\ fraction within $R_{\rm opt}$ is about $65\%$, while it is $\sim 75-80\%$ within $R_{\rm out}$, these finding coincides with the results reported in \citet{GS21b}.
				
				\begin{figure*}
					\begin{center} 
						\includegraphics[angle=0,height=6.5truecm,width=18.0truecm]{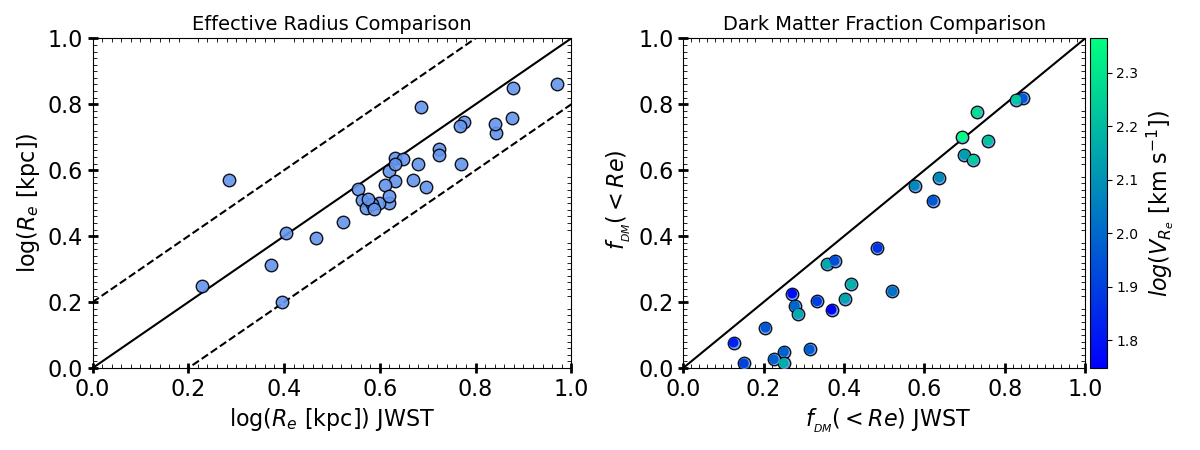}  
						
						\caption{{\em Left panel:} Comparison of previously measured $R_{\rm e}$ values of the KROSS sub-sample with the new JWST  $R_{\rm e}$. The solid black line represents the one-to-one relation, while the dashed black lines indicate a scatter of 0.2 dex. {\em Right panel:} Dark matter fraction within $R_{\rm e}$ using JWST photometry. Objects are color-coded based for their circular velocities within $R_{\rm e}$. } 
						\label{fig:FdmRe-JWST}
					\end{center}
				\end{figure*}
				
				\begin{figure*}
					\begin{center}		
						\includegraphics[angle=0,height=2.0truecm,width=2.5truecm]{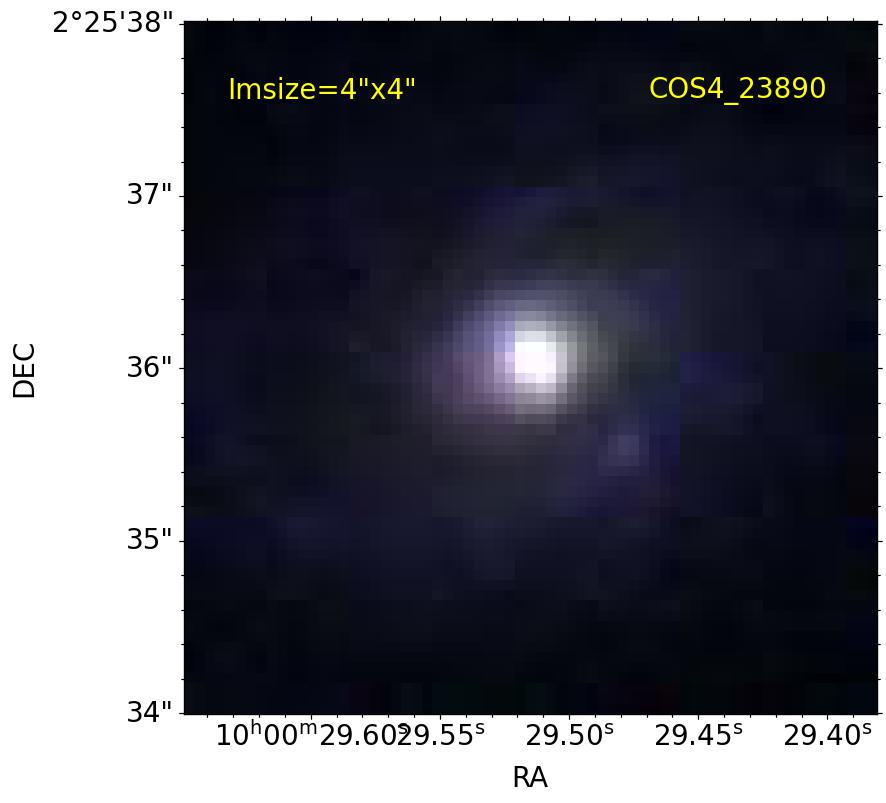}  
						\includegraphics[angle=0,height=2.0truecm,width=2.5truecm]{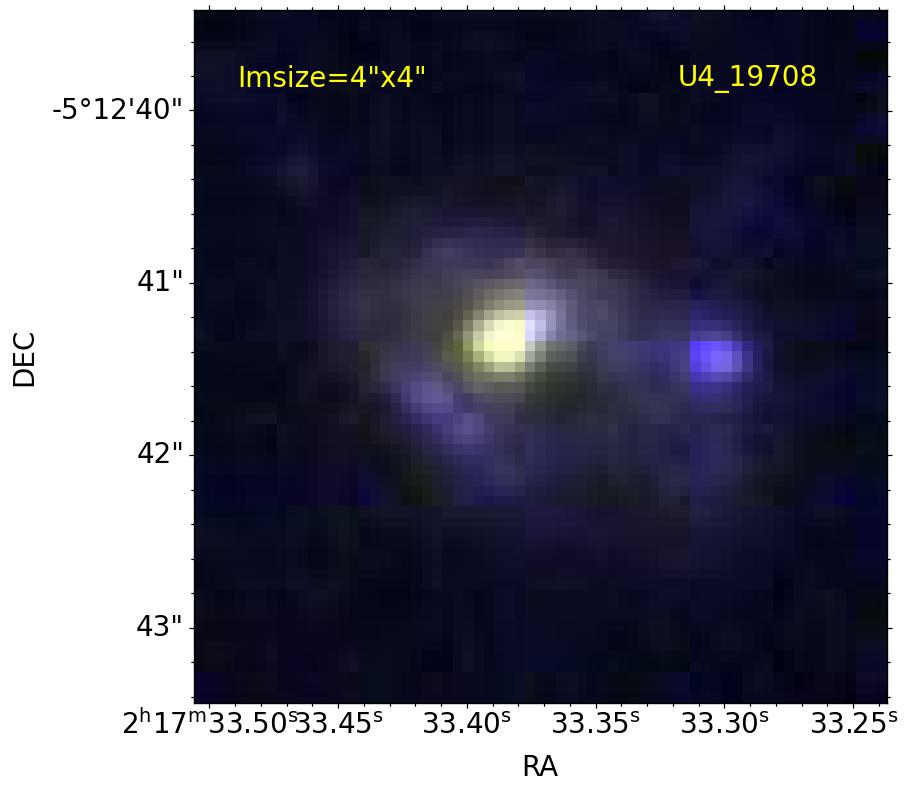}  
						\includegraphics[angle=0,height=2.0truecm,width=2.5truecm]{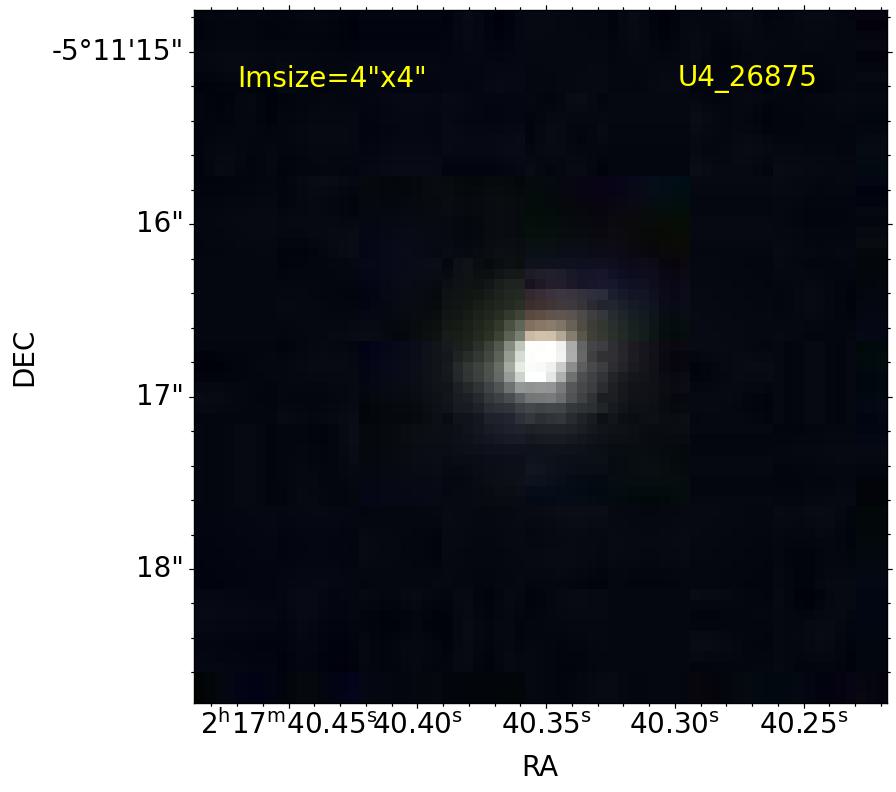}  
						\includegraphics[angle=0,height=2.0truecm,width=2.5truecm]{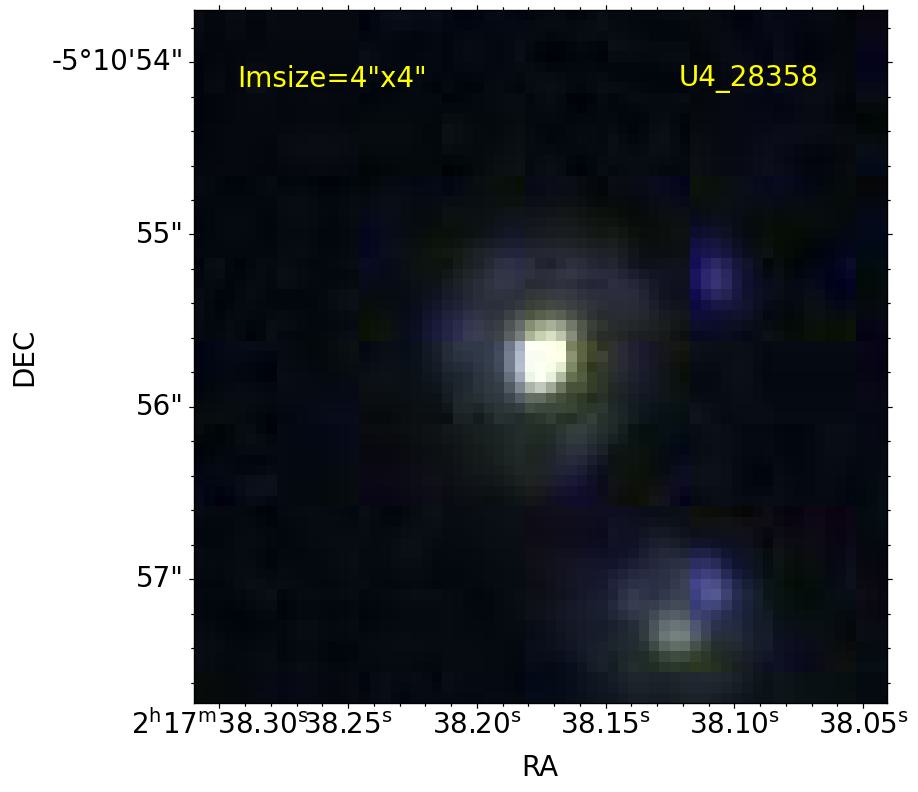}  
						\includegraphics[angle=0,height=2.0truecm,width=2.5truecm]{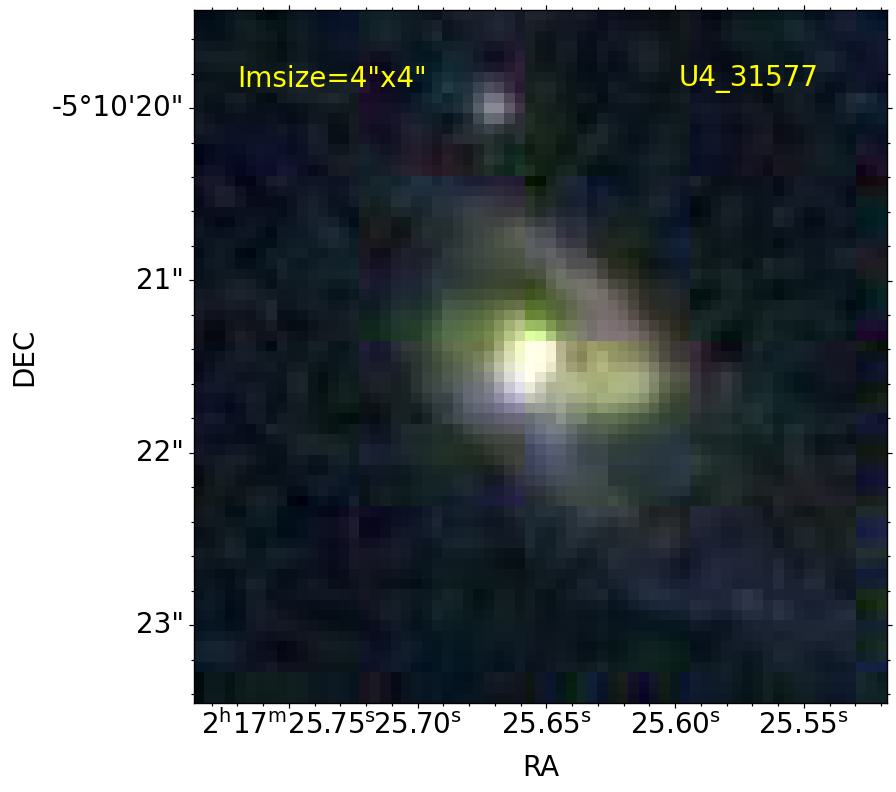}  
						\includegraphics[angle=0,height=2.0truecm,width=2.5truecm]{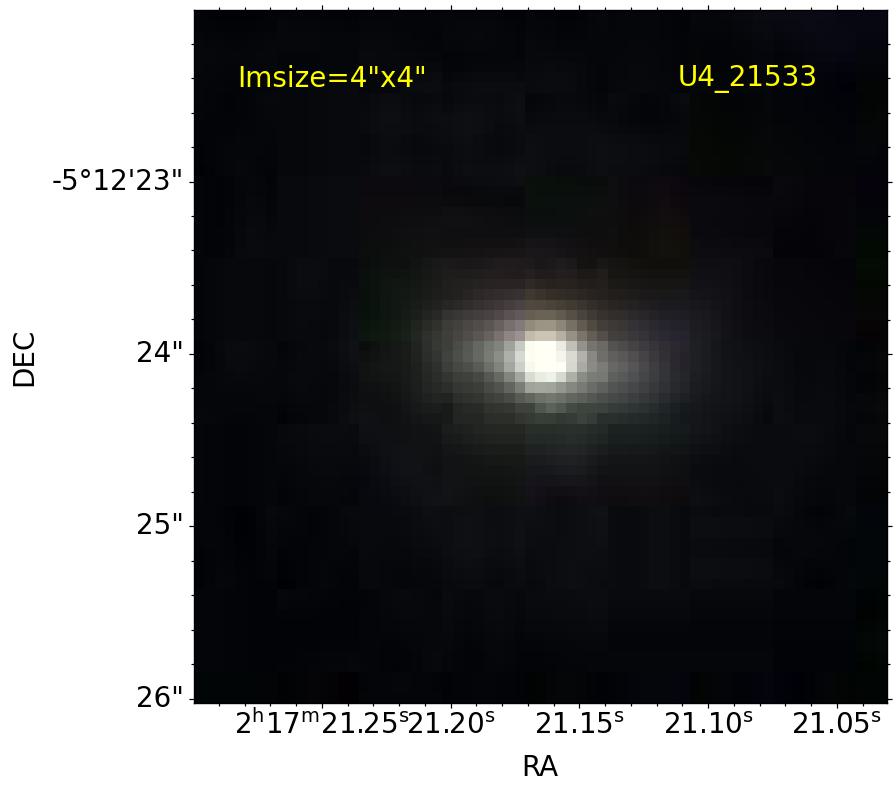}

						\includegraphics[angle=0,height=2.0truecm,width=2.5truecm]{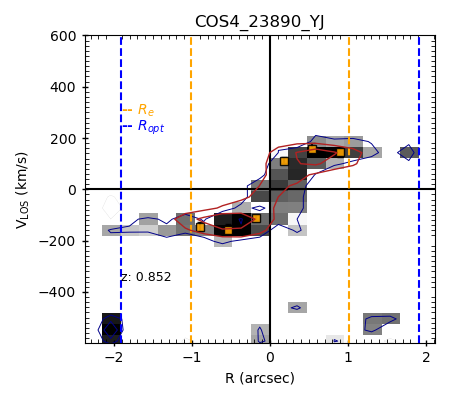}  
						\includegraphics[angle=0,height=2.0truecm,width=2.5truecm]{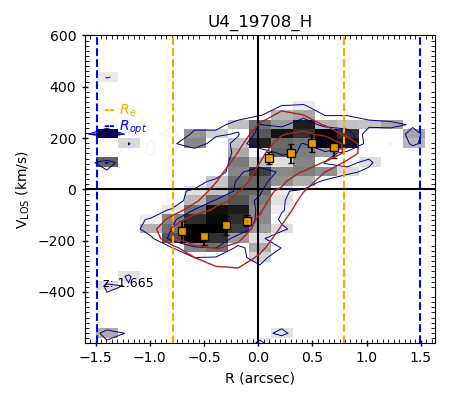}  
						\includegraphics[angle=0,height=2.0truecm,width=2.5truecm]{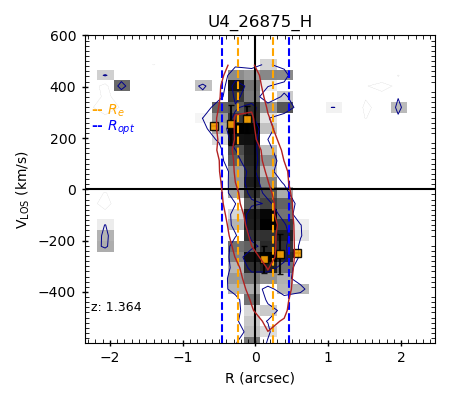}  
						\includegraphics[angle=0,height=2.0truecm,width=2.5truecm]{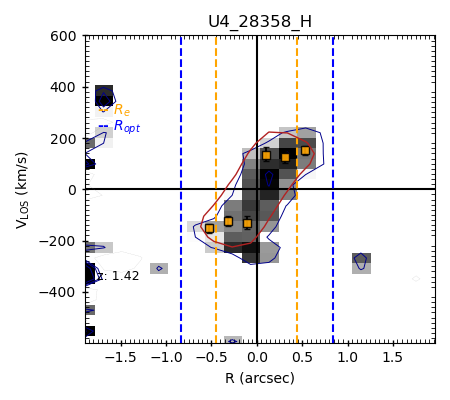}  
						\includegraphics[angle=0,height=2.0truecm,width=2.5truecm]{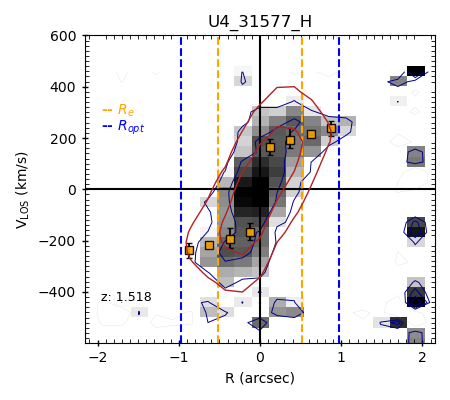}  
						\includegraphics[angle=0,height=2.0truecm,width=2.5truecm]{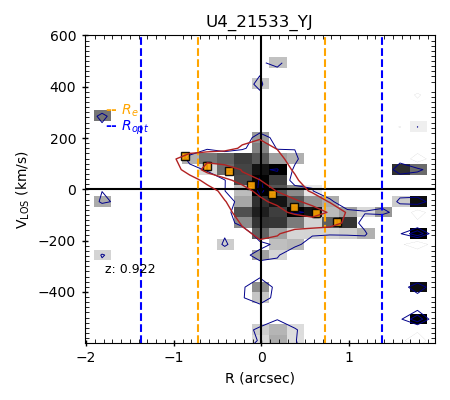}   
						
						\caption{Hubble Space Telescope images and RCs of massive galaxies containing high dark matter fraction.  That is, galaxies those are standing odd in $f_{_{\rm DM}} - M_{\rm star} $ relation.}
						\label{fig:massive_gal}
					\end{center}
				\end{figure*}
				
				\paragraph{Improved $R_e$ measurements:} We had the opportunity to improve the measurement of $R_{\rm e}$ for a subset of KROSS sample through the latest observations from the James Webb Space Telescope (JWST). This particular subset of galaxies is located within the COSMOS field \citep{Skelton2014}, which has recently been observed by the COSMOS-WEB team \citep{COSMOS-Web}. For these targets JWST/NIRCam sizes in the filters F115W, F150W, F277W and F444W were recovered via S$\acute{e}$risc profile fitting using GALFITM \citep{GALFITM2013} with the same methodology outlined in \citet{Martorano2023} on mosaics publicly available in the Dawn JWST Archive\footnote{ https://dawn-cph.github.io/dja/index.html}\citep{Valentino2023}. In this study, we employ the $R_{\rm e}$ values obtained from the F115W band, which closely corresponds to the rest-frame $H\alpha$-wavelength observed at $z\sim0.85$. 

				In Figure~\ref{fig:FdmRe-JWST}, the left panel illustrates the comparison between the new JWST-derived $R_{\rm e}$ and the previously measured $R_{\rm e}$ adopted from KROSS parent catalog. We observe that the JWST $R_{\rm e}$ values are higher than the previous measurements but remain within a scatter of 0.2 dex. Although the difference is small, these newly improved $R_{\rm e}$ measurements are crucial for investigating the dark matter fraction, especially within $R_{\rm e}$. Therefore, we have estimated the dark matter fraction using the updated $R_{\rm e}$ values, as shown in the right panel of Figure~\ref{fig:FdmRe-JWST}. We observe that galaxies with low dark matter fractions (<20\%) have been pushed toward higher values. In this specific sub-sample, none of the galaxies have a dark matter fraction below $20\%$.


			\end{appendix}

		\end{document}